\documentclass[preprint,12pt,authoryear, table]{elsarticle}

\usepackage{amssymb}
\usepackage{amsmath}
\usepackage{textcomp}
\usepackage{stfloats}
\usepackage{verbatim}
\usepackage{graphicx}
\usepackage{booktabs}
\usepackage{multirow}
\usepackage{caption} 
\usepackage{etoolbox,siunitx}
\usepackage{url}
\usepackage{tabularray}
\usepackage{longtable}
\usepackage{hyperref}
\usepackage{threeparttable}
\usepackage{tikz}
\usetikzlibrary{fit,tikzmark,calc}
\usepackage{pifont}
\newcommand{\cmark}{\textcolor{green!80!black}{\ding{51}}}
\newcommand{\xmark}{\textcolor{red}{\ding{55}}}
\usepackage[nolist]{acronym}
\usepackage{pdflscape}
\usepackage{amsmath}

\UseTblrLibrary{booktabs}

\newacro{MVDR}[MVDR]{Minimum Variance Distortion-less Response}
\newacro{GEV}[GEV]{Generalized Eigenvalue Decomposition}
\newacro{MCWF}[MCWF]{Multi-Channel Wiener Filter}
\newacro{SDW-MWF}[SDW-MWF]{Speech Distortion Weighted MWF}
\newacro{SISDR}[SI-SDR]{Scale Invariant Signal-to-Distortion Ratio}
\newacro{STFT}[STFT]{Short-Time Fourier Transform}
\newacro{DNN}[DNN]{Deep Neural Network}
\newacro{SNR}[SNR]{Signal-to-Noise-Ratio}
\newacro{FLOPs}[FLOPs]{Floating Point Operations}
\newacro{SIR}[SIR]{Signal-to-Interference Ratio}
\newacro{ASR}[ASR]{Automatic Speech Recognition}
\newacro{IBM}[IBM]{Ideal Binary Mask}
\newacro{IRM}[IRM]{Ideal Ratio Mask}
\newacro{WLM}[WLM]{Wiener-Like Mask}
\newacro{SDR}[SDR]{Signal-to-Distortion Ratio}
\newacro{STOI}[STOI]{Short-time Objective Intelligibility}
\newacro{MSE}[MSE]{Mean-Squared Error}
\newacro{WER}[WER]{Word Error Rate}
\newacro{WPE}[WPE]{Weighted Prediction Error}
\newacro{MISO}[MISO]{Multiple Input Single Output}
\newacro{DFL}[DFL]{Deep Feature Loss}
\newacro{DASR}[DASR]{distant automatic speech recognition}

\usepackage{xcolor}
\definecolor{maroon}{HTML}{F26035}
\definecolor{yellow}{HTML}{FDBC42}
\definecolor{darkred}{RGB}{156, 39, 33}
\definecolor{darkblue}{RGB}{31, 90, 153}
\definecolor{forestgreen}{rgb}{0.13, 0.55, 0.13}
\definecolor{olmoDarkBlue}{HTML}{012e59}
\definecolor{olmoBlue}{HTML}{265ed4}
\definecolor{olmoLightBlue}{HTML}{012e59}
\definecolor{olmoTeal}{HTML}{00d5ff}
\definecolor{olmoYellow}{HTML}{ffbb00}
\definecolor{olmoOrange}{HTML}{ff9100}
\definecolor{maroon}{HTML}{F26035}
\definecolor{yellow}{HTML}{FDBC42}
\definecolor{lavender}{HTML}{734f96}
\definecolor{darkergrey}{HTML}{444444}
\definecolor{midgrey}{HTML}{e6eded}
\definecolor{ai2pink}{HTML}{f0529c}%
\definecolor{ai2midpink}{HTML}{fad3e5}
\definecolor{ai2lightpink}{HTML}{fbecf3}
\definecolor{ai2midwhite}{HTML}{f2e5d9}
\definecolor{ai2offwhite}{HTML}{fbf4ee}
\definecolor{ai2green}{HTML}{0fcb8c}
\definecolor{ai2lightgreen}{HTML}{e7f9f3}
\definecolor{ai2darkgreen}{HTML}{105257}
\definecolor{ai2purple}{HTML}{B932EB}
\definecolor{ai2lightpurple}{HTML}{f7e8fc}
\definecolor{neutralEight}{HTML}{343434}
\definecolor{neutralFive}{HTML}{838383}
\definecolor{neutralThree}{HTML}{bebebe}
\definecolor{neutralOne}{HTML}{dedede}
\definecolor{lightgrey}{HTML}{fafcfc}

\usepackage{multirow, multicol, makecell}
\usepackage{adjustbox}
\usepackage{listings}
\usepackage{textcomp}

\definecolor{codegreen}{rgb}{0,0.6,0}
\definecolor{codegray}{rgb}{0.5,0.5,0.5}
\definecolor{codepurple}{rgb}{0.58,0,0.82}
\definecolor{backcolour}{rgb}{0.95,0.95,0.92}

\usepackage{inconsolata}
\lstdefinestyle{mystyle}{
    backgroundcolor=\color{backcolour},   
    commentstyle=\color{codegreen},
    keywordstyle=\color{magenta},
    numberstyle=\tiny\color{codegray},
    stringstyle=\color{codepurple},
    basicstyle=\ttfamily\footnotesize,
    breakatwhitespace=false,         
    breaklines=true,                 
    captionpos=b,                    
    keepspaces=true,                 
    numbers=left,                    
    numbersep=5pt,                  
    showspaces=false,                
    showstringspaces=false,
    xleftmargin=14pt,
    framexleftmargin=14pt,
    showtabs=false,                  
    tabsize=2
}

\colorlet{punct}{red!60!black}
\definecolor{background}{HTML}{EEEEEE}
\definecolor{delim}{RGB}{20,105,176}
\colorlet{numb}{magenta!60!black}

\lstdefinelanguage{json}{
    basicstyle=\normalfont\ttfamily,
    numbers=left,
    numberstyle=\tiny\color{codegray},
    xleftmargin=14pt,
    framexleftmargin=14pt,
    stepnumber=1,
    numbersep=8pt,
    showstringspaces=false,
    breaklines=true,
    frame=lines,
    backgroundcolor=\color{backcolour},
    literate=
     *{0}{{{\color{numb}0}}}{1}
      {1}{{{\color{numb}1}}}{1}
      {2}{{{\color{numb}2}}}{1}
      {3}{{{\color{numb}3}}}{1}
      {4}{{{\color{numb}4}}}{1}
      {5}{{{\color{numb}5}}}{1}
      {6}{{{\color{numb}6}}}{1}
      {7}{{{\color{numb}7}}}{1}
      {8}{{{\color{numb}8}}}{1}
      {9}{{{\color{numb}9}}}{1}
      {:}{{{\color{punct}{:}}}}{1}
      {,}{{{\color{punct}{,}}}}{1}
      {\{}{{{\color{delim}{\{}}}}{1}
      {\}}{{{\color{delim}{\}}}}}{1}
      {[}{{{\color{delim}{[}}}}{1}
      {]}{{{\color{delim}{]}}}}{1},
}

\usepackage{rotating}

\usepackage[many]{tcolorbox}
\tcbuselibrary{listings}

\lstdefinelanguage{none}{}{}

\newtcblisting{gbox}[1][lightgray]{
  colback=#1,
  listing only,
  listing options={language=none,basicstyle=\ttfamily,columns=flexible},
}

\journal{Computer Speech and Language}

\begin{document}

\begin{frontmatter}



\title{Recent Trends in Distant Conversational Speech Recognition: A Review of CH\-iME-7 and 8 DASR Challenges} %% Article title

% Recent Advances (Trends) in Distant Conversational Speech Recognition: A comprehensive review of CH\-iME-7 and 8 DASR Challenges
% 


\author[1]{Samuele Cornell}
\author[2]{Christoph Boeddeker}
\author[3]{Taejin Park}
\author[3]{He Huang}
\author[4]{Desh Raj}
\author[5]{Matthew Wiesner}
\author[6]{Yoshiki Masuyama}
\author[1]{Xuankai Chang}
\author[8]{Zhong-Qiu Wang}
\author[9]{Stefano Squartini}
\author[5]{Paola Garcia}
\author[1]{Shinji Watanabe}
%% Author name

\affiliation[1]{
    organization={
            Carnegie Mellon University},
            country={USA},
}

\affiliation[2]{
    organization={
            Paderborn University},
            country={Germany},
}

\affiliation[3]{
    organization={
            NVIDIA},
            country={USA},
}

\affiliation[4]{
    organization={
            Meta},
            country={USA},
}

\affiliation[5]{
    organization={
            Johns Hopkins University},
            country={USA},
}

\affiliation[6]{
    organization={
            Mitsubishi Electric Research Laboratories},
            country={USA},
}

\affiliation[8]{
    organization={
            Southern University of Science and Technology},
            country={China},
}

\affiliation[9]{
    organization={
            Università Politecnica delle Marche},
            country={Italy},
}

%% Abstract around 250 words is ok
\begin{abstract}
The CH\-iME-7 and 8 distant speech recognition (DASR) challenges focus on multi-channel, generalizable, joint automatic speech recognition (ASR) and diarization of conversational speech. 
With participation from 9 teams submitting 32 diverse systems, these challenges have contributed to state-of-the-art research in the field.
This paper outlines the challenges' design, evaluation metrics, datasets, and baseline systems while analyzing key trends from participant submissions.
From this analysis it emerges that:
1) Most participants use end-to-end (e2e) ASR systems, whereas hybrid systems were prevalent in previous CH\-iME challenges. This transition is mainly due to the availability of robust large-scale pre-trained models, which lowers the data burden for e2e-ASR.
2) Despite recent advances in neural speech separation and enhancement (SSE), all teams still heavily rely on guided source separation, suggesting that current neural SSE techniques are still unable to reliably deal with complex scenarios and  different recording setups. 
3) All best systems employ diarization refinement via target-speaker diarization techniques. Accurate speaker counting in the first diarization pass is thus crucial to avoid compounding errors and CHiME-8 DASR participants especially focused on this part. 
4) Downstream evaluation via meeting summarization can correlate weakly with transcription quality due to the remarkable effectiveness of large-language models in handling errors. On the NOTSOFAR-1 scenario, even systems with over 50\% time-constrained minimum permutation WER can perform roughly on par with the most effective ones (around 11\%).
5) Despite recent progress, accurately transcribing spontaneous speech in challenging acoustic environments remains difficult, even when using computationally intensive system ensembles.

\end{abstract}

\end{frontmatter}



%%%%%%%%%%%%%%%%%%%%%%
\section{Introduction}
\label{sec:intro}
%%%%%%%%%%%%%%%%%%%%%%

Today, current state-of-the-art (SotA) automatic speech recognition (ASR) systems are well known to be capable of obtaining performance on par or superior to humans on several widely used benchmark datasets~\citep{godfrey1992switchboard, paul1992design, panayotov2015librispeech} featuring close-talk single speaker speech, mainly from telephone or audiobook recordings.
% However, tackling far-field conversational speech remains an arduous problem. For example, even in the AMI dataset~\citep{carletta2005ami}, which is now more than 20 years old and features an office meeting scenario with moderate complexity, the state-of-the-art results in terms of speaker-attributed word error rate (WER) are still higher than $\sim$20\%~\citep{kanda2022transcribe, Cornell2024icassp, park2024sortformer} with no oracle diarization.
% Yoshiki repolaces the above sentences.
However, far-field conversational speech recognition remains an arduous problem. For example, even in the AMI dataset~\citep{carletta2005ami}, developed in 2005 and featuring an office meeting scenario, SotA results in speaker-attributed word error rate (WER) are still above $\sim$20\%~\citep{kanda2022transcribe, Cornell2024icassp, park2024sortformer} without oracle diarization. 

As such recent work~\citep{szymanski2020we} has raised doubts if such widely used benchmark datasets are still adequate today, calling for a more comprehensive approach that can also, importantly, assess generalization capability instead of focusing only on one domain of interest. 
% If benchmark datasets from a narrow sets of application scenarios are always used for evaluation, such procedure will inevitably favor techniques with low bias, but with high variance, i.e. more prone to break under domain shifts in inference.
% Yoshiki repolaces the above sentences.
If benchmark evaluation persistently focuses on a narrow, fixed set of scenarios/domains, it inadvertently promotes techniques that are tailored to such domains, potentially compromising their robustness to domain shifts and their real-world effectiveness. 
%If benchmark evaluation persistently focuses on narrow scenarios, the technical advancements will inherently drive into techniques dedicated to the specific application and reduce robustness against domain shifts.
% evaluation protocol ? oracle diarization
%In fact, we can argue this is a problem as old as statistics itself, intimately tied to the bias-variance trade-off: if benchmark datasets from a narrow sets of application scenarios are always used for evaluation, such procedure will favor techniques with low bias, but with high variance, i.e. more prone to break under domain shifts in inference. 
% Yoshiki repolaces the sentences.
In fact, in practical applications, domain shifts are inevitable due to the inherent unpredictability of real-world conditions and human behavior.
Moreover, flexibility is often highly desired, e.g., being able to deal with different deployment platforms and services, recording hardware, and so on. %Consequently, some techniques designed for specific benchmark datasets could not be optimal for real-world applications, where robustness and flexibility are among the most important features. 
%Moreover, it is also desirable , from a commercial point of view, of being able to have a flexible transcription system.  
% This problem is amplified for distant ASR (DASR) of meeting scenarios as the number of possible factors of variation increases compared to close-talk ASR.
% In detail, multi-party conversations introduce additional factors including recording setup (e.g. the array configuration and having a single or multiple devices), the number of speakers and their behavior (mainly static or can move significantly), the acoustic scenario (noise and reverberation), the setting (informal vs. formal), and so on.
% These factors are typically uncontrollable and also affect the speaking style as well as the rate for paralinguistic phenomena such as interruptions, fillers, stutters, repairs, and so on. 
% These factors, in turn, contribute to overlapping speech, which requires specialized techniques to address effectively.
% Yoshiki repolaces the above sentences.
This issue is more pronounced for distant ASR (DASR) for meeting scenarios, where the number of variability factors is significantly higher compared to close-talk ASR. 
 In addition to acoustic conditions (e.g., variations in noise and reverberation), multi-party conversations~\citep{li2014overview, haeb2019speech} introduce complexities in recording setups (e.g., array configurations and single versus multiple devices), speaker dynamics (e.g., static versus mobile participants), and contextual settings (e.g., informal versus formal environments). 
 These factors are typically uncontrollable and also affect the speaking style and the rate of paralinguistic phenomena such as interruptions, fillers, stutters, repairs, backchannel responses and so on which in turn contribute to overlapped speech.
%as the number of possible factors of variation increases compared to close-talk ASR. Aside from the acoustic environment (e.g. less or more noise and reverberation), multi-party conversations have additional factors in the recording setup (e.g., array configuration or the use of single versus multiple devices), the number of speakers as well as their behavior (mainly static or can move significantly) and the setting (informal versus formal). These factors are typically uncontrollable and also affect the speaking style and the rate of paralinguistic phenomena such as interruptions, fillers, stutters, repairs, and so on which contribute to the presence of overlapped speech .
This diversity of conditions, and the presence of overlapped speech, necessitate the need for specialized techniques, complicating the development of DASR systems for multi-speaker scenarios.
%Moreover, in conversational/meeting scenarios, precise utterance segmentation is a challenging task by itself. This is again different from, again, virtual assistant applications where keyword spotting can be employed, and it is known when the virtual assistant is talking as well as its content, thus simplifying dealing with overlapped speech. These segmentation errors can have a significant effect on the ASR component and this phenomenon is often overlooked: e.g., in the aforementioned benchmark datasets, oracle segmentation is considered in most of the literature, leading to optimistic results. To make things worse, most applications often require speaker-attribution of the words, thus needing diarization on top of ASR, adding another layer of complexity and difficulty. 
%Again, this latter is also not needed in dyadic conversations with virtual assistants. 
For example, even reliable segmentation of speech in conversational and meeting scenarios itself presents a challenge.
Segmentation errors in these contexts can severely affect ASR performance~\citep{novitasari2022improving}, a phenomenon often overlooked in ``pure'' ASR research. Many widely used ASR benchmark datasets in the literature assume oracle segmentation~\citep{ardila2020common, panayotov2015librispeech, chen2021gigaspeech, wang2021voxpopuli, hernandez2018ted}, leading to overly optimistic result. Furthermore, most applications require speaker-attribution of the words, thus needing diarization together with ASR, adding another layer of significant complexity and difficulty. 
% maybe add DIHARD stuff here
Despite these challenges, we argue that addressing the complexities of spontaneous ``speech-in-the-wild'' data also offers unique opportunities. 
Due to its richness, spontaneous conversational data can arguably be considered the next frontier in speech processing, particularly for the development of better ``human-like'' conversational agents~\citep{defossez2024moshi, fang2024llama} that can engage in more nuanced, natural and context-aware communication with possibly multiple speakers.
After all, dialog modeling and transcription, much like ASR and text-to-speech (TTS), are closely intertwined problems that can be viewed as complementary and drive progress in both domains~\citep{defossez2024moshi, cornell2024generating}. %dialog modeling and dialog transcription are intimately related problems, and can be seen as dual like ASR and TTS.  
Successfully tackling conversational speech opens up novel applications in healthcare~\citep{zhang2021leveraging}, government, education, customer service, and beyond, where flexibility and robustness in transcription systems are critical. 
Moreover, with the rise of large language models (LLMs), new possibilities for speech-enabled machine interactions have emerged, such as meeting summarization and retrieval~\citep{zhong2021qmsum, fang2024llama} to name just two. 
This paves the way for speech-enabled assistants that can better integrate into dynamic real-world environments. 
\subsection{Previous challenges and datasets}
\label{sec:previous}
%%%%%%%%%%%%%%%%%%%%%%%%%%%%%%%%%%%%%%%%%%%%%%%%%

\begin{table}[]
    \centering
     \footnotesize
    \setlength{\tabcolsep}{0.01em}
      \caption{Notable robust ASR datasets and challenges since the 2000s. We categorize as ``real-world'' only datasets and challenges strictly featuring fully real-world recorded data.}
    \label{tab:challenges}
   \begin{tabular}{|ccccccc|}  
    \cline{2-7}
    \multicolumn{1}{c|}{} & \multirow{2}{*}{Year} & Real & Long & Multi & Far & Multi \\

     \multicolumn{1}{c|}{} &  & World & Form &  Speaker & Field  & Domain \\
    \hline
   % SwitchBoard~\citep{godfrey1992switchboard} & 1992 & \cmark & \cmark & \cmark & \xmark  & \xmark \\
\rowcolor{ai2offwhite}    Santa Barbara~\citep{du2000santa} & 2000 & \cmark & \cmark & \cmark  & \cmark  & \cmark \\
    Aurora2-6~\citep{hirsch2000aurora} & 2000/06 & \xmark & \xmark  & \xmark &  \cmark &  \cmark \\
\rowcolor{ai2offwhite}    RT evaluations~\citep{garofolo2002nist} & 2002/09 & \cmark & \cmark  & \cmark &  \cmark &  \cmark \\
    ICSI~\citep{janin2003icsi} & 2003 & \cmark & \cmark & \cmark & \cmark & \xmark \\
\rowcolor{ai2offwhite}    Fisher~\citep{cieri2004fisher} & 2004 & \cmark & \cmark & \cmark & \xmark  & \xmark \\
    AMI~\citep{carletta2005ami} & 2005 & \cmark & \cmark & \cmark & \cmark & \xmark \\
 \rowcolor{ai2offwhite}   Pascal~\citep{cooke2010monaural} & 2006 & \xmark & \xmark & \cmark & \xmark  & \xmark \\
    CHIL~\citep{mostefa2007chil}  & 2007 & \cmark & \cmark & \cmark & \cmark & \xmark \\
 \rowcolor{ai2offwhite}   CH\-iME Corpus~\citep{christensen2010chime} & 2010 & \xmark & \xmark &  \xmark &  \cmark & \xmark \\
    Mixer 6 Speech~\citep{brandschain2010mixer} & 2010 & \cmark & \cmark & \cmark & \cmark  & \xmark \\
 \rowcolor{ai2offwhite}   CH\-iME-1~\citep{barker2013pascal} & 2011 & \xmark & \xmark & \xmark & \cmark  & \xmark \\
    COSINE~\citep{stupakov2012design} & 2012 & \cmark & \cmark & \cmark & \cmark  & \xmark \\
 \rowcolor{ai2offwhite}     Sheffield Wargames~\citep{fox2013sheffield} & 2013 & \cmark & \cmark & \cmark & \cmark  & \xmark \\
    REVERB~\citep{kinoshita2013reverb} & 2013 & \xmark & \xmark & \xmark & \cmark & \xmark \\
 \rowcolor{ai2offwhite}     CH\-iME-2~\citep{vincent2013second} & 2013 & \xmark & \xmark & \cmark  & \cmark & \xmark  \\
    DIRHA~\citep{cristoforetti2014dirha} & 2014 & \xmark & \xmark & \xmark &  \cmark & \xmark \\
 \rowcolor{ai2offwhite}     CH\-iME-3~\citep{barker2017third} & 2015 & \xmark & \xmark & \cmark & \xmark & \xmark  \\
    CH\-iME-4~\citep{vincent20164th} & 2015 & \xmark & \xmark & \cmark & \xmark & \xmark  \\
 \rowcolor{ai2offwhite}     ASpIRE~\citep{harper2015automatic} & 2015 & \cmark & \cmark & \xmark & \cmark & \xmark  \\
    CH\-iME-5~\citep{barker2018fifth} & 2018 & \cmark & \xmark & \cmark & \cmark  & \xmark  \\
 \rowcolor{ai2offwhite}     VOiCES~\citep{richey2018voices} & 2018 & \xmark & \xmark & \xmark & \cmark & \xmark  \\
     DiPCo~\citep{van2019dipco} & 2019 & \cmark & \cmark & \cmark & \cmark & \xmark \\
  \rowcolor{ai2offwhite}    CH\-iME-6~\citep{watanabe2020chime} & 2020 & \cmark & \cmark & \cmark & \cmark & \xmark  \\
    Aishell-4~\citep{fu2021aishell} &  2020 & \cmark & \cmark  & \cmark  &  \cmark & \xmark  \\
 \rowcolor{ai2offwhite}     AliMeeting~\citep{yu2022m2met} &  2020 & \cmark & \cmark  & \cmark  &  \cmark & \xmark  \\
    LibriCSS~\citep{chen2020continuous} &  2020 & \xmark & \cmark  & \cmark  &  \cmark & \xmark  \\
  \rowcolor{ai2offwhite}    \textcolor{black}{Ego4D}~\citep{grauman2022ego4d} & 2022 & \cmark & \cmark & \cmark &  \cmark &  \cmark \\
    MISP~\citep{Wang2023misp} & 2022 & \cmark & \cmark & \cmark &  \cmark &  \cmark \\
  \rowcolor{ai2offwhite}    CH\-iME-7 DASR~\citep{cornell2023chime} & 2023 & \cmark & \cmark & \cmark & \cmark & \cmark  \\
     CH\-iME-8 DASR~\citep{cornell2024chime} & 2024 & \cmark & \cmark & \cmark & \cmark  & \cmark\\
 \rowcolor{ai2offwhite}     CH\-iME-8 NOTSOFAR-1~\citep{vinnikov2024notsofar} & 2024 & \cmark & \cmark & \cmark &  \cmark  & \xmark \\
    CH\-iME-8 MMCSG~\citep{zmolikova2024chime} & 2024 & \cmark & \cmark & \cmark & \cmark  & \xmark\\
    \hline
\end{tabular}
  %  \footnotetext{ddddd}
\end{table}

%%%%%%%%%%%%%%%%%%%%%%%%%%%%%
Over the past 25 years, the emergence of new datasets and challenges has been a catalyst for the advancement of research toward robust ASR. For the reader's convenience, we summarize them in Table~\ref{tab:challenges} alongside some of their ``high-level'' characteristics.  

In fact, at the turn of the millennia, Gaussian mixture model (GMM) hidden Markov model (HMM) ASR technology~\citep{rabiner1989tutorial, lee1990overview, gales2008application} reached maturity and inventions such as GMM universal background model (UBM)~\citep{reynolds2000speaker} opened the way for data-driven diarization.
These advances raised interest in the automatic transcription of meeting scenarios captured with far-field microphone arrays.
The first decade of the 2000s saw the release of the Santa Barbara Corpus, AMI~\citep{carletta2005ami}, ICSI~\citep{janin2003icsi} and CHIL~\citep{mostefa2007chil} corpora. 
All of these feature real-world conversational speech captured with far-field microphone devices. In particular, among these, as mentioned, AMI is still widely used to this day. 
At the same time, research in the telephone speech domain continued, with the collection of Fisher~\citep{cieri2004fisher} contributing greatly in this latter direction. % and challenges toward robust ASR for telephone applications such as Aurora-2~\citep{hirsch2000aurora}.
During these years, the National Institute of Standards and Technology (NIST) was instrumental in leading robust ASR and diarization research by organizing several Rich Transcription (RT) evaluation campaigns~\citep{garofolo2002nist, fiscus20072003,  garofolo2004rich, fiscus2006rich, Fiscus2007} that ran from 2002 to 2009 almost every year. 
These RT evaluation campaigns challenged participants to create robust ASR and diarization systems in different domains, from telephone speech to broadcast news, and also meeting scenarios. Importantly, RT evaluation campaigns established foundations on standard scoring procedures, file formats (such as the rich transcription time-marked, segment time mark, conversation time mark etc.) and tools that are still widely used today. 
Together with the NIST RT campaigns, the Aurora challenges, which ran from the 2nd (2000) to the 5th (2006), were key in spurring research toward robust ASR during this period. Each challenge explored different aspects and scenarios focusing on single speaker and non-long-form input: hands-free telephone speech (Aurora-2 and 4), car scenarios (Aurora-3) and car and hands-free telephone scenarios together (Aurora-5). 
%\textcolor{black}{RICH transcription nist evaluation effort. and darpa ears. INSERT HERE} %In the next decade, 

Over the next decade, the transition to data-driven approaches accelerated dramatically, driven by the rise of deep learning and the increasing availability of data. Deep neural network (DNN)-based models started to replace traditional statistical models such as GMMs. This shift was particularly evident in automatic speech recognition (ASR) acoustic modeling. By the end of the 2010s, ASR in industry relied mainly on end-to-end frameworks~\citep{prabhavalkar2023end}, including transducers~\citep{gravessequence}, connectionist temporal classification (CTC)~\citep{graves2006connectionist}, and attention-based approaches~\citep{Chorowski2015}, rather than the traditional hybrid DNN-HMM approach inherited from GMM-HMM.  

During this period, the CH\-iME (computational hearing in multisource environments) challenges emerged as a ``natural'' successor to Aurora challenges played a key role in driving robust ASR research forward.
% The early CH\-iME challenges, up to CH\-iME-3, focused on synthetic data scenarios.
% This focus on synthetic data stems from the significant labor and cost associated with recording and, especially, annotating real-world multi-speaker interactions, which require complex setups involving close-talk microphones for each speaker. 
% Moreover, synthetic data, while lacking real-world complexities, offer significant advantages such as precise control over signal-to-noise ratio (SNR) and the ability to disentangle the performance for specific components of transcription pipelines, such as front-end processing and acoustic modeling. The CH\-iME corpus challenge~\citep{christensen2010chime}, CH\-iME-1~\citep{barker2013pascal} concentrated on binaural data in a simulated domestic environment, while CH\-iME-2~\citep{vincent2013second}, 3~\citep{barker2015third} expanded to more complex scenarios featuring multi-channel audio and increasingly challenging noise conditions.
% Similarly to these first CH\-iME challenges, other notable efforts such as the REVERB Challenge~\citep{kinoshita2013reverb}, DIRHA~\citep{cristoforetti2014dirha}, and VOiCES~\citep{richey2018voices} datasets also focused on partially synthetic scenarios.
The early CH\-iME challenges, up to CH\-iME-4, focused on single-speaker scenarios under controlled conditions with synthetic data. 
The motivation was simple, as synthetic data allowed for much less labor and cost for collecting data compared with recording and annotating real-world multi-speaker interactions.
Moreover, synthetic data offered significant advantages such as precise control over signal-to-noise ratio (SNR) and the ability to disentangle the performance for specific components of transcription pipelines, such as front-end processing and acoustic modeling.
CH\-iME-1~\citep{barker2013pascal} and CH\-iME-2~\citep{vincent2013second} concentrated on binaural data in domestic environment~\citep{christensen2010chime}.
Then, CH\-iME-3~\citep{barker2015third} and CH\-iME-4~\citep{vincent20164th} expanded to more complex scenarios featuring multi-channel audio and more various noise environments.
Similarly to these previous CH\-iME challenges, other notable efforts such as the REVERB Challenge~\citep{kinoshita2013reverb}, DIRHA~\citep{cristoforetti2014dirha}, and VOiCES~\citep{richey2018voices} datasets also mainly focused on single-speaker scenarios again with fully or partially synthetic data.

The later CH\-iME-5~\citep{barker2018fifth} and CH\-iME-6~\citep{watanabe2020chime} challenges moved beyond single-speaker data in artificial environments. Compared to CH\-iME-4 focusing on single-speaker reading speech, CH\-iME-5 and 6 feature a complex real-world multi-speaker scenario consisting of dinner party conversations recorded with far-field microphones. Another initiative in this direction was the 2015 ASpIRE challenge~\citep{harper2015automatic} which was focused on speech recognition in noisy/reverberant environments using recordings from the Mixer 6 corpus~\citep{brandschain2010mixer}. 

The focus on real-world data has recently gained momentum since 2020 with the release of even more datasets and challenges emphasizing spontaneous, multiparty conversational speech. 
This, again, has been driven by recent breakthroughs in self-supervised and weakly supervised training, along with increased availability of computing and training data as well as growing interest from industry as technology matures. 
The latter is also further fueled by the rapid advancements in LLMs which, as mentioned, enable more applications for conversational speech technologies such as meeting summarization. 
Prominent examples of this trend are \textcolor{black}{Ego4D}~\citep{grauman2022ego4d}, a recently collected dataset of egocentric videos collected through smart glasses, conversations recordings on distant microphone arrays collected for MISP challenges~\citep{Wang2023misp}, office meeting datasets such as Aishell-4~\citep{fu2021aishell} and the Alimeeting dataset (collected for M2MeT ICASSP 2022 challenge~\citep{yu2022m2met}).

The past two CH\-iME-7 and 8 challenges are also aligned with this research direction: the proposed CH\-iME-7 and 8 DASR challenges~\citep{cornell2023chime, cornell2024chime} (C7-8DASR), as well as the concurrent ``twin'' CH\-iME-8 NOTSOFAR-1~\citep{vinnikov2024notsofar} and CH\-iME-8 MMCSG~\citep{zmolikova2024chime} challenges, all feature real-world meeting scenarios captured by far-field microphone arrays. 
A key distinguishing feature of the C7-8DASR challenges lies in their strong emphasis on multi-domain generalization and array-agnostic processing, as participant systems are evaluated across diverse scenarios with drastically different recording setups.
This contrasts with CH\-iME-8 NOTSOFAR-1, which also focuses on meeting transcription but instead assume that knowledge about the deployment domain and the array device is available. The main motivation for having both challenges in CH\-iME-8 was to compare how ``generalist'' transcription systems compare with specialized ones and how the former can be adapted once the target domain and device are known. This question, in particular, will be addressed here in Section~\ref{ssec:c8dasr_notsofar1_results}.

\subsection{Main Contributions}\label{ssec:contrib}

This work is an extension of our previous CH\-iME-7 DASR (C7DASR)~\citep{cornell2023chime} and CH\-iME-8 DASR (C8DASR)~\citep{cornell2024chime} works, where we presented the motivations, datasets, rules, and baseline systems for C7-8DASR challenges.
The primary objective of this paper is to provide a unified perspective on these two challenges by analyzing and comparing participant submissions, identifying key trends, and highlighting the most effective techniques across both iterations. Additionally, we draw comparisons with the concurrent CH\-iME-8 NOTSOFAR-1 challenge which focuses on a specific office-scenario setting with known microphone array configuration. This comparison is possible because CHiME-8 NOTSOFAR-1 was included as one of the four scenarios considered in C8DASR. 
This paper extends our previous work by offering significant novel contributions in the following areas:
%The goal of this work is to provide a unified perspective on these two challenges by analyzing and comparing participant submissions, identifying key trends, and highlighting the most effective and less effective techniques across both iterations, the previous CH\-iME-6 challenge and the CH\-iME-8 NOTSOFAR-1 challenge with which C8DASR shares an portion of evaluation data. 

\begin{itemize}
   % \item We offer a comprehensive and unified perspective over these two challenges in relation to the previous 
    \item We analyze, summarize and discuss a total of 32 participant submissions from the C7-8DASR and CH\-iME-8 NOTSOFAR-1 challenges.  
    \item We compare results across the past two CH\-iME challenges that focus on meeting transcription: C7DASR and subsequent C8DASR and CH\-iME-8 NOTSOFAR-1 twin challenges. 
    \item In addition to WER-based metrics and diarization metrics, such as Jaccard error rate~\citep{ryant2019second}, we also investigate downstream evaluation on meeting summarization on the NOTSOFAR-1 scenario. %and alternative metrics based on bidirectional LM that can better capture semantic meaning and are more agnostic to spelling and text normalization errors. 
    %To this latter end, we propose a novel metric derived from SemDist~\citep{kim2021semantic}, but suitable for long-form multi-speaker ASR evaluation. 

    %\item We present, compare and contextualize results across C7DASR, C8DASR, CH\-iME-6, and %CH\-iME-8 NOTSOFAR-1.  we extend SemDist to also include continuous meetings settinsg. 
    % \item We identify and discuss the organizational difficulties and scientific limitations we encountered during the C7-8DASR challenges.  %and raise awareness.  %for some key problems 
     % are general enough to benefit the broader speech research community. %and outline possible solutions to improve future iterations.  
     % providing recommendations to improve future iterations
   % \item we highlight the limitations and challenges encountered in organizing C7-8DASR challenges 
\end{itemize}
%Furthermore, this paper discusses limitations of the current approaches and metrics.
%Our goal is to try to get insights into how much progress has been made in distant ASR and diarization systems in order to get possible promising future work directions.

%We aim to offer deeper insights into the progress made in distant ASR and diarization systems and to explore how the lessons learned can guide future research in this domain. 
%The goal of this work is to provide a unified  

\subsection{\textcolor{black}{Main Findings}}

\textcolor{black}{This paper presents several key findings from analyzing the submitted systems:}

\begin{itemize}

    \item \textcolor{black}{Array and domain agnostic systems can achieve competitive performance. Cross comparison between CHiME-8 DASR and the concurrent NOTSOFAR-1 challenge (Section~\ref{ssec:c8dasr_notsofar1_results}) revealed that generalizable, array-agnostic transcription systems can perform comparably to systems specifically tailored to known recording setups and deployment domains.}
    \item \textcolor{black}{Transition to end-to-end ASR with large-scale data pre-trained models (Section~\ref{ssec:sys_description_asr}). Unlike CHiME-6, which predominantly featured hybrid ASR systems, nearly all CHiME-7/8 participants employ end-to-end ASR frameworks. This transition was enabled by the availability of robust large-scale pre-trained models (e.g., WavLM~\citep{chen2022wavlm}, Whisper~\citep{radford2022robust}), which significantly reduced the data burden for training competitive end-to-end (e2e) systems.}
   
     \item \textcolor{black}{DNN-based front-end speech separation methods still struggle with far-field conversational speech. All participating teams relied on guided source separation (GSS)~\citep{boeddeker2018front} and, when a DNN-based front-end was used, it was used together with GSS for initialization or refinement (Section~\ref{ssec:sys_description_separation})}. 

     \item \textcolor{black}{Accurate diarization and, especially, speaker counting are critical. All top-performing systems employed diarization refinement via target-speaker voice activity detection (TS-VAD) techniques~\citep{medennikov2020target} (Section~\ref{ssec:sys_description_diarization}). Accurate speaker counting in the initial diarization pass is crucial, as errors propagate catastrophically through the subsequent separation and recognition stages (Section~\ref{ssec:joint_c78dasr_results}). The CHiME-8 DASR challenge, in particular, drove innovations in robust speaker counting methods to handle diverse scenarios (Section~\ref{ssec:speaker_counting}).}
     %This suggests that current neural separation methods remain unable to 

     \item \textcolor{black}{Weak correlation between transcription accuracy and summarization. Downstream evaluation via LLM-based meeting summarization in the NOTSOFAR-1 scenario (Section~\ref{ssec:summarization_results}) revealed a weak correlation with transcription quality. This demonstrates the remarkable effectiveness of large language models in handling transcription errors and suggests that if precise transcription is not required, end-to-end meeting summarization may be a promising research direction.}

     \item \textcolor{black}{Generalization across scenarios remains difficult. Balancing performance across all scenarios proved challenging (Section~\ref{ssec:joint_c78dasr_results}). Systems optimized for CHiME-6 often performed worse on Mixer 6 and vice versa. This trade-off was evident even in the CHiME-8 DASR baseline systems, where re-tuning to accommodate NOTSOFAR-1 degraded performance on other scenarios.}
     %The scenarios ranked consistently from most to least difficult as CHiME-6 $>$ DiPCo $>$ Mixer 6, with CHiME-6's multi-room setting and colloquial speech presenting the greatest challenge. Even the best-performing systems achieved tcpWER above 30\% on CHiME-6, meaning nearly one in three words is incorrectly transcribed.}
    
    \item \textcolor{black}{Multi-channel mechanisms remain, in general, under-explored. Most systems relied on simple ensembling techniques to fuse information across microphones rather than native multi-channel processing (Section~\ref{ssec:sys_ensembling_multi_channel_tta}). The effectiveness of sophisticated multi-channel diarization and separation techniques in complex conversational scenarios remains largely unexplored.} 
    %Notable exceptions include the NeMo baseline's multi-channel speaker embeddings and channel clustering approaches. 
    
    \item \textcolor{black}{Limited success of test-time adaptation and LLM integration. Several teams explored test-time adaptation (TTA) techniques for both diarization  (Section~\ref{ssec:sys_ensembling_multi_channel_tta}) and ASR, as well as LLM-based post-processing (Section~\ref{ssec:lm_sys_asr}). However, these approaches yielded only marginal improvements despite significant computational overhead, raising questions about their practical value in real-world applications.}
    
    %However, these approaches yielded only marginal improvements despite significant computational overhead, raising questions about their practical value in real-world applications. Only one team (STCON C8) experimented with LLM-based hypothesis rescoring, achieving only 0.5\% absolute reduction in macro-tcpWER.}
     %\item Generalization across scenarios remains difficult.
     %\item Multichannel mechanisms remain underexplored.
     %\item Limited success of test-time adaptation and LLM integration.
     
\end{itemize}

\subsection{Outline}\label{ssec:outline}
This paper is structured as follows: in Section~\ref{sec:motivation} we introduce the C7-8DASR challenges and discuss the motivations and goals of these two challenges in the context of the previous CH\-iME challenges. 
In Section~\ref{sec:description} we describe in detail the two challenge datasets, evaluation tracks, and baseline systems. 
Section~\ref{sec:systems_description} presents an overview of the submissions of the participants and their design choices for the systems. 
In Section~\ref{sec:results} we present, analyze and discuss the results of the two challenges with the goal of identifying common trends and techniques that appear to be particularly effective. 
C8DASR systems are also compared and discussed within the ``twin'' CH\-iME-8 NOTSOFAR-1 challenge, and meeting summarization as a potential downstream evaluation task is also considered.  
Finally in Section~\ref{sec:conclusion} we draw conclusions, outline the limitations of these two challenges, and propose directions for future research and evaluation campaigns.

%in Section~\ref{sec:motivation} we give an historical overview of the datasets and challenges for robust ASR proposed since the beginning of the 21st century. Subsequently, ,  ones and present in detail evaluation tracks, metrics and rules. 

\section{DASR challenges motivation}\label{sec:motivation}

%The CH\-iME-7 and CH\-iME-8 DASR challenges (C7-8DASR in the following) are built directly as a continuation of the previous CH\-iME-5/6 challenges. 
%The focus is still on far-field, possibly multi-array/device transcription of long-form meetings but C7-8DASR importantly expand the breadth of evaluation and make generalization their `first principle. 
%The goal is to spur research towards methods that can work well across different array topologies; (2) are capable of handling variable numbers of speakers in each session; (3) account for linguistic differences between dinner party scenarios (CH\-iME-6 and DiPCo) versus interviews (Mixer 6); and (4) can effectively handle diverse acoustic conditions.
%Such variability reflects real use-cases where cross-domain generalization capability is  desirable. 

%The CH\-iME-7 and CH\-iME-8 DASR challenges (C7-8DASR) build directly upon the CH\-iME-5 and CH\-iME-6 challenges, maintaining the focus on far-field transcription of long-form meetings through joint ASR and diarization. 
%However, it brings several novelties motivated by recent trends in the speech processing field such as the increasing availability of pre-trained ``foundation'' models as well as open source datasets. 
%The goal is to encourage participants to explore new research directions compared to the previous CH\-iME-6 challenge but keeping at the same time also a strong emphasis on practical application scenarios. 

The C7-8DASR challenges build on the previous CH\-iME-6 challenge and thus focus on far-field transcription of long-form meetings through ASR and diarization. 
These new challenges emphasize generalization as a core principle and introduce several innovations inspired by recent trends in the speech processing field, including the growing availability of pre-trained ``foundation models'' and open-source datasets.
%However, these new challenges adopt generalization as "first principle`` and introduce several innovations inspired by recent trends in the speech processing field, such as the growing availability of pre-trained ``foundation'' models and open-source datasets.

The main goal is to encourage participants to explore new research directions beyond those tackled in the CH\-iME-6 challenge while, at the same time, keeping a strong emphasis on practical application scenarios. 
%This balance aims to foster cutting-edge advancements and, at the same time, ensure that the developed solutions can be relevant to real-world use cases.

\subsection{Robustness and generalization}

Compared to previous CH\-iME challenges, C7-8DASR significantly broaden the scope of evaluation, prioritizing generalization as the core principle. The C7-8DASR challenges aim to foster research into transcription systems that are robust to:

\begin{itemize}
    \item changes in the recording setup/array topologies,
    \item variable meetings duration and with different number of speakers,
    \item varying acoustic conditions,
    \item linguistic and para-linguistic differences between diverse scenarios, such as ``colloquial'' dinner party conversations, office meetings and interviews.
\end{itemize}
\noindent
This emphasis on variability mirrors real-world applications, where the ability to generalize across domains is essential and control on the recording setup (e.g. through proprietary devices) is not always possible as it is inherently less flexible and more expensive. 

As such, C7-8DASR challenges ``stress-test'' the robustness and versatility of the participant's systems on $4$ different ($3$ for C7DASR) datasets that feature spontaneous conversational speech and are highly diverse in recording setup, setting, duration, and number of speakers. 
These scenarios are CH\-iME-6, DiPCo, Mixer 6 Speech (MX6) and NOTSOFAR-1 (added in C8DASR). %and will be described in detail in Section~\ref{sec:core_scenarios}. 
Their ``high-level'' characteristics are summarized in Table~\ref{tab:dset_overview}, while a detailed description for each of them is in Section~\ref{sec:core_scenarios}. 
As can be seen, they are very diverse: on one hand, systems have to deal with multiple arrays and long meetings with $4$ participants (CH\-iME-6); on the other, NOTSOFAR-1 has only a single array, features up to $8$ participants, and a very short duration. This inter-scenario variability reflects real-world deployment scenarios and significantly complicates the diarization speaker counting task.

\begin{table}[htbp]
\centering
\footnotesize
\caption{Summary of the characteristics for the C7-8DASR challenges core scenarios. As the main focus is on generalization, the scenarios considered are highly diverse. NOTSOFAR-1 was included only in C8DASR.}
\label{tab:dset_overview}
\setlength{\tabcolsep}{0.1em} % Increased spacing between columns
\begin{tabular}{lccccc} % Simplified column alignment
\toprule   
 \textbf{Scenario} & \textbf{Setting} & \makebox[0pt]{\textbf{Number of}} & \textbf{Recording} &  \textbf{Tot.} & \textbf{Duration} \\
 & & \textbf{Speakers} & \textbf{Setup} & \textbf{Mics} & \textbf{\textcolor{black}{(minutes)}} \\ 
\midrule  % Added midrule for better visual separation
\rowcolor{ai2offwhite} CHiME-6 & dinner party & 4 & 6 linear arrays & 24 & \textcolor{black}{$\sim$120 to 150} \\ 
 DiPCo & dinner party & 4 & 5 circular arrays & 35 & \textcolor{black}{$\sim$33 to 47} \\ 
\rowcolor{ai2offwhite} Mixer 6 Speech& \,\,\,\,1-to-1 interview & 2 & 10 heterogeneous devices & 10 & \textcolor{black}{$\sim$25} \\ 
 NOTSOFAR-1\,\,\, & office meeting & 4-8 & 1 circular array & 7 & \textcolor{black}{$\sim$10} \\ 
\bottomrule
\end{tabular}
\end{table}

%This is the main feature that sets C7-8DASR apart from the other contemporary challenges such as for example the CHi

\subsection{Availability of large-scale pre-trained models}

In the last 4 years, there has been growing research towards large-scale pre-trained ``foundation'' speech models, either trained via self-supervised learning (SSL) or weakly supervised objectives. 
These are increasingly available to speech researchers and have demonstrated exceptional effectiveness in numerous downstream applications~\citep{yang2021superb}.
Notable examples are Whisper~\citep{radford2022robust}, OWSM~\citep{peng2023reproducing}, WavLM~\citep{chen2022wavlm}, HuBERT~\citep{hsu2021hubert} and wav2vec 2.0~\citep{baevski2020wav2vec} to name just a few. 
Their success is so indisputable that in many speech processing applications nowadays the de facto standard approach is to use these as feature extractors or, more directly, via fine-tuning. 
This procedure allows to implicitly leverage, in a cost-effective way, their massive pre-training data, thus leading to a more robust system. 

%While for monaural, pre-segmented single-talker speech data, this procedure is rather straightforward, in the multi-channel, multi-speaker, long-form setting it is still unclear how these foundation models can be fully exploited.
%Can we leverage them to also improve diarization or separation ? Can they be extended to perform target speaker ASR or multitalker ASR even in challenging conditions ? How can we deal with multiple channels as these models are mostly monaural ? 
%These are some of the research questions we wanted to address by introducing this novelty in the DASR challenges. 

For monaural, pre-segmented, single-talker speech data, integrating these models is relatively straightforward.
However, in the multi-channel, multi-speaker, long-form setting, fully exploiting their potential remains an open question.
%Can these models also improve diarization or separation tasks?
%Could they be extended to handle target speaker ASR or multi-talker ASR under challenging conditions?
%How can we adapt these predominantly monaural models effectively process multichannel audio? 
%Addressing these questions has been a key motivation for incorporating foundation models into DASR challenges. 

Another significant advantage of allowing pre-trained models is the lower entry barrier for challenge participants, particularly those with limited resources. By relying on pre-trained models, training becomes faster and more efficient, eliminating the need for many training epochs. In many cases, modest amounts of fine-tuning data are sufficient along with few (1 or 2 epochs) training iterations on such data, making participation in the DASR challenges more accessible to a broader community compared to CH\-iME-6. 
For example, as also explained later in Section~\ref{ssec:accessibility} the C7DASR baseline system is much faster to train compared to the CHiME-6 one. 
The full list of allowed pre-trained models is available on the challenge website\footnote{\href{https://www.chimechallenge.org/challenges/chime8/task1/rules\#external\_models}{https://www.chimechallenge.org/challenges/chime8/task1/rules\#external\_models}}.

%\footnote{available: \href{https://www.chimechallenge.org/challenges/chime8/task1/rules\#external\_models}{https://www.chimechallenge.org/challenges/chime8/task1/rules\#external\_models}}

%Another advantage of allowing such pre-trained models is that they actually lower the entrance barrier for new participants and especially those with limited resources, since training is often more efficient and quicker when these models are leveraged, and the need for large-scale training is eliminated, as modest fine-tuning data can suffice. 
%While for monaural, pre-segmented data this is rather straightforward, in the multi-channel, long-form setting 

\subsection{Leveraging open-source external datasets}

CH\-iME-5 and 6 challenges had quite strict restrictions on the training material allowed, which prevented research in important directions.
In C7-8DASR, the amount and diversity of allowed external data sources\footnote{\href{https://www.chimechallenge.org/challenges/chime8/task1/data\#external\_datasets}{https://www.chimechallenge.org/challenges/chime8/task1/data\#external\_datasets}} are significantly expanded
to include not only popular open-source speech datasets such as LibriSpeech~\citep{panayotov2015librispeech} but also noise~\citep{fonseca2021fsd50k, snyder2015musan} and room impulse responses~\citep{jeub2009binaural} datasets. 
This opens up the possibility of exploring also the use of techniques that heavily rely on synthetic data for pre-training such as neural speech separation and enhancement (SSE), neural speaker extraction~\citep{vzmolikova2019speakerbeam, boeddeker2023ts}, or end-to-end neural diarization (EEND)~\citep{fujita2019end}.
C7-8DASR succeeded partially in this direction, as these techniques were explored by some participants, but with marginal improvements over guided source separation (GSS)~\citep{boeddeker2018front}. Details are reported in Section~\ref{sec:description}.

% These techniques, despite being quite promising, were not considered in the previous CH\-iME-5 and 6 challenges due to the aforementioned restrictions. 
% Both techniques are arguably underexplored for real-world meeting scenarios applications. 

\subsection{More comprehensive evaluation for meeting transcription}\label{ssec:tcpwer}

Another limitation of the CH\-iME-6 challenge is that it used the concatenated minimum permutation word error rate (cpWER)~\citep{watanabe2020chime} as the ranking metric. 
cpWER does not pose explicit requirements for participants to also produce reasonable utterance-level segmentation, as it only cares about speaker attribution.
On the other hand, having reasonable segmentation at the utterance level is desirable in real-world applications since, among other things, it allows for quick audio retrieval for verification of the transcription.
As such, in C7DASR we proposed a new metric called diarization-attributed WER (DA-WER)~\citep{cornell2023chime} that computed the concatenated WER by using the speaker-assignment permutation that minimizes the diarization error rate (DER) rather than, as in cpWER, the permutation that directly minimizes the WER.  
Since C7DASR featured three scenarios, each with different duration and number of recordings and the focus was on probing robustness and generalization, the final ranking was given by macro-averaging DA-WER across all scenarios.  
DA-WER encouraged participants to also produce reasonable segmentation at the utterance level, while at the same time, in the C7DASR description paper~\citep{cornell2023chime}, we raised awareness of the fact that this was not an optimal solution and better metrics were needed. 

Our call was answered the same year and, in the following C8DASR, we instead adopted the time-constrained cpWER (tcpWER), recently proposed in the Meeteval toolkit~\citep{von2023meeteval}.
tcpWER incorporates temporal constraints into Levenshtein distance computation by restricting substitutions or correct mappings between reference and hypothesis words to those within a specified time collar (e.g., 5 seconds).
Utterances are segmented into word-level units, with segment lengths proportional to word length.
The collar is set wide enough to account for potential inaccuracies in the word boundary estimation (in C8DASR, a generous collar of 5 seconds is employed).
This temporal constraint makes tcpWER much more sensitive to segmentation errors than DA-WER.
An alternative time-constrained measure is Asclite~\citep{fiscus2006multiple}, which includes a time penalty but disregards speaker attribution, making it unsuitable for certain applications.
% tcpWER makes use of temporal constraints in the Levenshtein distance computation and thus is much more sensitive to segmentation errors than DA-WER.
% Moreover, compared to other measures that take into account segmentation and speaker attribution, such as Asclite~\citep{fiscus2006multiple} with time penalty, tcpWER has the significant advantage that it does not require word level, but only utterance segmentation.
For this reason, and to be able to cross-compare between the two challenges, in Sections~\ref{sec:results} we will use tcpWER instead of DA-WER even when analyzing C7DASR submissions.

\subsection{Accessibility}\label{ssec:accessibility}

Building a SotA diarization and robust ASR pipeline for meeting transcription is an inherently difficult task that requires a team to have expertise in different speech processing areas. 
As such, the entry bar for successfully participating in the CHiME-5 and CH\-iME-6 challenges was quite high. 
For example, running successfully the Kaldi~\citep{povey2011kaldi} baseline of the CH\-iME-6 challenge could easily take more than a week if no large-scale computing infrastructure with many CPUs is available. As said, in C7-8DASR, at least this was reduced due to the possibility of leveraging large-scale pre-trained models. For example, the C7-8DASR ESPnet~\citep{watanabe2018espnet} baseline ASR, which uses WavLM as a feature extractor, can be trained in less than three days on $2$ NVIDIA \textcolor{black}{RTX 3090 GPUs}. Moreover, both baselines use a GPU-accelerated implementation~\citep{Raj2022GPUacceleratedGS} of GSS that allows, in most computing infrastructures, to massively speed up the inference time compared to the CH\-iME-6 Kaldi baseline and the official implementation of GSS.
In C8DASR, we also provided another baseline system based on the C7DASR NVIDIA NeMo team submission~\citep{park2023chime}, which was built instead with the NeMo toolkit~\citep{nemo_2019}. 
This baseline also exploits pre-trained models for ASR and a GPU-based implementation of GSS. 
By offering two baseline systems that utilize two widely adopted speech processing toolkits, our goal was to make participation more accessible to a broader range of researchers. 
%Having two baseline systems covering two of the most used speech processing toolkits could enhance the challenge accessibility further. 

%Moreover, the baseline made use of a GPU-friendly GSS implementation. 

Significant work was also done to make data preparation easier. C7-8DASR feature different core scenarios, each with its own annotation format and directory structure, potentially adding significant additional workload for the participants to parse and prepare each dataset. 
Thus, in C7DASR we provided a script that prepared all the data from the different scenarios to be available in a single common format which follows the convention of CH\-iME-6 and DiPCo. 
Expanding on this, in C8DASR, we developed a dedicated toolkit: \texttt{chime-utils}\footnote{available: \href{https://github.com/chimechallenge/chime-utils}{https://github.com/chimechallenge/chime-utils}},
designed to streamline the process even more. With just a single line of code, \texttt{chime-utils} can automatically download and prepare CH\-iME-6, DiPCo, MX6 and NOTSOFAR-1 datasets.
Additionally, \texttt{chime-utils} offers a suite of tools to facilitate integration with popular speech processing toolkits, including ESPnet~\citep{watanabe2018espnet}, Kaldi~\citep{povey2011kaldi}, NeMo~\citep{nemo_2019} and k2~\citep{poveyk2} via Lhotse~\citep{zelasko2021lhotse}. It also includes scoring scripts that seamlessly interface with Meeteval~\citep{von2023meeteval} and Pyannote~\citep{bredin2020pyannote} as well as various utilities including datasets statistics computation (Table~\ref{tab:class_stats} was produced using \texttt{chime-utils}). 
This effort aims to benefit the broader speech processing community beyond the CH\-iME challenges by making datasets like CH\-iME-6, DiPCo, MX6 and NOTSOFAR-1 more accessible and easier to experiment with.

\section{DASR challenges description}\label{sec:description}

\subsection{Core scenarios analysis}\label{sec:core_scenarios}
%As mentioned,  while  is reported in Table~\ref{tab:class_stats}.
In the following sub-sections~\ref{sssec:chime6}-\ref{sssec:notsofar1}, we describe the C7-8DASR core scenarios in detail, while here we discuss their high-level characteristics.
In Table~\ref{tab:class_stats} we report a summary of their statistics including the percentage of overlapped speech, silence, and single-speaker speech.
These statistics are computed by using the manual annotated \textcolor{black}{ground-truth} utterance-level segments.
\textcolor{black}{It is important to note that the characteristics between \texttt{dev} and \texttt{eval} splits can differ substantially for some scenarios, which may affect system development and generalization. For CHiME-6, the \texttt{dev} split has significantly higher overlapped speech ratio (43.5\%) compared to \texttt{eval} (26.7\%), while silence is lower (13.1\% vs. 21.3\%). These differences stem from the fact that different dinner party sessions can have very different interaction dynamics. For NOTSOFAR-1, the \texttt{dev} split has notably higher silence ratio (15.6\%) and lower overlapped speech (16.7\%) compared to \texttt{eval} (5.6\% and 29.6\% respectively). These distributional differences can affect hyperparameter tuning, but they also reflect real-world deployment challenges where conversational dynamics vary unpredictably across sessions and environments. In this sense, the \texttt{eval} set serves as a proxy for potentially unseen deployment scenarios, testing system robustness.}

%It is evident that among the 4 scenarios, Mixer 6, which consists of one-to-one interviews is the one with the lowest percentage of overlapped speech 

%However, it is important to note that the quality and characteristics of these annotations are not consistent across all four scenarios. For instance, the CHiME-6 segmentation is generally less precise and tightly defined compared to that of DiPCo or NOTSOFAR-1. Therefore, any comparison of these values across the four scenarios should be made with caution.

%In addition to these four core scenarios \texttt{train} sets, as said, for training and data augmentation, participants could also leverage external open-source data sources from a predefined list\footnote{\href{https://www.chimechallenge.org/challenges/chime8/task1/data\#external\_datasets}{https://www.chimechallenge.org/challenges/chime8/task1/data\#external\_datasets}} as well as propose new ones during the first month of the challenges. 

%This list includes popular clean speech datasets such as LibriSpeech~\citep{panayotov2015librispeech} but also noise-only~\citep{fonseca2021fsd50k, snyder2015musan} and room impulse responses~\citep{} datasets. 
%Since the challenge features 4 different datasets, that have different annotation formats and different folder structure, 
%a significant effort was made to make 
% mention that annotation is made the same as wella s the same structure with chime utils describd later 
%To simplify challenge participation 

\begin{table}[htbp]{
\centering
\footnotesize
\caption{C7-8DASR core datasets statistics overview. 
We report the number of utterances, speakers, and sessions, as well as silence (sil), single-speaker speech (1-spk) and overlapped speech (ovl) ratios over the total duration. 
}\label{tab:class_stats}
\setlength{\tabcolsep}{0.2em}
%\adjustbox{max width=\linewidth}{
\begin{tabular}{@{}llrrrrSSS@{}}
\toprule   
 \textbf{Scenario} & \textbf{Split} & \textbf{Size (\textcolor{black}{hh:mm})} & \textbf{Utts} & \textbf{Spk.} & \textbf{Sess.} & \textbf{sil (\%)} & \textbf{1-spk (\%)} & \textbf{ovl (\%)}   \\
\midrule
\multirow{3}{*}{\textbf{CH\-iME-6}} & \multirow{1}{*}{train} &  \multirow{1}{*}{40:05} &  79967   & \multirow{1}{*}{32} & \multirow{1}{*}{16} & 22.6  & 52.7  & 24.7  \\
% &  & &  (78640) & & & (38.8) & (50.2) & (11.0) \\
% \cmidrule{2-9}
& \multirow{1}{*}{dev} &  \multirow{1}{*}{4:27} &  7437  & \multirow{1}{*}{8} & \multirow{1}{*}{2} & 13.1  & 43.4  & 43.5  \\
% &  & &  (11020) & & & (24.1) & (54.2) & (21.7) \\
% \cmidrule{2-9}
& \multirow{1}{*}{eval} &  \multirow{1}{*}{5:12} &  11028  & \multirow{1}{*}{8} & \multirow{1}{*}{2} & 21.3  &  52.0  &  26.7  \\
% &  & &  (81890) & & & (38.8) & (49.2) & (12.0) \\
\midrule

\multirow{3}{*}{\textbf{DiPCo}} & \multirow{1}{*}{train} &  \multirow{1}{*}{1:12} & 1379  & \multirow{1}{*}{8} & \multirow{1}{*}{3} & 8.3 & 72.0 &  19.6 \\
% &  & &  (78640) & & & (38.8) & (50.2) & (11.0) \\
% \cmidrule{2-9}
& \multirow{1}{*}{dev} &  \multirow{1}{*}{1:31} &  2294  & \multirow{1}{*}{8} & \multirow{1}{*}{2} &  7.4 & 61.9  & 30.6 \\
% &  & &  (11020) & & & (24.1) & (54.2) & (21.7) \\
% \cmidrule{2-9}
& \multirow{1}{*}{eval} &  \multirow{1}{*}{2:36} & 3405  & \multirow{1}{*}{16} & \multirow{1}{*}{5} & 9.4  &  65.7 & 24.9 \\

\midrule
\multirow{5}{*}{\textbf{Mixer 6}}  & train calls &  36:09 & 27280  & 81 & 243 & {--}   & {--}  &    {--}  \\
 & train intv & 26:57  &  29893  & 77 & 189 & {--} &  {--} &  {--}    \\
  & train & 6:13  &   3785 & 19 & 24 & 8.6 & 73.3  &  18.0    \\
& dev &  8:56 & 5903 & 22 & 35 & 8.4 & 72.1 &  19.5   \\
& eval & 5:45 & 5115  & 18 & 23 &  2.4 & 83.6 & 13.9 \\
\midrule
\multirow{4}{*}{\shortstack{\textbf{NOTSOFAR-1}}} & \multirow{1}{*}{train} &  \multirow{1}{*}{14:43} & 101301  & \multirow{1}{*}{14} & \multirow{1}{*}{379} & 6.0 & 62.3 &  31.7  \\
% &  & &  (78640) & & & (38.8) & (50.2) & (11.0) \\
% \cmidrule{2-9}

 & \multirow{1}{*}{train\_sc} &  \multirow{1}{*}{53:43} & 139913  & \multirow{1}{*}{14} & \multirow{1}{*}{526} & 5.9  & 62.4  & 31.7 \\

& \multirow{1}{*}{dev} &  \multirow{1}{*}{13:25} & 24238  & \multirow{1}{*}{11} & \multirow{1}{*}{130} & 15.6  &  67.7  & 16.7 \\
% &  & &  (11020) & & & (24.1) & (54.2) & (21.7) \\
% \cmidrule{2-9}
 & \multirow{1}{*}{eval} &  \multirow{1}{*}{16:29} & 38662  & \multirow{1}{*}{12} & \multirow{1}{*}{160} & 5.6  & 64.7  & 29.6  \\
\bottomrule
\end{tabular}
}
\end{table}

\begin{figure}[h]
    \centering
    \includegraphics[width=0.8\linewidth]{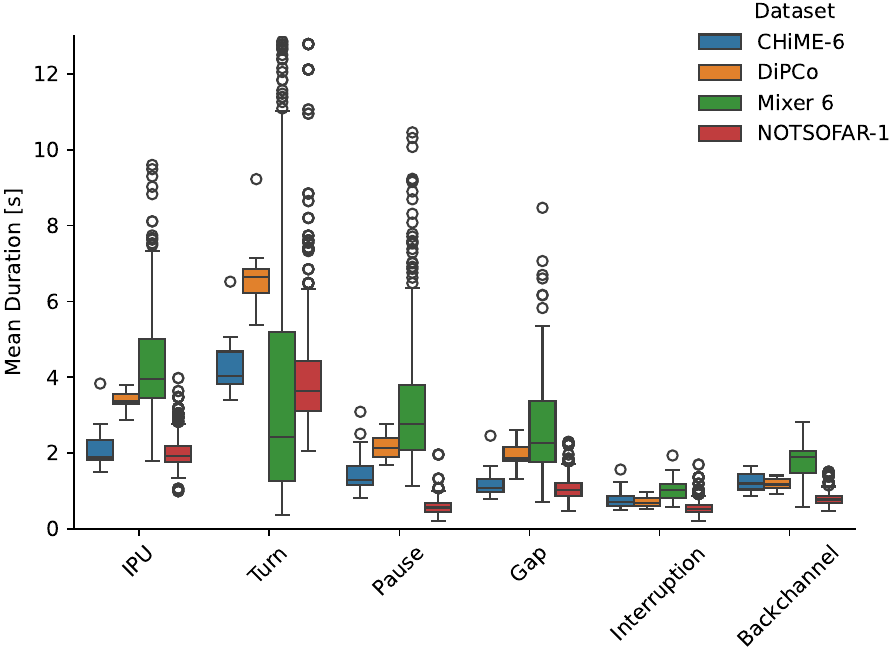}
    \caption{Mean duration distribution of turn taking events for each C7-8DASR scenario as obtained on the whole \texttt{train}, \texttt{dev} and \texttt{eval} splits.}
    \label{fig:turn_taking_stats}
\end{figure}

In Figure~\ref{fig:turn_taking_stats} we report, for each scenario, the distribution for the mean duration of interpausal units (IPU), turns, pauses, gaps, interruptions and backchannels. These turn-taking events are defined in~\citep[Figure 3]{nguyen2023generative} as:
\begin{itemize}
    \item IPUs: single utterances i.e. continuous speech segments from a single speaker without significant internal pauses ($\le 0.5$\,\textcolor{black}{seconds}). 
    \item Pauses: pauses ($> 0.5$\,s) that occur within single utterances belonging to the same speaker. 
    \item Gaps: pauses that instead occur within single utterances belonging to different speakers.
    \item Interruptions: overlapped speech where a second speaker begins talking near the conclusion of the first speaker's utterance.
    \item Backchannels: usually brief overlapping vocalizations (e.g., ``mm-hmm'', ``yeah'') produced by a second speaker within the primary speaker's utterance. 
    \item Turns: span of contiguous utterances (with pauses in between) from the same speaker without interruptions. 
\end{itemize}
% \begin{itemize}
%  \item interpausal units (IPUs): continuous speech (utterance) without pauses. 
%     \item Pauses: pauses between IPUs from the same speaker.
%     \item Turns: spans of IPUs and pauses from the same speaker.   
%     \item Gaps: pauses between IPUs of different speakers.
%     \item Interruptions: overlaps by a different speaker that occur at the end of an IPU. 
%     \item Backchannels: overlaps by a different speaker which occurs within an IPU. % \end{itemize}
Again, we used the ground-truth annotation for each dataset and computed these metrics for each session, considering all \texttt{train}, \texttt{dev} and \texttt{eval} splits together. 
These metrics are complementary with what is observed in Table~\ref{tab:class_stats} and offer a different perspective.
All 4 scenarios exhibit characteristics typical of spontaneous conversational speech, such as utterances on average around $2$ to $4$ seconds long and the presence of short interruptions and backchannel responses. Despite being short, these occur frequently and thus lead to a non-negligible percentage of total speech duration being overlapped as observed in Table~\ref{tab:class_stats}. %responses that however occur often and thus lead to a non-negligible percentage of speech being overlapped. 
NOTSOFAR-1 and CHiME-6 tend to have very similar statistics across all event types, while among all scenarios, Mixer 6 is the one that exhibits the largest variance in gaps, turns, IPUs and pause durations. This is largely due to the fact that it consists of only two speaker interactions and thus there is more opportunity for a single speaker to produce longer speech segments without interruptions. Likewise, longer pauses can be obtained when having fewer speakers.

\begin{figure}[h]
    \centering
    \includegraphics[width=0.7\linewidth]{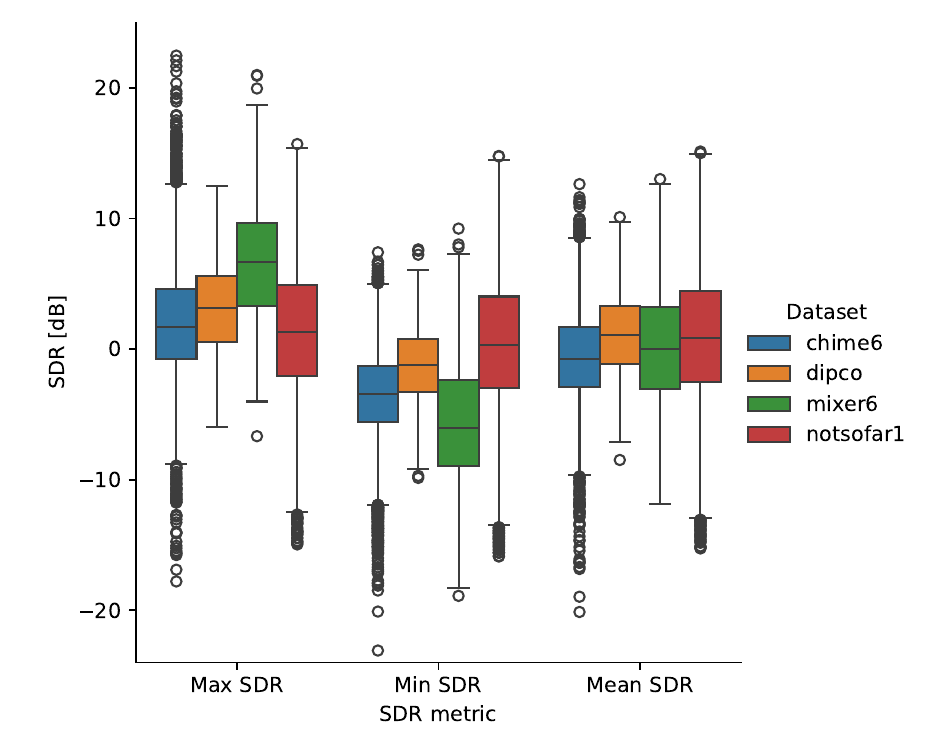}
    \caption{Distribution of mean, maximum and minimum SDR as obtained across microphones for each utterance. We report statistics for each of the 4 scenarios separately.}
    \label{fig:sdr_stats}
\end{figure}

In Figure~\ref{fig:sdr_stats} we instead analyze the signal-level amount of noise and interfering speech for each scenario.
In detail, we report, for each scenario, the distribution of different signal-to-distortion ratio (SDR)~\citep{vincent2006performance} statistics as computed for each utterance across all \texttt{train}, \texttt{dev} and \texttt{eval} splits.
In detail, we used ground-truth diarization to compute the SDR for each utterance from each far-field device with respect to the close-talk microphone of the speaker of interest.
Then, we compute for each utterance the maximum, the minimum, and the average SDR obtained across all microphones available and plot statistics over all each of the 4 datasets. 
SDR was computed across the whole utterance. Since datasets such as DiPCo present significant synchronization issues between far-field and close-talk signals, we used a filter with 4096 taps (instead of the default 512) for SDR.
As this modification alone was insufficient to address the synchronization problems, we also computed SDR for different time-shifts of the reference close-talk signal. Specifically, we considered offsets of -8192, -6144, -4096, -2048, 0, +2048, +4096, +6144, and +8192 samples. The final SDR value for each utterance was determined by selecting the maximum SDR value obtained across these nine offset positions.

%The 4 scenarios can have very different distributions for these metrics. 
%For example, DiPCo has fairly consistent SDR across all devices (the \emph{min}, \emph{max}, \emph{mean} SDR are similar), but the SDR is also lower compared to the other scenarios due to the fact that the devices are always far from the speakers. 
%NOTSOFAR-1 instead also exhibits consistent SDR across all microphones (and this is obvious since they belong to the same device) but the SDR distribution is centered around $0$\,dB instead
%indicating a better overall signal quality even if the variance is quite large. 
%Similarly for CHiME-6 the average SDR is distributed around $0$\,dB but, being multi-room and multi-device (previous Table~\ref{tab:dset_overview}, Section~\ref{sec:description}) has a large gap between the \emph{max} and \emph{min} SDR that can be obtained for each utterance. 
%This suggests that, for most utterances, there exists a microphone or a subset of microphones in which the speaker of interest has less interference and noise. 
%Similarly for Mixer 6, we have such channel selection problem. 
%The gap for the latter is even more noticeable and amounts to around $10$,dB due to the fact that some devices are placed far from the speakers and can be close to some noise sources, such as the ventilation system.

\noindent
The 4 scenarios exhibit distinctly different distributions for these metrics.
For example, DiPCo and NOTSOFAR-1 shows relatively consistent SDR across all devices (with close \emph{min}, \emph{max}, and \emph{mean} SDR values across devices). On the other hand, for  CHiME-6, being multi-room and multi-device has instead a significant disparity between \emph{max} and \emph{min} SDR obtainable for each utterance. This suggests that for most utterances, certain microphones or microphone subsets capture the speaker of interest with reduced interference and noise.
%NOTSOFAR-1 also demonstrates consistent SDR across all microphones (an expected outcome since they belong to the same device), but its SDR distribution is centered around 0\,dB, indicating better overall target speaker signal quality (but there is substantial variance).
Mixer 6 also presents a channel selection challenge, but with an even more pronounced gap of approximately 10\,dB. This substantial difference comes from the varied placement of microphone devices. Some are positioned at considerable distances from speakers and may be relatively near noise sources such as ventilation systems.

%To conclude, the mean SDR values presented in Figure~\ref{fig:sdr_stats}, along with their substantial variance across all scenarios, . These metrics effectively highlight the difficulties that must be overcome in far-field multi-speaker recognition tasks, where signal quality can vary dramatically depending on microphone positioning, room acoustics, and the dynamic nature of conversational speech. The significant differences in channel quality emphasize the importance of developing robust systems capable of operating effectively across diverse acoustic environments and microphone configurations.
In general, the mean SDR values in Figure~\ref{fig:sdr_stats} and crucially their significant variance (observed for all scenarios in Figure~\ref{fig:sdr_stats}) are good indicators of the difficulty of these challenges and, more broadly, far-field multi-speaker ASR. The significant difference in channel quality emphasizes the importance of the research directions that DASR challenges aim to foster. Only a robust system capable of operating effectively across diverse acoustic environments and microphone configurations can tackle this issue, whereas ad hoc systems will fail.

\subsubsection{CH\-iME-6}\label{sssec:chime6}
CH\-iME-6~\citep{watanabe2020chime} consists of recording ``sessions'' of different dinner parties between friends.
Each dinner party is attended by $4$ participants and takes place each time in a different home environment. Participants can move freely in the kitchen, living, and dining rooms. Thus, at a particular time, they may be scattered across different rooms (e.g., kitchen and living room).

In the home environment, there are also $6$ Kinect array devices with $4$ microphones each to capture the speech of the participants.
Close-talk microphones are also made available for annotation and training purposes.
However, these close-talk recordings still contain significant noise and cross-talk.
Due to the particular ``casual'' dinner party setting, CH\-iME-6 is characterized by highly informal conversations with fast turn-taking and high environmental noise, for example, cooking or dining together.
% Close-talk microphones are also made available for annotation and training purposes.
% These are, however, on-speaker binaural microphones and thus still contain significant noise and cross-talk.

It is worth mentioning that CH\-iME-6 and CH\-iME-5 share the exact same data, but with one important difference. CH\-iME-6 addresses the severe inter-array synchronization issue of CH\-iME-5 data. 
This synchronization issue was due to the compounding effect of packet losses and clock drift and led to inter-array signals misalignment of even several seconds in some instances. 
In CH\-iME-6 this misalignment is reduced to a few thousand samples on average as the data are resynchronized using the video from each Kinect (available only to CH\-iME-6 organizers). 
However, this misalignment is still quite significant as it amounts on average to several thousands of samples, thus posing a not-so-trivial challenge to the participants.
% it thus poses challenges to the participants.
% how much does it reflect real world applications ? 
%, posing challenges for any participant. 

%data has drastically improved the significant inter-array misalignment problem  FILLME 
% talk about misalignment 

In particular, in the C7DASR and C8DASR challenges, we provided universal evaluation map (UEM) files to correct an issue of the previous CH\-iME-6 challenge. In CH\-iME-6, in each session, the very first minute was included by mistake in the evaluation. 
This contained a speaker enrollment section that was not annotated, but was scored nonetheless, thus leading to an overestimation of insertion errors. 
% here we rescore the results using the UEM. so that we can cross compare all systems between the three challenges. 

CH\-iME-6 consists of a total of $24$ recording sessions: $16$ \texttt{train} (training), $2$ \texttt{dev} (development), and $2$ \texttt{eval} (evaluation). These three splits have no speakers in common, and each session is about 2 to 2 and a half hours long.  
In C7DASR, two training sessions (\texttt{S19} and \texttt{S20}) with no overlapping speaker-ids were moved to evaluation to increase the amount of evaluation data. C8DASR returns to the original CH\-iME-6 split as we did not observe, in C7DASR, this expansion providing any significant additional insight.  
Thus, in the remainder of this paper, we consider the \texttt{eval} split of CH\-iME-6 and C8DASR. This includes the results presented in Section~\ref{sec:results} for the CH\-iME-6 scenario. Importantly, such choice allows us to consistently compare systems in CH\-iME-6, C7DASR, and C8DASR challenges.

\subsubsection{Dinner Party Corpus (DiPCo)}\label{sssec:dipco}
The dinner party corpus (DiPCo)~\citep{van2019dipco} is directly inspired by CH\-iME-5/6 and, as the name implies, also features a dinner party scenario between 4 participants. Compared to CH\-iME-5/6, where dinner party takes place in a home environment with possibly multiple rooms, in DiPCo all recording sessions take place in a single room shared between all sessions.
The participants' speech is captured using $5$ far-field devices, each with a 7-mic circular array (6+1 microphone in the center) scattered throughout the room.
Again, closed-talk on-speaker lapel microphone signals are also provided for annotations and training purposes. Compared to the binaural microphones used in CH\-iME-5/6 these have significantly less cross-talk and noise. However, they present severe misalignment with far-field microphone signals, with offsets exceeding 1000 samples. This misalignment is also often noncausal, meaning the close-talk signal can precede the far-field one. 

DiPCo consists of $10$ sessions originally partitioned into $5$ for \texttt{dev} and $5$ for \texttt{eval} without overlapping speaker-ids but, as said, with the same room and microphone placement shared across all recordings. 
C7DASR keeps this original split, which lacks an adaptation/training set. As a result, during C7DASR many participants used part of the \texttt{dev} to adapt/fine-tune their systems. As a result, in C8DASR we made an explicit \texttt{train} split, also made by using part of the original \texttt{dev} set, again splitting to ensure that there are no overlapping speakers.  
Since the same room and mic placement is shared across all recordings, in some way, DiPCo can be considered a ``best-case'' scenario when it is possible to fine-tune/adapt a system with respect to a specific environment. This seldom happens for real-world applications, but it is reasonably possible in some domains (e.g. in-vehicle applications). 

Compared to CH\-iME-6, DiPCo is characterized by arguably less noise and less overlapped speech (as is evident in Table~\ref{tab:class_stats}) because the setting is more formal and the activities of the participants are more limited (e.g., no cooking).  
Moreover, it does not present the synchronization issues of CH\-iME-6, all the signals from all the arrays are sample-synchronized. 

%originally divided as $5$ \texttt{dev} and $5$ \texttt{eval} 

%However, compared to CH\-iME-6 all recordings take place in a single room, features arguably less informal conversations and, on average, have higher SNR. 
%Recordings are captured by  and by close-talk on person lapel microphones. These latter have far-less cross-talk and noise than CH\-iME-6 ones. 
%DiPCo originally consists of $10$ sessions: $5$ \texttt{dev} and $5$ \texttt{eval}. 
%As said in Section~\ref{sec:motivation}, we further split the original \texttt{dev} set into a \texttt{train} (3 sessions) and a \texttt{dev} partition ($2$ sessions) with approximately the same total duration. 

\subsubsection{Mixer 6 Speech (MX6)}\label{sssec:mx6}

The Mixer 6 Speech corpus (MX6)~\citep{brandschain2010mixer, cornell2023chime} consists of 1425 recording sessions carried out between 2009 and 2010 among 594 native English speakers. 
Each recording session consists of mostly two-party conversations between an interviewer and a subject, and is divided into
4 parts: i) entry questions, ii) an interview, iii) transcript reading, and iv) a phone call.

The speech is captured by 10 different heterogeneous far-field recording devices scattered across the room where the session takes place. These include arrays (e.g. an Acoustimagic Array, with a single signal provided after internal processing), but also commercial single-microphone devices (RODE NT6) and even multi-purpose media recording devices (Panasonic Camcorder).  
As such, it presents a significantly different recording setup from CH\-iME-6 and DiPCo. 
Close-talk microphones are also available for the subject and interviewer. However, the subject lapel microphone has significant crosstalk in some sessions while a headset microphone is only available for the interviewer. The recording sessions are conducted in two different rooms called ``LDC'' and ``HRM''. 
In C7-8DASR, we consider only the interview portion for evaluation as it is the only multi-talker conversational speech portion in each session. The average duration is $\sim$ 14 minutes and is carried out by the interviewer in such a way as to elicit an informal conversation. 

Note that the original MX6 release~\citep{brandschain2010mixer} was not transcribed other than a small section of read speech spoken by a single speaker. A portion ($\sim$8.9 hr) of the telephone call transcripts were released as part of the ASpIRE Challenge~\citep{harper2015automatic}.  
A significant contribution of C7DASR, led by Matthew Wiesner and Desh Raj, was to produce annotations for 450 interview portions of the total 1425 conversations in order to obtain at least \texttt{dev} and \texttt{eval} splits with full annotation of respectively 41 and 23 sessions.
The split is performed using the NACHOS toolkit~\footnote{\url{https://github.com/m-wiesner/nachos}} to ensure that there is no overlap between the interviewer, the subject, or the room between the two sets. 
In C8DASR, the development annotation is divided into \texttt{ train} and \texttt{dev} (see Table~\ref{tab:class_stats}) portion to provide participants with a small annotated training set for adaptation purposes (same rationale as done for DiPCo). 

%Human-annotated time-marked transcriptions are available 
The annotation for the interviewer was obtained semi-automatically using Whisper~\citep{radford2022robust} \texttt{large} on the close-talk interviewer lapel microphone followed by forced alignment using Kaldi~\citep{povey2011kaldi} toolkit.
Subsequently, this annotation is manually checked and corrected. 
Note that full conversation annotation for the remaining 975 sessions is still partial, with only the subject speech available. 
This data, together with the call portion, is made available for participants as an additional two training sets: \texttt{train\_calls} and \texttt{train\_intv}. Note that in Table~\ref{tab:class_stats} silence and overlapped speech statistics for these splits are not reported as the annotation is partial. 

\subsubsection{NOTSOFAR-1}\label{sssec:notsofar1}

The natural office talkers in settings of far-field audio recordings (NOTSO\-FAR-1)~\citep{vinnikov2024notsofar} introduced in the CH\-iME-8 NOTSOFAR-1 challenge consists in  $315$ recordings of office meetings. Each meeting is very short ($\sim 6$\,mins) and is attended by $4$ up to $8$ speakers. The conversation is mediated by a professional actor whose job is to guide it around a certain topic. %As such, similarly to AMI, NOTSOFAR-1 offers, for each meeting, some metadata information which can be exploited for downstream task evaluation or for error analysis. This metadata includes the topic as well as  
%meeting characteristics summarized as hashtags: 
%\begin{itemize}
%    \item 
%\end{itemize}
The same meeting is captured by up to $7$ commercially available far-field array devices. These include $4$ tabletop circular array devices with $7$ microphones each and $3$ linear array devices. For the latter, however, only monaural signals after in-device acoustic front-end processing (AFE) is made available. This presents some issues as the AFE pipeline could suppress some speakers, but nonetheless it is also a use-case that can happen in practical scenarios. 
Note that NOTSOFAR-1 data is shared between CH\-iME-8 DASR and CH\-iME-8 NOTSOFAR-1 challenges, and these single channel data was originally conceived for NOTSOFAR-1 challenge single channel track. In CH\-iME-8 DASR, the single-channel recordings are available only for training purposes (\texttt{train\_sc} in Table~\ref{tab:class_stats}), as the focus is strictly on multi-channel. 
Importantly, while each meeting is captured by many devices, in NOTSOFAR-1 the signals captured by each device are considered as a single, independent ``session'' in order to constrain participants to focus also on single-array application scenarios, which are arguably the most common. 
As such, the total number of sessions, in Table~\ref{tab:class_stats}, is 1195 despite having only $315$ unique meetings. 

Again, as with the other corpora, on-person close-talk headset devices with low cross talk are made available for all \texttt{train} and \texttt{dev} splits. 
NOTSO\-FAR-1 shares similarities with AMI, but distinguishes itself by featuring more speakers and a more dynamic conversational setting. Transcriptions are of high quality, with word-level alignment provided, as well as useful metadata such as special tags for filler words. 
Transcriptions are obtained using a multi-judge system by using two different transcribers without any automatic aid, in order to reduce the risk of introducing bias in the annotation procedure.  
%were obtained with 
%The main challenge of NOTSOFAR-1 is the fact that it is highly diverse compared to the other three corpora. It is the only one with single array and the duration is very short while the number of speakers can be quite high. As we will see in Section~\ref{sec:baseline} this latter characteristic makes it very difficult to correctly estimate the number of speakers while keeping, at the same time, decent performance on CH\-iME-6. 

% quality of transcriptions multi judge filler words and so on... 

%However, for inference, participants are bounded to use only one device only at a time
%such as particularly noisy meeting
%In total, there are $32$ unique participants captured in 30 different meeting rooms.

%and most are started by a professional actor whose task is to ``jumpstart'' and guide the conversation around a certain topic. 
%Each meeting is captured by up to $7$ commercially available far-field array devices. These include $4$ tabletop circular devices with $7$ microphones each and $3$ linear array devices. 
%In addition to far-field devices on-person close-talk headset devices with low cross talk are made available for \texttt{train} and \texttt{dev} splits. 
%Ground truth annotation is available in the form of JSON files. Such annotation %also includes word level alignment for each speaker utterance as well as metadata %such as the meeting topic. Transcriptions were obtained with a multi-judge system %as described in~\citep{vinnikov2024notsofar}. 
%A more detailed description is available in~\citep{vinnikov2024notsofar}. 
\subsection{Rules}\label{ssec:rules}
%Rules are largely the same as the previous C7DASR challenge. The main rationale is to discourage automatic or manual domain identification or any use of a-priori information from the scenarios. Among such information is knowledge about array type and topology. 

The main rationale of the C7-8DASR rules is to prevent specialized approaches that, e.g., use domain identification to tackle one scenario at a time or leverage a-priori information such array geometry and thus ``by design'' are limited in their generalization capability.
A truly generalist, versatile system can almost always be fine-tuned or improved by exploiting some domain-specific knowledge (e.g. array geometry), while the contrary is far more challenging to realize. 
As such, participants are forbidden to use manual or automatic domain identification or any information on the recording setup (included the total number of microphones) or number of speakers. They need to produce a single array-topology agnostic system for all scenarios. 

For training purposes, only data from the \texttt{train} splits of the core scenarios can be used together with the external data sources allowed. %\footnote{\href{https://www.chimechallenge.org/challenges/chime8/task1/data\#external\_datasets}{https://www.chimechallenge.org/challenges/chime8/task1/data\#external\_datasets}}. 
%Similarly, only external pre-trained models as listed on the website\footnote{available: \href{https://www.chimechallenge.org/challenges/chime8/task1/rules\#external\_models}{https://www.chimechallenge.org/challenges/chime8/task1/rules\#external\_models}} can be leveraged. 
We encouraged participants to propose additional external pre-trained models and data sources during the first month of the challenge. Up to 8 pre-trained models and 5 external data sources as proposed by participants were added during C7DASR. 
A detailed description of the rules is available on the challenge website\footnote{see \href{https://www.chimechallenge.org/challenges/chime8/task1/rules}{https://www.chimechallenge.org/challenges/chime8/task1/rules}}.

\subsection{Evaluation tracks}\label{ssec:tracks}

Since C7-8DASR challenges are designed to be a continuation of the CH\-iME-6 challenge, the evaluation setup is also inherited. Participants must perform joint ASR and diarization of different meeting scenarios using far-field recording devices and produce a transcription with speaker attribution and segmentation at the utterance level. 
As CH\-iME-6, C7DASR featured another optional ``acoustic robustness'' track, where oracle diarization information could be used. The motivation for such a track was to probe to what extent the error was due to diarization or, instead, to other components such as the ASR system and/or the particular acoustic front-end separation technique employed. 
In C8DASR, this latter optional track was removed, in order to keep the evaluation data fully blind for the newly added NOTSOFAR-1 scenario. 
Another important difference compared to CH\-iME-6 is that in all tracks no restriction is made on the language model used (LM), provided that it is trained only on the allowable training datasets. 
This was motivated by the fact that the CH\-iME-6-constrained LM track was too restrictive and practically forced many participants to adopt hybrid deep neural network (DNN)-HMM ASR models. 
%, to see if they can be exploited to improve performance and to what extent.
In C8DASR, an additional, optional track was proposed that instead allowed also the use of some pre-trained external LLMs. 
This direction was motivated by recent work that found that LLMs are effective, in particular, for diarization post-processing~\citep{park2023enhancing, wang2024diarizationlm}. 
However, this track received limited participation during the challenge, with only one team submitting and reporting very marginal improvements compared to the main track. 
On the other hand, in a recent work~\citep{ogawa2024applying}, it is shown how Llama 2~\citep{touvron2023llama} can be used effectively to improve transcription results on the C7DASR challenge. The authors found that by refining the ASR model’s N-best list, Llama 2 can significantly improve results, primarily due to its ability to model long-context, inter-utterances dependencies.

Additionally, in the C8DASR and the CH\-iME-8 NOTSOFAR-1 challenges, a ``jury award special mention'' was introduced to recognize teams that developed efficient and practically viable transcription systems. This initiative aimed to encourage research into approaches that did not rely heavily on ensembles, iterative inference schemes, or test-time adaptation. Although this track also received limited interest, one team succeeded in creating a more practical system with excellent results. These findings will be discussed further in Section~\ref{sec:results}.

%Moreover, at the same time, in C8DASR and together with CH\-iME-8 NOTSOFAR-1 challenge, a ``jury award special mention'' was considered for the team that produced a transcription system that was efficient and more practically interesting, to encourage research towards approaches that did not rely necessary on ensembles, iterative inference schemes and/or test-time adaptation. 
%This however received again limited interest, but, in particular, one team succeed in creating a more practically viable system with veryt good results. Again, this will be discussed further in Section~\ref{sec:results}.

%when used to refine an ASR model N-best list, mainly thanks through long-context modeling. 
% mainly thanks to its long context. 

%However, another team did a subsequent study on this. This we will explain more in detail in Section~\ref{sec:results}. 

% and by the many recent advancements in e2e ASR which can have an implicit language model. 

% mention failed evaluation tracks of c8dasr ? 
% mention about jury prize ? meh LLM failed jury prize only NTT did it. 

\subsubsection{Annotation and text normalization}\label{ssec:normalization}

Participants' systems have to produce, for each core scenario, a JSON SEGment-wise Long-form Speech Transcription annotation (segLST)~\citep{von2023meeteval} file. 
An short example of the content of this file for the CH\-iME-6 scenario is given in Listing~\ref{lst:annotation}. 

\begin{lstlisting}[language=json, firstnumber=1, basicstyle=\footnotesize, caption=JSON SEGment-wise Long-form Speech Transcription annotation (segLST) for two utterances. Each utterance is a dictionary with several attributes., escapechar=*, label={lst:annotation}]
{*“*end_time":   76.340,
 *“*start_time": 73.500,
 *“*words":      *“*let's do lunch",
 *“*speaker":    *“*P08",
 *“*session_id": *“*S02"},
{*“*end_time":   76.560,
 *“*start_time": 75.300,
 *“*words":      *“*okay",
 *“*speaker":    *“*P02",
 *“*session_id": *“*S02"}
\end{lstlisting}

For each transcribed utterance, the start, end, and an arbitrary (but consistent through the meeting) speaker-id tag must be inferred together with the words.
The \texttt{session\_id} is known and corresponds to the meeting session. 
The segLST format is inherited from the CH\-iME-5 challenge directly, but it differs from the fact that \texttt{end\_time} and \texttt{start\_time} are in seconds (instead of the \texttt{hh:mm:ss} format used in CH\-iME-5/6) to make the annotation easier to use. % handy for audacity 

Note that this format is also used for the annotation of the C7-8DASR core scenarios \texttt{train} and \texttt{dev} splits.
As stated in Section~\ref{ssec:accessibility}, the annotation and the directory structure are made consistent across all scenarios to make the data easier to parse and prepare. 
For the \texttt{train} and \texttt{dev} split, additional entries may be present, such as \texttt{location} (for CH\-iME-6) or \texttt{word\_alignment} (for NOTSOFAR-1) depending on the metadata available for each dataset. 
In detail, for \texttt{train} and \texttt{dev} splits, we make available to participants two separate JSON segLST annotations: one containing all the original dataset metadata and without text normalization, and another one, which has the same exact format (with only the base entries as in Listing~\ref{lst:annotation}) participants need to produce in inference and has text normalization applied. 

Text normalization differs between C7DASR and C8DASR.
C7DASR builds on CH\-iME-6 text normalization: punctuation is removed, and all characters are converted to lower case. In addition, nonverbal speech sounds such as ``uhm'', ``umh'' etc. are assigned to ``hmmm'' to be consistent across all datasets. 
%expands it as it features more scenarios, each with different transcription convention.  
% correcting mistakes from CH\-iME-6 challenge 
In C8DASR, text normalization is more sophisticated and instead builds on the Whisper text normalization pipeline, but with crucial modifications.
First, it was modified to be idempotent. The original Whisper text normalization gives inconsistent results if applied more than once on the same text. We argue that while this can be acceptable for data preprocessing, it is not ideal for scoring: text normalization should be a straightforward and well-defined mapping that can, in principle, also be applied manually. 
Secondly, we remove number normalization (i.e. ``two dollars'' will not be converted to 2\$) so that the end result is more verbatim.
Compared to C7DASR normalization, nonverbal speech sounds such as ``uhm'', ``uhhh'', ``ah'' are completely removed, and common abbreviations are expanded (e.g. ``goin'' to going)~\footnote{Some examples are available here: \href{https://github.com/chimechallenge/chime-utils/blob/main/tests/test\_normalizer.py}{https://github.com/chimechallenge/chime-utils/blob/main/tests/test\_normalizer.py}}. 
This procedure ensures that the task is well defined: the text is more consistent among the scenarios. Moreover, it avoids biasing the final results by filler sounds (``uhm'', ``uhhh' etc.)  which otherwise would be among the most common words. This same text normalization was adopted by the concurrent CHiME-8 NOTSOFAR-1 challenge, as the goal was to be able to compare systems across the two. 

For these reasons, in the following sections, we will report all results (including those of C7DASR) always using C8DASR text normalization when scoring participant submissions. 

% text normalization is applied to the text one utterance at a time and not for concatenated utterances. 

\subsection{Baseline Systems}\label{ssec:baselines}

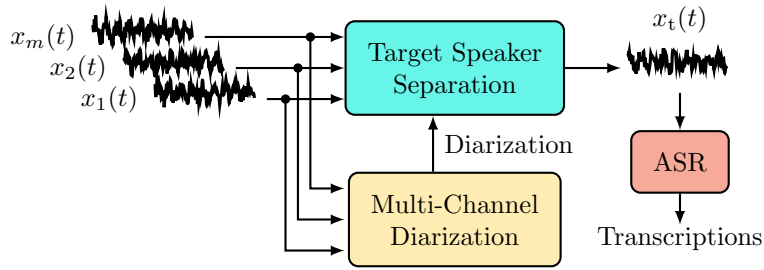
\begin{figure}
    \centering
    % \documentclass[crop,tikz]{standalone}

% \usetikzlibrary{fit,tikzmark,calc}

% \begin{document}
  
\begin{tikzpicture}[every node/.style={font=\footnotesize}, declare function={
      excitation(\t,\w) = sin(\t*\w);
      noise = rnd - 0.5;
     source(\t) = excitation(\t,200) + noise*3;
      filter(\t) = 1 - abs(sin(mod(\t, 500)));
      speech(\t) = 1 + source(\t)*filter(\t);
    }]

% \tikzset{>=stealth}
\tikzset{>=latex}
\tikzstyle{branch}=[{circle,inner sep=0pt,minimum size=0.3em,fill=black}]
\tikzstyle{block}=[
    draw,
    text depth=0pt,
    thick, 
    rectangle,
    text centered,
    minimum height=3.2ex,
    rounded corners=0.3em,
    fill=black!6,
    inner sep=0.7em,
    ]
\tikzstyle{arrow}=[{}-{>}, thick]

\tikzset{
  wav2/.pic={
    \begin{scope}[x=0.2em,y=0.07em]
      \coordinate (-west) at (-1,0);
      \coordinate (-east) at (18,0);
      \coordinate (-south) at (8.5,-8);
      \coordinate (-north) at (8.5,8);
      \foreach \x/\y in {0/1.2,1/2,2/4,3/7,4/4,5/2,6/1.2,7/2,8/1.2,9/2,10/4,11/7,12/4,13/7,14/1.2,15/2,16/5,17/1.2/2} {
          \draw[line width=0.13em] (\x,\y) -- (\x,-\y);
      }
    \end{scope}
  }
}

\tikzset{
wav/.pic={
            \begin{scope}[x=0.2em,y=0.1em]
                \coordinate (-west) at (-1,0);
                \coordinate (-east) at (18,0);
                \coordinate (-south) at (8.5,-8);
                \coordinate (-north) at (8.5,8);
                \draw[line width=0.1em, smooth] plot[domain=0:16, samples=144] 
                    (\x, {2*speech(\x)});
            \end{scope}
    }
}

\definecolor{tss}{HTML}{67f5e9}
\definecolor{mcd}{HTML}{ffecb2}
\definecolor{asr}{HTML}{f7aa99}

\node[block,align=center,fill=tss] (tss) at (0, 0) {Target Speaker\\Separation};
\node[block,align=center,anchor=north,fill=mcd] (mcd) at ($(tss.south)+(0,-1.8em)$) {Multi-Channel\\Diarization};

\pic (wavOne) at ($(tss.west) + (-6.1em,-1em)$) {wav};
\pic (wavTwo) at ($(tss.west) + (-7.1em,0)$) {wav};
\pic (wavThr) at ($(tss.west) + (-8.1em,1em)$) {wav};

\coordinate (tmp) at ($(tss.south) + (-0.7em,0)$);
\draw[arrow] (mcd.north-|tmp) -- node[right]{Diarization} (tmp);
\draw[arrow] (wavOne-east) -- (tss.west|-wavOne-east);
\draw[arrow] (wavTwo-east) -- (tss.west|-wavTwo-east);
\draw[arrow] (wavThr-east) -- (tss.west|-wavThr-east);

\draw[arrow] ($(tss.west|-wavOne-east)$) +(-1.9em,0) node[branch] {} |- ++($(mcd)-(tss)$);
\draw[arrow] ($(tss.west|-wavTwo-east)$) +(-1.5em,0) node[branch] {} |- ++($(mcd)-(tss)$);
\draw[arrow] ($(tss.west|-wavThr-east)$) +(-1.1em,0) node[branch] {} |- ++($(mcd)-(tss)$);

\node[anchor=east] () at ($(wavOne-west) + (0,0)$) {$x_1(t)$};
\node[anchor=east] () at ($(wavTwo-west) + (0,0)$) {$x_2(t)$};
\node[anchor=east] () at ($(wavThr-west) + (0,0)$) {$x_m(t)$};

\pic (wavOut) at ($(tss.east) + (2em,0em)$) {wav};
\node [above] () at (wavOut-north) {$x_{\mathrm{t}}(t)$};

\node[block,align=center,anchor=north,fill=asr] (asr) at ($(wavOut-south)+(0,-1.2em)$) {ASR};

\draw[arrow] (wavOut-south) -- (asr);
\draw[arrow] (tss) -- (wavOut-west);
\draw[arrow] (asr.south) -- node[below=0.2em]{Transcriptions} +(0,-1em);

\end{tikzpicture}
  
% \end{document}
    \caption{ESPnet and NeMo baseline systems high-level overview. This same scheme was adopted by almost all C7-8DASR participants and top performing systems in the ``twin'' CH\-iME-8 NOTSOFAR-1 challenge.}
    \label{fig:baseline_scheme}
\end{figure}

In C7DASR, as mentioned, a baseline system implemented with ESPnet~\citep{watanabe2018espnet} was provided\footnote{Available: \href{https://github.com/espnet/espnet/blob/master/egs2/chime7\_task1}{https://github.com/espnet/espnet/blob/master/egs2/chime7\_task1}}. 
In C8DASR, an additional baseline was added\footnote{Available: \href{https://github.com/chimechallenge/C8DASR-Baseline-NeMo}{https://github.com/chimechallenge/C8DASR-Baseline-NeMo}}, derived from the NVIDIA NeMo team submission to C7DASR. In C8DASR, the ESPnet baseline remained largely unchanged\footnote{Available: \href{https://github.com/espnet/espnet/blob/master/egs2/chime8\_task1}{https://github.com/espnet/espnet/blob/master/egs2/chime8\_task1}} except for some hyperparameters for the diarization component which were tuned to deal with the new NOTSOFAR-1 scenario.

The two baselines differ mainly in the diarization technique used. Both follow the pipeline illustrated in Figure~\ref{fig:baseline_scheme} consisting of multi-channel diarization followed by target speaker separation/extraction (TSE) via GSS~\citep{boeddeker2018front} and finally ASR on each separated utterance. 
This approach aligns with the strategies employed by the winning teams in the previous two CH\-iME-5 and CH\-iME-6 challenges~\citep{arora2020jhu, chen2020improved, medennikov2020stc}, where GSS has been the de-facto standard approach for front-end speech separation due to its effectiveness. 

\subsubsection{Target Speaker Separation}\label{sssec:tse_baselines}
Both baselines employ the same TSE pipeline consisting of channel selection using the envelope variance method (EV)~\citep{WOLF_EV_2014} followed by GSS. 

The inclusion of EV-based channel selection was a novel feature introduced in the C7DASR baseline.
The motivation for using channel selection was twofold. First, it allows for accelerating the inference process and/or saving GPU memory, as the particular GPU-based GSS implementation scales quadratically in the GPU memory requirements with the number of channels used. 
Secondly, it also allows for a marginal improvement in performance because microphones with lower quality signal are excluded. 
In our previous work~\citep{cornell2023chime} introducing the C7DASR baseline, we found that there is an optimal trade-off between these two effects that occurs for C7DASR, when 80\% of the microphones are retained.  
The choice of using EV instead of other channel selection measures~\citep{WOLF_EV_2014, cornell2021learning} was dictated mainly because it is a simple and computationally efficient measure which has been found to be quite effective in the CH\-iME-5/6 scenario in a recent work~\citep{cornell2021learning}, even compared to more sophisticated data-driven methods or ASR-based measures. 

GSS~\citep{boeddeker2018front} employs a spatial mixture model (SMM)~\citep{ito2016cACGMM} guided by diarization estimates to obtain an enhanced signal for an utterance.
An SMM is fitted for each utterance (according to the diarization estimate) using observations from the utterance boundaries plus some context which can span several tens of seconds. 
The diarization estimate is used to indicate potential source activity. After this fitting, the model's posterior is used for mask-based beamforming. The minimum variance distortionless response (MVDR)~\citep{capon_mvdr} beamformer is used followed by a-posteriori maximum SNR channel selection~\citep{erdogan2016improved} and blind analytic normalization (BAN)~\citep{warsitz2007blind} post-filtering. As such, GSS does not use any training data, as it is simply fitted at inference time on each utterance. Thereby, it does not suffer from the domain mismatch issues often encountered with supervised DNN methods. 
As will be shown in Section~\ref{sec:systems_description}, this same pipeline, comprising EV-based channel selection followed by GSS, was also adopted by almost all participants without significant modifications, further demonstrating its reliability and effectiveness.

% why it is effective ? discuss later. 

%We will see in Section~\ref{sec:description} that this same pipeline of EV-based channel selection followed by GSS was also adopted by almost all participants, confirming its effectiveness. 

%As said, regarding GSS, we employ the GPU-accelerated implementation from~\citep{}. 
%One important fact about the GSS implementation is that 
% important difference is the higher context in the GPU based implementation. 
% would be cool to do some ablations here. 

%GSS consists of multi-input multi-output weighted prediction error (MIMO-WPE)~\citep{} dereverberation, 
%These baselines are described in detail in the following. 

%It leverages a modified Pyannote diarization pipeline for diarization and, for separation, a GPU-based GSS implementation~\citep{}. 

\subsubsection{Diarization}\label{sssec:diar_baseline}

\begin{figure}
    \centering
    \includegraphics[width=0.6\linewidth]{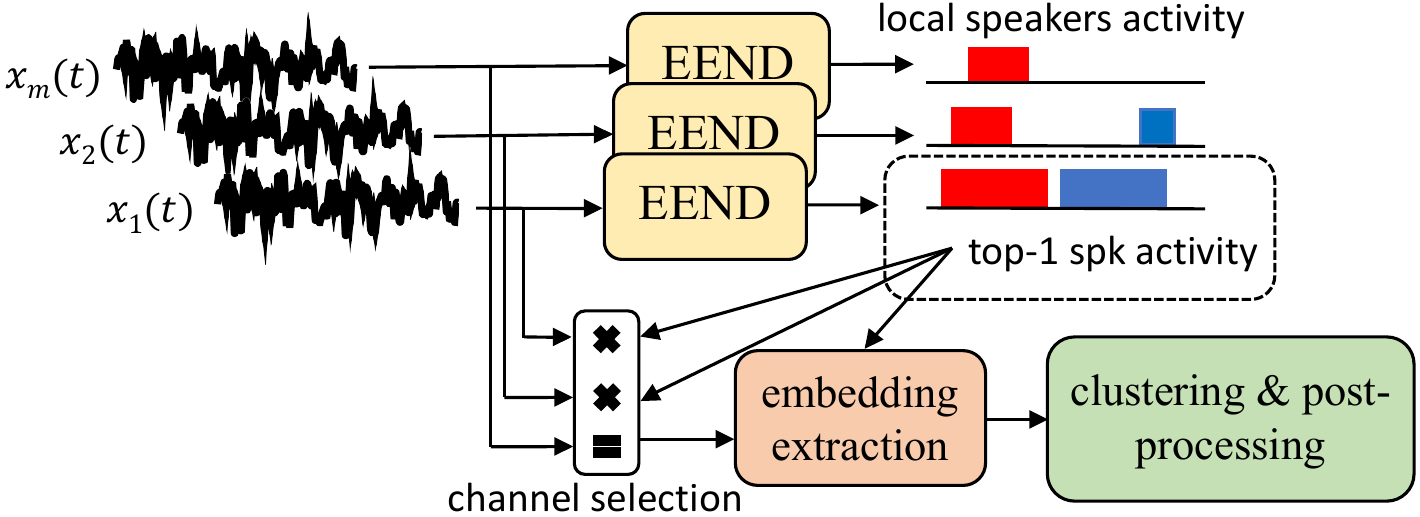}
    \caption{ESPnet baseline diarization pipeline scheme.}
    \label{fig:espnet_diar_scheme}
% \end{figure}
\vspace{1em}
% \begin{figure}
    \centering
    \includegraphics[width=0.6\linewidth]{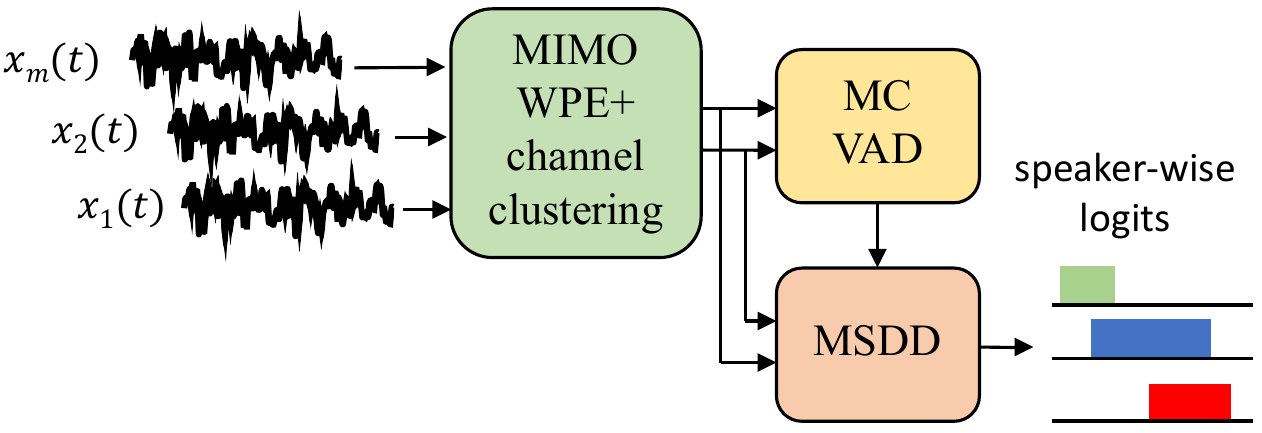}
    \caption{NeMo baseline diarization pipeline scheme.}
    \label{fig:nemo_diar_scheme}
\end{figure}

As said, the two baselines differ substantially in the diarization component employed. 
The ESPnet baseline diarization component is illustrated in Figure~\ref{fig:espnet_diar_scheme} while the NeMo one is in Figure~\ref{fig:nemo_diar_scheme}.

The ESPnet baseline diarization component is built on top of the Pyannote~\citep{bredin2020pyannote} diarization 2.1 pipeline~\citep{bredin2023pyannote} which is suitably modified to deal with multiple microphone inputs. 
The latter consists of a local EEND model~\citep{fujita2019end, bredin2021end} and the pre-trained ECAPA-TDNN~\citep{desplanques2020ecapa} speaker-id embedding model.
The local EEND model predicts speech activation for up to 3 concurrent speakers within a context of 5\,s and is applied to the whole meeting with a 0.5\,s stride and to every microphone channel. 
We use this model to select, for each 5\,s window, the channel with the highest total speech activity. This channel is then used to extract  ECAPA-TDNN speaker-id embeddings with the overlap-aware mechanism proposed in~\citep{bredin2021end, bredin2023pyannote}.
To mitigate false alarms, the speaker activity selection decision is made after logit thresholding and median filtering applied to each microphone channel.
This selection strategy is strongly needed as, especially in scenarios as CH\-iME-6, where diarization error rate (DER) between different microphones can vary by more than 10\%. 
We also experimented with ensembling the diarization pipeline results across arrays using DOVER-Lap~\citep{raj2021dover}. However we found this latter solution to have significantly worse performance on CH\-iME-6. 
Moreover, it is intrinsically significantly slower in inference, due to the fact that the whole pipeline must be run on every channel. 
In the proposed channel selection strategy instead, only the local EEND model is run on each channel, leading to a significant speed up, as computationally heavy operations, such as ECAPA-TDNN embedding extraction, are performed only for one channel.
%and, moreover, slightly worse performance than the proposed selection approach. 
The local EEND model is fine-tuned on the CH\-iME-6 \texttt{train} set (using all far-field channels only) using for validation and early stopping the MX6 \texttt{dev} set. 
Diarization output post-processing is employed in order to avoid too long utterances or too short utterances, both of which will hurt ASR performance. In detail, we merge utterances from the same speaker that are 0.5\,s seconds apart up to a maximum duration of 30\,s and with a target duration of around 10\,s. Utterances that are instead longer than 30\,s are split.
We tune hyper-parameters, such as clustering and local EEND model threshold, including this diarization post-processing, using the Optuna framework~\citep{akiba2019optuna} with macro JER (over MX6, CH\-iME-6 and DiPCo \texttt{dev} sets) as the optimizing criteria and tree-structured Parzen estimator~\citep{ozaki2020multiobjective} as the optimization method.
For the updated ESPnet C8DASR baseline these hyper-parameters are re-tuned by adding also a part of NOTSOFAR-1 \texttt{train} set (the \texttt{dev} set was not available when building the baseline).

The NeMo baseline diarization pipeline is based on a modified multi-scale diarization decoder (MSDD) technique~\citep{park2022multi} which is an improved variant of TS-VAD.  
First, as a pre-processing step, multi-channel dereverberation is performed via MIMO-WPE~\citep{yoshioka2012generalization} applied on 40\,s windows with 2\,s overlap. Next, to reduce the number of channels, channel clustering is performed using normalized maximum eigengap spectral clustering (NME-SC) clustering~\citep{park2019auto} on the spatial coherence matrix obtained from all microphone channels by averaging across the whole meeting. 
To identify speech and silent regions, a multi-channel VAD model (MC-VAD) is employed. This model is based on the pre-trained MarbleNet VAD model, which is then fine-tuned on a 3000 hours mix of CH\-iME-6 \texttt{train} data and synthetic data obtained from VoxCeleb 1\&2, LibriSpeech and MUSAN using the NeMo multi-speaker simulator~\citep{park2023property}.  
Such VAD model is applied on each clustered channel independently, with logits fused by a max operation over all channels for each timestep. 
Multi-resolution TitaNet~\citep{koluguri2022titanet} speaker-id embeddings are extracted on each speech segment detected by the VAD using 3\,s, 1.5\,s, and 0.5\,s all with 50\% overlap. 
The MSDD model is then initialized using NME-SC clustering over these multi-resolution TitaNet embeddings. To create multi-channel aware speaker-id embeddings, the TitaNet embeddings from each clustered channel belonging to the same segment are concatenated in this step. However, the channel with the lowest correlation is excluded. 
Then after a first diarization hypothesis, the MSDD model is applied on each clustered channel independently (thus not using multi-channel aware speaker embedding) with a context size of 15\,s and an hop-size of 3\,s and outputs a logit for each speaker each 0.05\,s using input embedding scale interpolation~\citep{park2023chime}. 
The final diarization output is derived, after logit thresholding, via majority voting over all clustered channels for each timestep. 
The particular MSDD model differs from the original MSDD model proposed in~\citep{park2022multi} as it employs a four layers Transformer network with an embedding size of 384 and feed-forward network hidden size of 2048 instead of long-short term memory networks. 
It is trained using the same 3000\,h data employed for the VAD model consisting of CH\-iME-6 and simulated conversations. 
As in the ESPnet baseline, here diarization post-processing is also employed. Specifically, too short silences or utterances are suppressed, and onset and offset padding is applied on each utterance in order to avoid truncated words. 
Hyper-parameters were tuned using the Optuna framework on the C7DASR (C8DASR for the C8DASR challenge) \texttt{dev} set using DA-WER as the optimizing criteria. 

%, which is here extended in order to deal more effectively with multiple channels in an array-agnostic manner. 
%The Pyannote diarization 2.1 pipeline consists in a local EEND model with a context of 5\,s

\subsubsection{Automatic Speech Recognition}\label{sssec:asr_baseline}

Both baselines employ a ``classical'' single-channel ASR back-end; i.e. no target-speaker ASR or multi-channel features are integrated into the ASR model. 

The ESPnet baseline ASR back-end is based on IRIS~\citep{chang2022end} and multi-IRIS~\citep{masuyama2022end} ASR back-ends. It integrates WavLM as a front-end for a Transformer~\citep{vaswani2017attention} attention-based encoder-decoder model. 
WavLM \texttt{large} is employed and its weights are kept frozen. Instead, its internal representation is fed to the Transformer encoder through layer-wise learnable softmax weights, a technique which is commonly employed~\citep{yang21c_interspeech, chang2022end, masuyama2022end, yang2024large}. 
The model consists of 12 encoder layers and 6 decoder layers for a total of only 30\,M trainable parameters as the pre-trained WavLM representation is leveraged.
It is trained using hybrid CTC/attention loss~\citep{Watanabe2017} on the full CH\-iME-6 \texttt{train} and MX6 \texttt{train\_intv} and \texttt{train\_calls} portions, using data from all microphone devices.  
Furthermore, data-augmented close-talk microphone data is created from both MX6 and CH\-iME-6. 
We follow the data augmentation pipeline developed for the Kaldi CH\-iME-6 baseline~\citep{watanabe2020chime}, where MUSAN~\citep{snyder2015musan} and SLR26~\citep{ko2017study} are employed to get noisy/reverberant signals from  close-talk microphone utterances. 
To obtain satisfactory performance, we found critical to also include GSS-enhanced data (obtained using oracle diarization) from the CH\-iME-6 \texttt{train} set. 
The model is trained for 5 epochs, and the weights from the best 3 checkpoints are averaged. Adam is used for optimization with a peak learning rate of 0.0001. A warmup schedule of 40k steps is employed followed by linear decay. SpecAugment~\citep{park2020specaugment} masking is performed during training on the WavLM representation after the learnable softmax operation.
%The entire training process takes approximately 18 hours on two A100 GPUs. 
Hybrid CTC attention decoding is then performed in inference. We use a beam size of 10 and a CTC weight of $0.3$. 

The NeMo baseline ASR model is based instead on the pre-trained NeMo Conformer~\citep{gulati2020conformer} transducer \texttt{XL} model\footnote{Available: \href{https://huggingface.co/nvidia/stt\_en\_conformer\_transducer\_xlarge}{https://huggingface.co/nvidia/stt\_en\_conformer\_transducer\_xlarge}}. 
The pre-trained model is fine-tuned using CH\-iME-6 and MX6 \texttt{train} subsets data after pre-processing with GSS using  ground-truth diarization. %This fine-tuning phase
%lasted for 35,000 updates with a batch size set to 128.
The model consists of 600\,M parameters and is updated for 35\,k steps with 128 batch size and a learning rate of $0.00001$. 
For decoding, adaptive expansion search is employed~\citep{kim2020accelerating} together with an external word-piece LM.
The LM is based on byte-pair-encoding (BPE) tokens using SentencePiece~\citep{kudo2018sentencepiece} and is constructed using the KenLM toolkit~\citep{heafield2011kenlm} by using text data from C7DASR \texttt{train} and dev \texttt{sets}.

%This LM is word-piece level Ngram language model with byte-pair-encoding (BPE) tokens using SentencePiece [23, 24] and KenLM [25, 26] toolkits from
%transcription of CH\-iME-7 train and dev sets. Token sets of our
%ASR and LM model were matched. To combine several N-gram
%models with equal weights we used OpenGrm library [27, 28].
%MAES decoding [29] was used for transducer.

% this will be very long probably different pages 

%Since the C7-8DASR challenges 
%An useful contribution of C7-8DASR challenges 

% we can probably only cite the C7DASR for results. 

% it will be useful to report difference of tuned models between the two baselines. 
% just recompute DER and JER on the three common scenarios. 

%Describe baselines and changes during the time. 
%Should I redo the plotting for the saved time of performing channel selection with GSS ? 
%Need to include also NVIDIA system description. 
%Should I compare top-k selection vs 

% the use of channel selection and GSS is a significant contribution of these challenge baselines as it is adopted by almost all participants. 

% not all systems for chime 8 dasr only ones significantly different.
% e.g. NTT submission is an interesting case study :=) 
% would be cool to have a separate table with NTT_2 e.g. when integrating with NOTSOFAR-1 system.

\section{\textcolor{black}{Analysis of Submitted Systems}}\label{sec:systems_description}

%A total of \textcolor{black}{33} participants' systems were C7-8DASR. 5 for C8DASR and 28 for C7DASR. Among the 28 C7DASR systems 
The two challenges attracted a total of 9 teams, in alphabetical order: BUT FIT~\citep{karafiat2023but}, IACAS-Thinkit~\citep{ye2023iacas}, NVIDIA-NeMo~\citep{park2023chime}, NPU~\citep{mu2023npu}, NTT~\citep{kamo2023ntt, kamo2024ntt}, Paderborn University~\citep{boeddeker2023multi}, STCON~\citep{prisyach2023stcon, mitrofanov2024stcon}, University of Cambridge~\citep{deng2023university} and USTC-NERCSLIP~\citep{wang2023ustc}. 
%An additional team in C8DASR was disqualified as it did not present a technical description paper, and therefore its entry was anonymized. 
Two teams: STCON and NTT, participated in both the C7DASR and C8DASR challenges. 
Other teams, such as USTC-NERCSLIP, NPU and BUT participated instead only in the C7DASR and the CH\-iME-8 NOTSOFAR-1 challenge. 
% mention later that many are hardcore chime participants from the 6th and 5th edition. 

A total of 32 systems were submitted for the two challenges: 5 for C8DASR and 27 for C7DASR. Among the C7DASR systems, 13 were submitted to the optional oracle diarization track and 14 to the main track. 
Due to the sheer number of submitted systems and due to the fact that most of them differ only by hyper-parameter choices, we will restrict our analysis here by considering mostly the best systems submitted by each team with few notable exceptions which we deem particularly interesting.

%In C7DASR 
%C8DASR saw only two teams participating, however the two submissions came from two of the strongest systems in C7DASR 

We summarize baseline and participants systems diarization and front-end components in Table~\ref{tab:dasr_sys_summary_frontend} and the back-end components (ASR and external LM) in Table~\ref{tab:dasr_sys_summary_backend}. 
Each table is divided into three panels: in the top panel, for reference, are the baseline systems which have been described before in Section~\ref{ssec:baselines}; in the middle one C8DASR submissions and in the last one C7DASR submissions. 
Within each panel, the team submissions are sorted in ascending order by tcpWER with best systems first. Challenge results will be summarized and discussed later in Section~\ref{sec:results}.

\begin{landscape}

\begin{table}[htbp]{
\centering
\scriptsize
\caption{Summary of diarization and front-end components for C7-8DASR baseline and best submitted systems. Top panel (\colorbox[gray]{0.9}{\vphantom{a}\smash{grey}}): baseline systems; middle panel (\colorbox{ai2offwhite}{\vphantom{a}\smash{yellow}}): C8DASR submissions; bottom panel (\colorbox{ai2lightpink}{\vphantom{a}\smash{pink}}): C7DASR submissions. Within each panel, participants systems are ordered according to macro-averaged tcpWER. We report only the best system for each team.}
\label{tab:dasr_sys_summary_frontend}
\setlength{\tabcolsep}{0.2em}  % Increased spacing between columns
\begin{tabular}{lccccccc}  % Simplified column alignment
\toprule   
System & \multicolumn{4}{c}{Diarization} & \multicolumn{2}{c}{Front-End} \\
&  &  &  &  &  \\ 
\cmidrule(l{2pt}r{2pt}){2-5}  \cmidrule(l{2pt}r{2pt}){6-7} 
 & Segmentation & Spk-id Emb.   & Refinement & Multi-Channel & Channel  & Separation  \\
 &     &  Extr. \& Clustering   &      & Mechanism    &  Selection  &   \\ 
 \midrule
 \rowcolor{neutralOne}
  \multirow{1}{*}{ESPnet} & Pyannote  & Pyannote  &  -  &  Top-1 Channel  &   &   \\ 
\rowcolor{neutralOne}   Baseline             & Segmentation  & Diarization 2.1  &  -  & VAD Selection  &  \multirow{-2}{*}{EV} & \multirow{-2}{*}{GSS} \\ 

\midrule
\cellcolor[gray]{0.9} \multirow{1}{*}{NeMo} & \cellcolor[gray]{0.9} MarbleNet  &  \cellcolor[gray]{0.9} Multi-scale TitaNet   &  \cellcolor[gray]{0.9} Transformer & \cellcolor[gray]{0.9} logit max pooling \& &  \cellcolor[gray]{0.9} &  \cellcolor[gray]{0.9} \\ 
 \cellcolor[gray]{0.9}  Baseline             &  \cellcolor[gray]{0.9} VAD  & \cellcolor[gray]{0.9} \&  NME-SC &  \cellcolor[gray]{0.9} MSDD &  \cellcolor[gray]{0.9} multi-channel TitaNet embeddings  & \multirow{-2}{*}{\cellcolor[gray]{0.9}EV} & \multirow{-2}{*}{\cellcolor[gray]{0.9}GSS} \\ 
   
\midrule
\midrule
\rowcolor{ai2offwhite}
% STCON tried to use CSS but did not help see paper 
 & TDNN  &  Ecapa-TDNN \& &  & DOVER-Lap and   &   &  GSS with \\ 
\rowcolor{ai2offwhite} &  stats-based  & wav2vec 2.0 AED multi-speaker OSD  &  &  TS-VAD &  & neural  \\ 
\rowcolor{ai2offwhite} \rowcolor{ai2offwhite}\rowcolor{ai2offwhite} \multirow{-3}{*}{STCON}  &   &  with UMAP+DBSCAN+GMM clustering&  \multirow{-3}{*}{NSD-MS2S} & posterior averaging  & \multirow{-3}{*}{EV} & refinement (G-TSE) \\ 

\midrule

\rowcolor{ai2offwhite} &   & EEND-VC \&   &  &  DOVER-Lap \&   & EV \& &  \\ 
\rowcolor{ai2offwhite} &   & ECAPA-TDNN &  &  channel clustering  & Brouhaha C$_{50}$ &   \\ 
\rowcolor{ai2offwhite} \multirow{-3}{*}{NTT} &  \multirow{-3}{*}{EEND-VC}  &  with NME-SC  & \multirow{-3}{*}{NSD-MS2S} & for speaker counting  &   &  \multirow{-3}{*}{GSS with SP-MWF} \\

%\midrule

%\multirow{3}{*}{NTT$_2$} &  \multirow{3}{*}{EEND-VC}  & EEND-VC \&   & \multirow{3}{*}%{-} &  DOVER-Lap \&   & EV \& & \multirow{3}{*}{GSS with SP-MWF} \\ 
%&   & ECAPA-TDNN &  &  channel clustering  & Brouhaha C$_{50}$ &   \\ 
% &   &  with NME-SC  &  & for speaker counting  &   &  \\

%STCON   &   &   & NSD-MA-MSE  &   &  EV  &  GSS+neural TSep refinement\\ 
%NTT   &   &   &   &  NSD-MA-MSE  &  EV  &  GSS (rank-1 Wiener filter) \\ 
\midrule
\midrule
\rowcolor{ai2lightpink}
IACAS- &  Pyannote  & Pyannote  &  &    &   & neural TSE for  \\ 
\rowcolor{ai2lightpink} Thinkit & Segmentation  & Diarization 2.1 & \multirow{-2}{*}{TS-VAD}  &  \multirow{-2}{*}{DOVER-Lap}  & \multirow{-2}{*}{-} &  GSS init   \\ 

\midrule

% actually uses multiple stages including also top 6 VAD based selection. 
\rowcolor{ai2lightpink}
USTC- &  Pyannote  & ECAPA-TDNN  &  &  NSD-MS2S   & EV + virtual & GSS  \\ 
\rowcolor{ai2lightpink}
NERCSLIP  & Segmentation  & + NME-SC &  \multirow{-2}{*}{NSD-MS2S} &  posterior averaging & subarray SINR &    \\ 

\midrule

\rowcolor{ai2lightpink} \multirow{1}{*}{NTT} &  EEND-VC  & EEND-VC  & \multirow{1}{*}{-} &  \multirow{1}{*}{DOVER-Lap}  & - & GSS  \\ 

\midrule
\rowcolor{ai2lightpink}
  & TDNN  & ECAPA-TDNN  &    & TS-VAD  &  &   \\ 
\rowcolor{ai2lightpink} \multirow{-2}{*}{STCON}  & stats-based  & + NME-SC  & \multirow{-2}{*}{TS-VAD}  &  posterior averaging &   \multirow{-2}{*}{EV}  &   \multirow{-2}{*}{GSS} \\ 

\midrule

\rowcolor{ai2lightpink} University of & Pyannote  & Pyannote  &    &  Top-3 Channel  &  &  \\ 
\rowcolor{ai2lightpink} Cambridge               & Segmentation  & Diarization 2.1  &   & VAD Selection  &   &  \\ 
\rowcolor{ai2lightpink} &   &   &  \multirow{-3}{*}{-}  &  + DOVER-Lap &  \multirow{-3}{*}{EV}    &   \multirow{-3}{*}{GSS} \\ 
\midrule

\rowcolor{ai2lightpink} & MarbleNet  & Multi-scale TitaNet  &  Transformer &  logit max pooling \& &   &   \\ 
\rowcolor{ai2lightpink} \multirow{-2}{*}{NVIDIA NeMo}     &  VAD  & \&  NME-SC &   MSDD &   Multi-channel TitaNet embeddings   & \multirow{-2}{*}{EV} &  \multirow{-2}{*}{GSS} \\ 
\midrule 
\rowcolor{ai2lightpink} & Pyannote  & Pyannote  &  -  &  Top-1 Channel  &  EV  &  GSS \\ 
 \rowcolor{ai2lightpink}        \multirow{-2}{*}{NPU}         & Segmentation  & Diarization 2.1  &  -  & VAD Selection  &   &  \\ 
\midrule 
\rowcolor{ai2lightpink} Paderborn & Pyannote  & Pyannote  & TS-VAD + &  TS-VAD &    &   \\ 
 \rowcolor{ai2lightpink}   University   & Segmentation  & Diarization 2.1  &  
   
  d-vector refinement & posterior averaging  & \multirow{-2}{*}{-}   &  \multirow{-2}{*}{GSS} \\ 

\midrule

% experimented with many things e.g. tried speakerbeam
\rowcolor{ai2lightpink} & Pyannote  & Pyannote  & -  & Top-1 Channel  &    &   \\ 
\rowcolor{ai2lightpink} \multirow{-2}{*}{BUT FIT} & Segmentation  & Diarization 2.1  &  - & VAD Selection  & \multirow{-2}{*}{EV}  & \multirow{-2}{*}{GSS}  \\ 
\bottomrule
\end{tabular}
}
\end{table}
\end{landscape}

\subsection{Front-end processing}

In Table~\ref{tab:dasr_sys_summary_frontend} participants system diarization techniques are broken down into:

\begin{itemize}
    \item \emph{Segmentation}: part of the diarization system used for VAD, overlapped speech detection (OSD), or multi-speaker segmentation. This includes EEND techniques~\citep{fujita2019end_lstm, fujita2019end, kinoshita2021advances, kinoshita2021integrating}, since these can be considered as a generalization of the VAD task (a VAD for each speaker). 
    \item \emph{Speaker-id embedding extraction and clustering}: technique used to extract speaker-id discriminative embeddings and clustering method used to determine the total number of speakers throughout the meeting. 
     \item \emph{Refinement}: diarization refinement techniques such as TS-VAD~\citep{medennikov2020target} if any is used.
      \item \emph{Multi-channel mechanism}: what mechanism/strategy is used to get predictions from multiple-microphones. A straightforward strategy is to just ensemble predictions from the same model applied to different channels.
\end{itemize}
\noindent
With front-end processing, we instead denote the components used for microphone array processing, which has to be array agnostic in the two challenges due to the difference, across scenarios, in recording setups.  
Front-end processing is broken down into \emph{Channel Selection} and  \emph{Speech Separation}. 
The two ESPnet and NeMo baselines, for example, employ EV and GSS for these two components. 

%Some participants focused more on front-end diarization 
%vast majority focused on both. 
%We break out and describe in detail more 

%As it can be seen from Table~\ref{}, some participants in the C7DASR challenge focused more on the back-end side and adopted

\subsubsection{Neural Speaker Extraction/Separation}\label{ssec:sys_description_separation}

From Table~\ref{tab:dasr_sys_summary_frontend} it is evident that all systems adopt GSS as the main separation component, thus following the same diarization\textrightarrow\,GSS\textrightarrow\,ASR pipeline scheme as adopted by the baseline systems and depicted in Figure~\ref{fig:baseline_scheme}. 
As said, this same scheme was also used by the best systems in the past CH\-iME-6 challenge. 
Few participant systems try to improve upon GSS, confirming its efficacy overall. 
For example, the C7DASR IACAS-Thinkit and the C8DASR STCON systems complement GSS with DNN-based neural TSE. 
However, while the IACAS-Thinkit system uses the DNN-based TSE to initialize GSS, the C8DASR STCON instead uses it for refinement: the TSE model takes the GSS output as an additional input to provide the target speaker cue. In the latter case, improvements over a GSS-only baseline where possible only after fine-tuning with the ASR loss. 
These facts outline the robustness of GSS and its effectiveness despite its simplicity. 
It is also worth mentioning that, in their C8DASR system submission paper~\citep{mitrofanov2024stcon}, the STCON team also reports that they experimented with a continuous speech separation (CSS)~\citep{chen2020continuous} approach similar to the one employed in the CHiME-8 NOTSOFAR-1 challenge baseline system~\citep{vinnikov2024notsofar}. However, they found that such an approach was too brittle, as the CSS model was prone to introduce speaker confusion errors in complex situations with fast turn-taking dynamics and low SNR, such as in CH\-iME-6 scenario.
%In detail, the IACAS-Thinkit system employs time-domain Speakerbeam~\citep{delcroix2020improving} to initialize GSS. 
%In their system description report~\citep{ye2023iacas} they outline the challenge in training such system and getting good generalization to the real-world scenarios featured in the C7DASR challenge. 
%They found that it is necessary to include in-domain simulated data by mixing together single-speaker utterances from C7DASR \texttt{train} set. 
 
%As such, they settle back for DNN-based TSE instead, as it does not require permutation resolution and is thus more robust, provided accurate diarization is performed.
%TF-GridNet~\citep{wang2023tf} and SpatialNet~\citep{changsheng2024} are employed for such task and are suitably modified to take in input a target speaker cue. 
%After experimenting with feeding as cue speaker-id embeddings or target speaker speech activities (as done in GSS), the most effective approach was to use the target-speaker time frequency mask estimated by GSS.
%Still, in order to surpass GSS-only performance, fine-tuning of the DNN TSE model in an e2e fashion with the ASR criterion was necessary.
%In addition to these two attempts to replace GSS, the C8DASR NTT submission instead tries to improve it using a multichannel Wiener filter (SP-MWF)~\citep{spriet2004spatially} instead of an MVDR beamformer. However, it is not clear how much improvement this modification brought, especially since the speech distortion coefficient was set to 0 (distortion-less constraint). 

\subsubsection{Channel Selection}

Similarly, regarding channel selection, almost all participants used the proposed baseline EV channel selection strategy. It was found to be quite effective despite its simplicity and has also the additional benefit of reducing inference time. 
The C8DASR NTT system combines it with the pre-trained Brouhaha~\citep{lavechin2023brouhaha} multi-task VAD model which can predict C$_{50}$ speech clarity index, which measures the amount of reverberation. 
The C7DASR USTC-NERCSLIP instead combines it with a more complicated strategy: EV scores for each microphone are computed, then the channels are sorted and grouped into ``virtual subarrays'' of 5 microphones each.
Notably, some teams, such as Paderborn University and NTT, did not use channel selection in their C7DASR submissions. However, NTT's next year C8DASR submission incorporated channel selection, indicating that this component is beneficial, particularly for excluding problematic channels.
After all, as outlined in Figure~\ref{fig:sdr_stats}, Section~\ref{sec:core_scenarios}, scenarios such as Mixer 6 and CHiME-6 inherently benefits from some explicit or implicit channel selection component due to their significant inter microphone channel SDR variability. 

\subsection{Diarization}\label{ssec:sys_description_diarization}

%Reliable diarization and speaker counting is a crucial component for meeting transcription, especially so as every participant rely on GSS for handling multiple speakers. 
%Thus, significant effort was spent in improving diarization accuracy. 
%The only exception are C7DASR NPU and BUT FIT systems which focused only on improving the ASR part and used the ESPnet baseline diarization system. 
%The University of Cambridge submission instead does some minor modifications which nonetheless bring notable improvement: these include the local EEND channel selection strategy, which is made more smooth by choosing the best channel not at every hop-size but for every window. DOVER-Lap ensembling across the diarization system ran over the top-3 best channels is also employed. 
%Other teams, such as Paderborn University and C7DASR IACAS-Thinkit (which ranks first with respect to tcpWER), adopt some of the ESPnet baseline diarization components, but improve it considerably by mostly adding a TS-VAD refinement step.
Reliable diarization and speaker counting are crucial components for meeting transcription, even more so since, as said before, all participants relied on GSS. Consequently, significant effort was dedicated to improve diarization accuracy. 
The only exceptions are the C7DASR NPU and BUT FIT systems, which focused solely on improving the ASR component and used the ESPnet baseline diarization system without modifications. 
In C7DASR, most of the other teams, with the exception of NeMo~\citep{park2023chime}, STCON~\citep{prisyach2023stcon}, mainly built upon the C7DASR baseline diarization system by introducing effective modifications~\citep{deng2023university} and/or also a TS-VAD module for refinement~\citep{boeddeker2023multi}.
The NTT system notably differs from all others as it does not rely on a classical segmentation plus clustering pipeline. Instead, it employs a modified end-to-end neural diarization with vector clustering (EEND-VC)~\citep{kinoshita2021integrating, kinoshita2021advances} which uses WavLM features, resulting in a simple and streamlined diarization pipeline.
A TS-VAD component appears in all top-performing C7-8DASR systems, confirming its effectiveness. Indeed, diarization performance heavily depends on TS-VAD's architecture and training data. 
C7DASR IACAS-Thinkit~\citep{ye2023iacas} and STCON teams use the original TS-VAD model~\citep{medennikov2020target}, while \textcolor{black}{Paderborn} University, USTC-NERCSLIP~\citep{wang2023ustc} and NeMo explore different architectures and speaker embeddings for conditioning. 
Among these, the USTC-NERCSLIP NSD-MA-MSE TS-VAD model~\citep{he2023ansd} proved to be very effective, leading to its adoption in subsequent C8DASR submissions by both STCON and NTT teams.

\subsubsection{Improvements in the speaker counting components in C8DASR}\label{ssec:speaker_counting}

In C8DASR, due to the increased difficulty of accurate speaker counting from the introduction of NOTSOFAR-1, both STCON~\citep{mitrofanov2024stcon} and NTT~\citep{kamo2024ntt} make substantial modifications regarding the core diarization component to make it more robust.  
For example, it is evident how the STCON team switched from a classical VAD+ECAPA-TDNN clustering pipeline for the initial diarization to a more complex one that can better deal with overlapped speech. The core component for their pipeline is a novel wav2vec 2.0 attention-based encoder-decoder (AED) model that extracts frame-wise speaker embeddings and can be used also to perform overlapped speech detection (OSD).
On the other hand, the NTT system still relies on EEND-VC but adds robustness via a multi-channel speaker counting component that leverages ECAPA-TDNN speaker embeddings extracted after GSS.

\subsection{Multi-Channel Mechanism, Ensembling and Test-Time Adaptation}\label{ssec:sys_ensembling_multi_channel_tta}

As it is evident from Table~\ref{tab:dasr_sys_summary_frontend}, most systems rely on ensembling techniques in order to fuse information across multiple microphones. 
As such, the efficacy of native multi-channel diarization techniques such as~\citep{horiguchi2022multi} remains largely under-explored for complex conversational scenarios. 
A notable exception is the C7DASR NeMo system, which leverages multi-channel information in the speaker embedding clustering phase by concatenating as explained in Section~\ref{sssec:diar_baseline}.

C7DASR NTT and IACAS-Thinkit systems explored test time adaptation (TTA) in their diarization pipelines through iterative pseudo-labeling. However, according to the NTT technical report, this approach yielded only marginal improvements while introducing substantial computational overhead, raising questions about its practical value in real-world applications.

%Interestingly, NTT also tried SSL test-time adaptation for the EEND-VC model using an approach similar to that of IACAS-Thinkit. However, they found that this SSL technique provided only marginal improvements~\citep{kamo2023ntt}. 
% mention enhanced speaker countring in C8DASR
% mention forced alignment here.
% NTT significantly departs from the baseline or the other submissions by instead relying on EEND-VC for diarization. 
% praise speed for this solution. 
% The TS-VAD system was modified to allow for handling a variable number of speakers unlike the original implementation~\citep{} by using backup enrollment embeddings~\citep{}

\subsection{Automatic Speech Recognition}\label{ssec:sys_description_asr}

In Table~\ref{tab:dasr_sys_summary_backend} we present a taxonomy of participants' ASR techniques. These are summarized and broken down into the following:

\begin{itemize}
    \item \emph{Features}: input features used for the ASR model. These can include latent representation from pre-trained models such as WavLM. 
    \item \emph{Model}: main DNN architecture used e.g. Conformer.  
    \item \emph{Criteria}: ASR criteria used, such as CTC+Attention, Attention-only, connectionist temporal classification (CTC)~\citep{graves2006connectionist} or transducer~\citep{gravessequence}. 
    \item \emph{Multi-channel}: if the ASR model uses some technique to fuse information from multiple-channels. 
    \item \emph{Text-time adaptation (TTA)}: what technique is used for unsupervised TTA. These include simple iterative pseudo-labeling, STAR~\citep{hu2024self} or single-utterance test-time adaptation (SUTA)~\citep{lin2022listen} for example.
    \item \emph{Speaker adaptation}: what technique, if any, is used for target speaker adaptation e.g. if the ASR accepts a cue for the target speaker. 
    \item \emph{Ensembling}: ensembling technique used e.g. Recognizer Output Voting Error Reduction (ROVER)~\citep{fiscus1997post} or others to combine the output of multiple systems. 
    \item \emph{External LM}: external language model (LM) used during ASR decoding.
\end{itemize}

\subsubsection{Input Features, Criteria and Architecture}

First, it can be noted that all participants leverage external pre-trained models in some way with WavLM being the most commonly used one.
Some exceptions are the NeMo baseline and the C7DASR NeMo team submission which use the pre-trained NeMo Conformer-transducer model and the C8DASR NTT submission which also use Whisper (it was allowed in C8DASR to be consistent with the concurrent CHiME-8 NOTSOFAR-1 challenge). 
Among all submissions the C7DASR USTC-NERCSLIP is also peculiar in the fact that it leverages an interesting combination of features from WavLM, ECAPA-TDNN and even a pre-trained Conv-TasNet speech enhancement model~\citep{luo2018convtasnet}.

%Among all submissions, system uses an interesting combination of input front-end by fusing representations from  which are integrated and fine-tuned together with the ASR.
%The C7DASR Paderborn submission~\citep{boeddeker2023multi} uses the ESPnet baseline ASR back-end without modifications, as they focus only on front-end and diarization techniques. 
%n C8DASR, since Whisper was allowed (in coordination with the CH\-iME-8 NOTSOFAR-1 challenge, which employs it in the baseline system), the NTT submission~\citep{kamo2024ntt} makes use of it, but only in combination with ESPnet and NeMo baseline ASR systems. 
%The C8DASR STCON submission~\citep{mitrofanov2024stcon} instead uses the same configuration as used for C7DASR~\citep{prisyach2023stcon}, but adds minor modifications in the ensembling techniques and experimented with STAR SSL adaptation. 
%C7DASR submission~\citep{prisyach2023stcon}, they compared the performance of various SSL features and found that WavLM significantly outperformed the others, delivering the best overall performance by a considerable margin.

Regarding the ASR models architecture, most participants use Transformer or Conformer based ASR models and CTC+Att or transducer-based criteria.
STCON C8DASR and C7DASR submissions also experiment with Kaldi-based DNN-HMM multi-stream TDNN-F~\citep{han2019state} and K2 Zipformer~\citep{yao2023zipformer} with CTC plus pruned stateless transducer~\citep{poveyk2}. 
The NTT C7DASR submission instead experimented with S4-based~\citep{gu2021efficiently} layers to augment the Conformer and Branchformer~\citep{peng2022branchformer} models. 
For the subsequent NTT C8DASR challenge, these S4 components were no longer incorporated into their approach, suggesting that their contribution to performance gains was marginal or debatable. 
Similarly to STCON, the University of Cambridge C7DASR submission~\citep{deng2023university} also includes alternative, more exotic, ASR criteria, such as a backward CTC+Attention, which operates on a right-to-left encoder representation, and the recently proposed label-synchronous neural transducer~\citep{deng2024label}. The rationale for this is that such diversity in criteria can enhance the effectiveness of ensemble techniques, as they can be complementary to each other. 
While the widespread use of multiple criteria and architectures for ensemble approaches makes direct comparisons challenging, the C7DASR IACAS-Thinkit WavLM-based CTC ASR system stands out for its notable simplicity and relative efficacy.

\subsubsection{Ensembling and Test-Time Adaptation Techniques}
Almost all submissions extensively use ensembling techniques from different ASR models, with ROVER being the most commonly used method. 
Notable exceptions are IACAS-Thinkit and BUT FIT submissions, as well as one submission from the NTT team for C8DASR, which focuses on lightweight inference and uses only the Whisper-M model.

Two teams use more complex ensembling strategies. 
The C8DASR STCON submission switches from ROVER as used for C7DASR to a more complicated fusion technique, involving lattice fusion and minimum Bayes risk decoding~\citep{kumar2004minimum} while the University of Cambridge's C7DASR submission adopts a two-pass approach where hypotheses from a CTC-based model are rescored and combined using the decoders of three other models. 
However, the wide usage of ROVER suggests that, despite its simplicity, it is still competitive with these more refined methods.

Concerning ASR TTA techniques, we observe the same trend as for diarization. Few systems make use of these techniques (C8DASR STCON, C7DASR IACAS-Thinkit, NTT and University of Cambridge). 
However, when ablations are reported in their respective technical description papers~\citep{kamo2023ntt}, they seem to offer small improvements despite their significant computational cost. This is why, for example, the C8DASR NTT system does not use TTA via pseudo-labeling anymore compared to their C7DASR system.

\subsubsection{Multi-Channel and Target Speaker Information}

The C7DASR NPU system~\citep{mu2023npu} is the only one that tries to incorporate into the ASR model multi-channel information via spatial features \textcolor{black}{such as cosine sine inter-channel phase differences (cosIPD)} so that the model can still recover the original speech signal if the GSS enhancement is imperfect. 
% It leverages spatial features from multiple channels via cosine+sine interchannel phase differences features together with post-GSS input. This latter serves as the main cue to disambiguate the target speaker. 
%Instead, regarding target-speaker information, this was not explicitly leveraged by any participant in both C7DASR and C8DASR challenges. On the other hand, it is also true that, due to the use of GSS, all ASR systems have some extent of target speaker capability. 
No participants explicitly leveraged target-speaker information in either the C7DASR or C8DASR challenges. However, we argue that all participants ASR systems inherently possess some target-speaker capability due to their use of GSS as pre-processing.
In fact, GSS-based TSE is far from perfect and, at least when looking at the baseline enhanced utterances, there are many instances where the background speakers are not well suppressed. Yet, the ASR model, if trained or fine-tuned on such data can learn that the target speaker is the one which is slightly more energetic and has the utterance ``centered'' regarding the segmentation. 
For this reason, as mentioned in Section~\ref{sssec:asr_baseline}, we found it crucial for both NeMo baseline and ESPnet baseline ASR components to include GSS-enhanced data into the training material. 
Similarly, all participants in the C7-8DASR challenges included GSS-enhanced data into the ASR training as specified in their technical reports. 

\subsubsection{Language Modeling}\label{ssec:lm_sys_asr}

%Maybe, rather unsurprisingly, the best submissions are the ones that also make use of neural LM models.
%C8DASR had an optional sub-track where participants could leverage external pre-trained LLMs. STCON C8DASR submission experimented with this direction, and tried to use, on top of their system, a fine-tuned Llama 2 \texttt{7B} model. This model was employed for hypothesis rescoring using a method similar as the one proposed by~\citep{ogawa2024applying}, where the LLM is used with a inter-utterance context.
%Llama 2 \texttt{7B} was fine-tuned on the whole C8DASR \texttt{train} text data augmented using LibriSpeech text.
%However, for C8DASR, they found this approach to bring very marginal improvements in tcpWER. This may be due to the fact that their ASR component is already quite strong and already employs several LMs, a 3-gram one and two neural ones, one Transformer-based and another ASGD Weight-Dropped (AWD) long-short term memory (LSTM)~\citep{} based.
%On the other hand, the C7DASR IACAS-Thinkit submission uses an inter-utterance Transformer-based LM model which they found particularly effective and similarly \citep{ogawa2024applying} report benefits on C7DASR challenge data with inter-utterance LLM processing. 

Maybe, rather unsurprisingly, most of the best performing systems (C8\-DASR STCON and NTT, C7DASR IACAS-Thinkit, NTT and STCON) are the ones that also make use of neural LM models.

C8DASR had an optional sub-track where participants could leverage external pre-trained LLMs. The STCON C8DASR submission experimented with this direction and employed a fine-tuned Llama 2 \texttt{7B} model on top of their system. This model was used for hypothesis rescoring using a method similar to the one proposed by~\citep{ogawa2024applying}, where the LLM is applied with inter-utterance context \textcolor{black}{i.e. using past context from the same speaker previous utterances (up to 1024 tokens in total).}
However, they found that this approach yielded very marginal improvements in tcpWER. This limited gain may be attributed to their already robust ASR component, which incorporates multiple language models: a 3-gram model alongside two neural models: one Transformer-based and another implemented using ASGD Weight-Dropped (AWD) long-short term memory (LSTM)~\citep{merity2018regularizing} networks.
Similarly, \citep{ogawa2024applying} shows small but more significant improvements on C7DASR challenge data by using Llama 2 \texttt{7B} for inter-utterance N-best list ASR hypothesis rescoring of the NTT C7DASR system. In contrast, the C7DASR IACAS-Thinkit submission reports significant improvements by using an inter-utterance Transformer-based LM model.
These observations suggest that the efficacy of integrating LLMs for ASR hypothesis post-processing remains debatable and requires further investigation, particularly when the ASR component is already strong, as this can lead to diminishing returns. In such cases, the computational resources required for LLM inference may outweigh performance improvements, especially in production environments where efficiency is crucial.

%In particular \citep{ogawa2024applying}, 
%As such, the efficacy of using LLMs for ASR hypothesis post-processing is still %debatable and likely depends on the ASR system configuration and performance. 
%is likely dependent on the ASR system configuration and performance. Systems that already employ sophisticated language modeling components may experience diminishing returns from additional LLM integration, while those with simpler architectures might benefit substantially. Furthermore,  where latency and deployment constraints are significant considerations. 

\begin{landscape}
\begin{table}[htbp]{
\centering
\footnotesize
\caption{Summary of back-end components for C7-8DASR baseline and best submitted systems. Top panel: baseline systems (\colorbox{neutralOne}{\vphantom{a}\smash{grey}}); middle panel: C8DASR submissions (\colorbox{ai2offwhite}{\vphantom{a}\smash{yellow}}); bottom panel (next page): C7DASR submissions (\colorbox{ai2lightpink}{\vphantom{a}\smash{pink}}). Within each panel, the participants' systems are ordered according to macro-averaged tcpWER. We report, for each team, only the best system configuration.}
\label{tab:dasr_sys_summary_backend}
\setlength{\tabcolsep}{0.2em}  % Increased spacing between columns
\begin{tabular}{lccccccccc}  % Simplified column alignment
\toprule   
System & \multicolumn{6}{c}{ASR}  \\
\cmidrule{2-8}
 & Features & Model  & Criteria  & Multi & TTA &  Spk & Ensembling &  External \\
  &  &   &  & Channel &  & Adapt  &  &  LM \\
 \midrule  
\rowcolor{neutralOne} ESPnet  &  &  &  &  &  &  &  &    \\

\rowcolor{neutralOne} Baseline &  \multirow{-2}{*}{WavLM} &  \multirow{-2}{*}{Transformer} &  \multirow{-2}{*}{CTC+Att.} & \multirow{-2}{*}{-}  & \multirow{-2}{*}{-} & \multirow{-2}{*}{-}  & \multirow{-2}{*}{-}  &   \multirow{-2}{*}{-} \\ 

 \midrule

 \rowcolor{neutralOne} NeMo  &  &   &  &  &  &   &  & word-piece \\
\rowcolor{neutralOne} Baseline & \multirow{-2}{*}{Fbank}  & \multirow{-2}{*}{Conformer}  & \multirow{-2}{*}{Transducer} & \multirow{-2}{*}{-} & \multirow{-2}{*}{-} & \multirow{-2}{*}{-} & \multirow{-2}{*}{-} &  N-gram \\ 

 \midrule
 \midrule
\rowcolor{ai2offwhite}  &  WavLM &   Uconv-Conformer & CTC+Att.  &  -  & STAR & -   &   &    \\

\rowcolor{ai2offwhite}  &  WavLM &  E-branchformer  & CTC+Att.  &  -  & STAR & -  &   &  \\

\rowcolor{ai2offwhite}  &  WavLM &  ZipFormer & CTC+Transducer  & -  & - & - &     &   \\

\rowcolor{ai2offwhite} \multirow{-4}{*}{STCON}  &  WavLM &  MS TDNN-F & phoneme HMM  &  -  & - & - &  \multirowcell{-4}{N-best \\ lattice fusion with \\   MBR decoding}   &  \multirowcell{-4}{word-piece 3-gram  + \\   Transformer + \\ AWD-LSTM} \\

\midrule

\rowcolor{ai2offwhite}  &  Fbank &  Whisper-M & Att.  &   &  &  &    &   \\

\rowcolor{ai2offwhite} &  Fbank &  Whisper-L & Att.  &   &  &  &     &   \\

 \rowcolor{ai2offwhite} &  Fbank &  NeMo Conformer & Transducer  &   &  &  &     &   \\

\rowcolor{ai2offwhite} \multirow{-4}{*}{NTT}  &  WavLM &  Transformer & Transducer  & \multirow{-4}{*}{-}  & \multirow{-4}{*}{-} & \multirow{-4}{*}{-} &  \multirow{-4}{*}{ROVER}    & \multirow{-4}{*}{Transformer}  \\

\bottomrule

\end{tabular}
}
\end{table}
\end{landscape}

\begin{landscape}
\begin{table}[htbp]{
\centering
\scriptsize
\setlength{\tabcolsep}{0.2em}  % Increased spacing between columns
\begin{tabular}{lccccccccc}  % Simplified column alignment
\toprule

\rowcolor{ai2lightpink}  IACAS-  & \multirow{3}{*}{WavLM} &  &  &   &  pseudo &   &  & word-piece N-gram + \\

\rowcolor{ai2lightpink}  Thinkit  &  WavLM & Transformer  &  &  & label &  &  & Transformer   \\
 \rowcolor{ai2lightpink}  &  &  \multirow{-3}{*}{-}  & \multirow{-3}{*}{CTC}  & \multirow{-3}{*}{-}  & fine-tuning & \multirow{-3}{*}{-} & \multirow{-3}{*}{-}   & (inter-utterance)   \\

\midrule  

 \rowcolor{ai2lightpink} & WavLM + & Conformer  & CTC+Att & & & &   &   \\
\rowcolor{ai2lightpink}  USTC-  & ECAPA-TDNN + ConvTasNet &   &  &  &  &  & ROVER  &   \\

\rowcolor{ai2lightpink} NERCSLIP  & wav2vec 2.0 + & Conformer  & CTC+Att &  &  &  &  &   \\
\rowcolor{ai2lightpink}  & ECAPA-TDNN + ConvTasNet &  &  & \multirow{-4}{*}{-}   &  \multirow{-4}{*}{-} &  \multirow{-4}{*}{-} &   & \multirow{-4}{*}{-}  \\

\midrule  

\rowcolor{ai2lightpink} & WavLM  &  Conformer+S4 & CTC+Att. &   & pseudo &   &   &  \\
 
\rowcolor{ai2lightpink} NTT  & WavLM &  Branchformer+S4 & CTC+Att. &  - & label & - & ROVER  &  Transformer   \\

\rowcolor{ai2lightpink}  & WavLM  & Branchformer  &  Transducer &  & fine-tuning &   &   &   \\

\midrule  

\rowcolor{ai2lightpink}  &  WavLM &   Uconv-Conformer & CTC+Att.  &   &   &   &   &   \\
 
\rowcolor{ai2lightpink}  &  WavLM &  E-branchformer  & CTC+Att.  &   &   &   &   &   \\

\rowcolor{ai2lightpink}  &  WavLM &  ZipFormer & CTC+Transducer  &   &   &   &   &   \\

\rowcolor{ai2lightpink}   \multirow{-4}{*}{STCON}  &  WavLM &  MS TDNN-F & phoneme HMM  & \multirow{-4}{*}{-}  & \multirow{-4}{*}{-}  &  \multirow{-4}{*}{-}  & \multirow{-4}{*}{ROVER}  &  \multirowcell{-4}{word-piece 3-gram  + \\ Transformer + \\ AWD-LSTM} \\

\midrule

\rowcolor{ai2lightpink}  & WavLM & -  & CTC &  &  &  &   &   \\

\rowcolor{ai2lightpink}  University of   & WavLM & Transformer  & CTC+Att. &  &  &  &  two-pass  & word-level  \\

\rowcolor{ai2lightpink}  Cambridge  & WavLM & Transformer  & CTC+Att.  & - & SUTA & - &  decoder-only  & N-gram  \\
 \rowcolor{ai2lightpink}   &  &   & backwards &  &  &  & rescoring  &   \\
 
\rowcolor{ai2lightpink}  & WavLM & Transformer  & LS-Transducer &  &  &  &   &   \\

%\multirowcell{4}{two pass \\ N-best \\ decoder-only \\ rescoring}
\midrule

\rowcolor{ai2lightpink}   NVIDIA  & Fbank & Conformer  & Transducer & - & - &  - &  & word-piece \\
\rowcolor{ai2lightpink}   NeMo &  &   &  &  & &  & ROVER  & N-gram \\ 
\rowcolor{ai2lightpink}    &  &   &  &  & &  &  &  \\

\midrule  

\rowcolor{ai2lightpink}  & WavLM +  & Transformer  & CTC+Att. &  MFCCA+   & - & implicit via & ROVER & Transformer  \\

 \rowcolor{ai2lightpink} \multirow{-2}{*}{NPU} & cosIPD &   &  & CGCS+FGCS & - & GSS ch. input & - & -  \\

\midrule  
\rowcolor{ai2lightpink}  Paderborn  & WavLM & Transformer & CTC+Att. & - & - & - & - & -  \\
\midrule

\rowcolor{ai2lightpink} \multirow{1}{*}{BUT FIT}  & WavLM & Transformer  & CTC+Att. & - & - & - & \multirow{1}{*}{-} &   -  \\
  %& WavLM & Conformer  & CTC+Att. & - & - & - &   & - \\
 % & WavLM & Conformer  & Transducer & - & - & - &     & - \\

\bottomrule

\end{tabular}
}
\end{table}

\end{landscape}

% diarization - front-end - ASR 

% three main schemes, components may be omitted, e.g. sep+ASR = target speaker/multi-speaker ASR
% diarization -> sep -> ASR 
% sep -> diar -> ASR
% sep -> ASR -> diar 

% design space

% all participants use 2. 
% comparison with NOTSOFAR-1 baseline. 
% run NOTSOFAR-1 baseline on DiPCo ? 

%\subsubsection{Pseudo-Labeling and Self-Training}
% mention also USTC recent paper ? 

% make a section on its own on this. 
% mention about Mixer 6 
% mention about impracticality 

\section{Challenges Results}\label{sec:results}

%In this section, we present and discuss the results of C7DASR and C8DASR. As mentioned in Section~\ref{sec:description}, due to the large number of submitted systems (up to 3 submissions per track were allowed for each team), we focus on and report results only for the best-performing systems for each team in terms of scenario-wise macro-averaged tcpWER.
In this section, we present and discuss the results of C7DASR and C8DASR. Due to the large number of submitted systems (up to 3 submissions per track were allowed for each team), we focus on and report results only for each team's best-performing system, as measured by scenario-wise macro-averaged tcpWER (as described in Section \ref{sec:description}).
One notable exception is the NTT submission to C8DASR, which includes a system designed with a greater emphasis on efficiency, aligned with the C8DASR jury award mechanism outlined in Section~\ref{ssec:tracks}. This system is referred to as NTT$_{small}$ in the following discussion. Note that despite being among the most efficient and practically oriented systems in both challenges, it still had a real time factor (RTF) of $2.46$ (but without any optimization). %Despite the fact it was on efficiency it must be noted that %This system had a real-time-factor of 
As highlighted in Sections~\ref{ssec:tcpwer} and \ref{ssec:tracks}, the results presented here are computed using the C8DASR text normalization strategy prior to scoring. 
Consequently, these results differ from those available on the official C7DASR challenge website\footnote{\href{https://www.chimechallenge.org/challenges/chime7/task1/results}{www.chimechallenge.org/challenges/chime7/task1/results}}, which used a different normalization and metric (DA-WER). This choice was made to ensure the results are more directly comparable between C8DASR and C7DASR but also with the CHiME-8 NOTSOFAR-1 challenge, which uses the same ranking metric (tcpWER) and the same C8DASR text normalization protocol. 

%In this section, we present and discuss C7DASR and C8DASR results.
%As said, in previous Section~\ref{sec:description}, due to the number of systems submitted we focus here and report results only for the best ones in terms of scenario-wise macro-averaged tcpWER. 
 %One notable exception, is the NTT submission to C8DASR, which includes a system that focuses more on efficiency, due to the C8DASR jury award mechanism (mentioned in Section~\ref{ssec:tracks}). This system is denoted as NTT$_{small}$ in the following. 
%As previously mentioned in Sections~\ref{ssec:tcpwer} and \ref{ssec:tracks}, results are computed using the C8DASR text normalization strategy prior to scoring and thus can differ from ones available in the official challenge website. This choice again was made such that the results can be more directly comparable. 

\begin{landscape}
   \begin{figure}
    \centering
    \includegraphics[width=\linewidth]{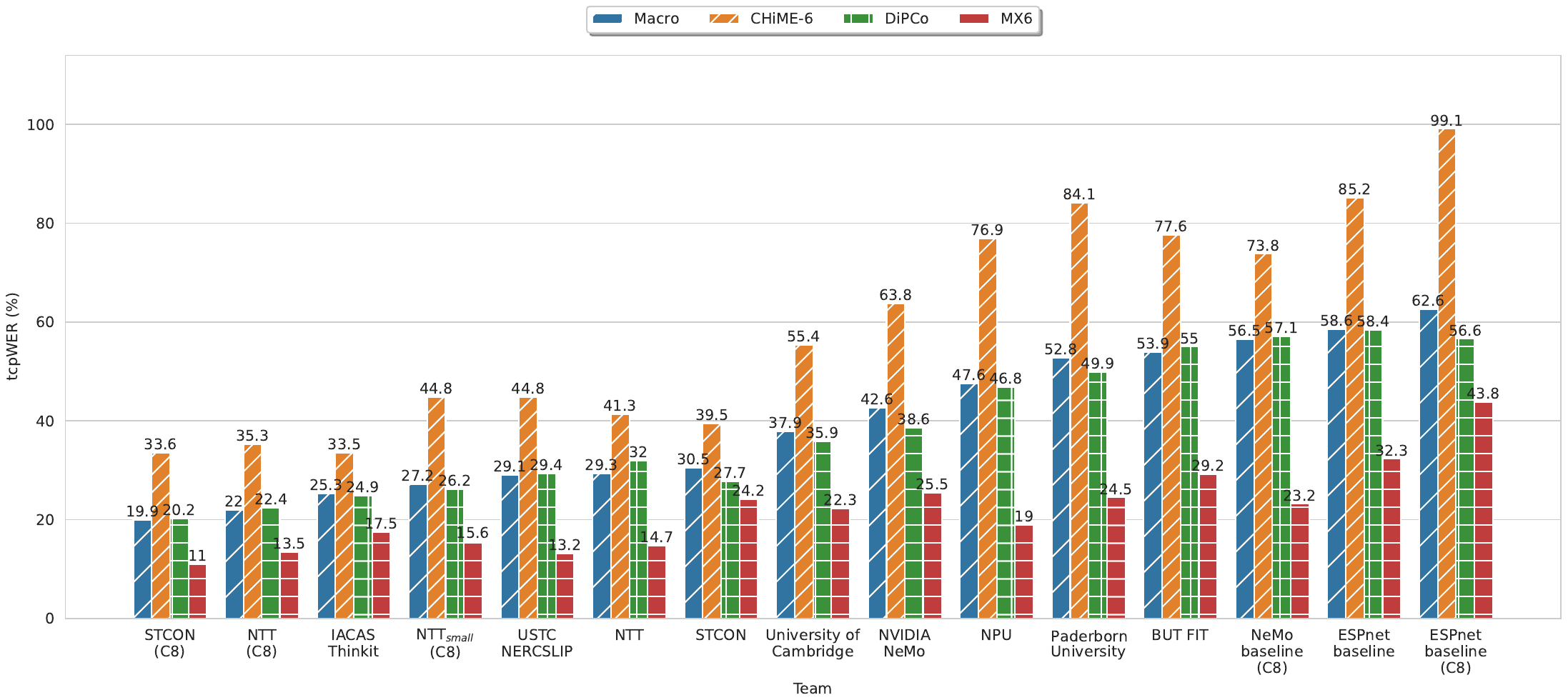}
    \caption{tcpWER (\%) for each C7DASR core scenarios (CH\-iME-6, DiPCo, MX6) as well its macro average (Macro) for both C7DASR and C8DASR systems. C8DASR systems are denoted by (C8).}
    \label{fig:c78dasr_results}
\end{figure} 
\end{landscape}

\subsection{Joint CH\-iME-7 and 8 DASR Results}\label{ssec:joint_c78dasr_results}

Since C8DASR includes an additional scenario compared to C7DASR (the NOTSOFAR-1 dataset), it allows for a comparative analysis of submitted systems across both challenges.
Figure~\ref{fig:c78dasr_results} shows the tcpWER results for the core scenarios of C7DASR, along with their macro-average (the C8DASR ranking metric, as explained in Section~\ref{ssec:tcpwer}), for systems submitted to both challenges.
Across all systems and baselines, the tcpWER figures consistently reveal that the scenarios, ranked from most difficult to easiest, are CH\-iME-6, DiPCo, and MX6. This ranking aligns with expectations: MX6 involves only two participants in a relatively static interview setting, whereas DiPCo and CH\-iME-6 feature four speakers that dynamically interact in more complex environments. CH\-iME-6 is particularly challenging due to its multi-room setting and highly colloquial speech, resulting in significantly higher tcpWER figures for all participants.
Even for the best performing systems, the tcpWER for CH\-iME-6 remains above 30\%. This means that nearly one in three words is incorrectly transcribed, despite advances in ASR and diarization technology and the fact that almost every participant used ASR systems ensembles. 

Achieving balanced performance across all three scenarios is a significant challenge. For instance, some systems, such as the C7DASR STCON submission, 
\textcolor{black}{rank among the top C7DASR systems on CHiME-6 but shows significantly higher tcpWER on MX6}. Conversely, systems like the C7DASR NPU submission achieve low tcpWER on MX6 but perform worse on CH\-iME-6. Others, such as the C8DASR submissions and the C7DASR IACAS-Thinkit submission, demonstrate more balanced performance overall. 
This difficulty underscores the main motivation behind these challenges: while it is relatively straightforward to optimize a system for a specific scenario like CH\-iME-6 or MX6, developing a system or technique capable of robust performance across all scenarios remains very challenging.
Another example of this is the C7-8DASR ESPnet baseline: the C8DASR version, due to the fact that needs to handle the NOTSOFAR-1 scenario, performs worse than the C7DASR one on C7DASR core scenarios.
In this sense, it is notable that STCON and NTT C8DASR submissions are still able to improve over C7DASR best systems even if they need to also tackle the new NOTSOFAR-1 scenario introduced in C8DASR. 
% consider adding a plot with JER for each scenario

\begin{figure}
    \centering
    \includegraphics[width=0.7\linewidth]{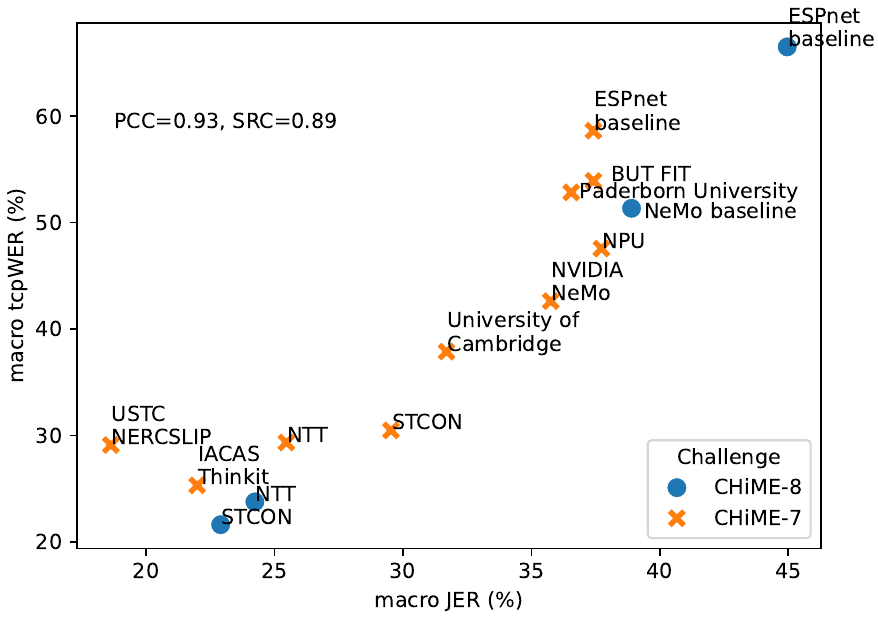}
    \caption{tcpWER (\%) vs. JER (\%) for C7DASR and C8DASR systems (only top system for each team). Both are macro-averaged across C7DASR core scenario (CH\-iME-6, DiPCo, MX6). Pearson correlation coefficient (PCC) and Spearman rank correlation (SRC) are also reported.}
    \label{fig:c78_macro_jer_tcpwer}
\end{figure}

In Figure~\ref{fig:c78_macro_jer_tcpwer} macro-averaged tcpWER versus macro-averaged Jaccard error rate (JER)~\citep{ryant2019second} is reported for the same C7DASR and C8DASR systems. The macro-average is taken across all three C7DASR scenarios. 
JER is computed by considering a start and end-point collar of 250\,ms, since the manual annotation of the utterance boundaries can have minor inconsistencies across the three datasets and is inherently imperfect due to human error~\citep{fiscus2007rich, ryant2019second}.
%JER is preferred over diarization error rate (DER) because it better accounts for speaker counting errors, which are particularly critical~\citep{cornell2024chime}. Errors in speaker counting can significantly impact downstream components of the pipeline, such as TS-VAD refinement, thereby affecting overall system performance.
%JER is used instead of diarization error rate (DER) as it better accounts for speaker counting errors, which are more crucial, as they can affect the rest of the pipeline (e.g. TS-VAD refinement) in a serious manner.  

\begin{figure}
    \centering
    \includegraphics[width=0.7\linewidth]{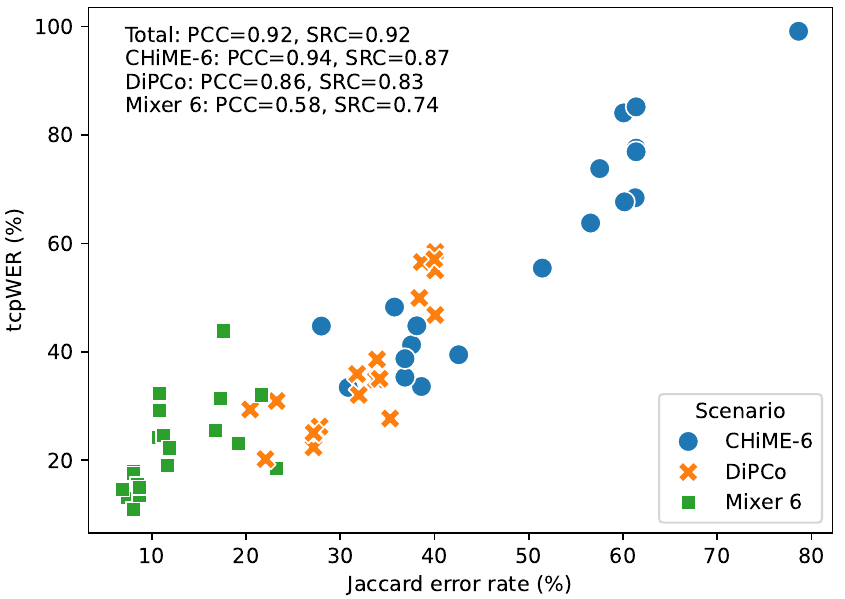}
    \caption{tcpWER (\%) vs. JER (\%) for C7DASR and C8DASR systems (all submissions are included for a total of 22 systems). Figures are computed for each core scenario separately (CH\-iME-6, DiPCo, MX6). Pearson correlation coefficient (PCC) and Spearman rank correlation (SRC) are also reported for all the scenarios jointly (Total) and also separately for each scenario.}
    \label{fig:c78_jer_tcpwer_correlation}
\end{figure}

\begin{figure}
    \centering
    \includegraphics[width=0.7\linewidth]{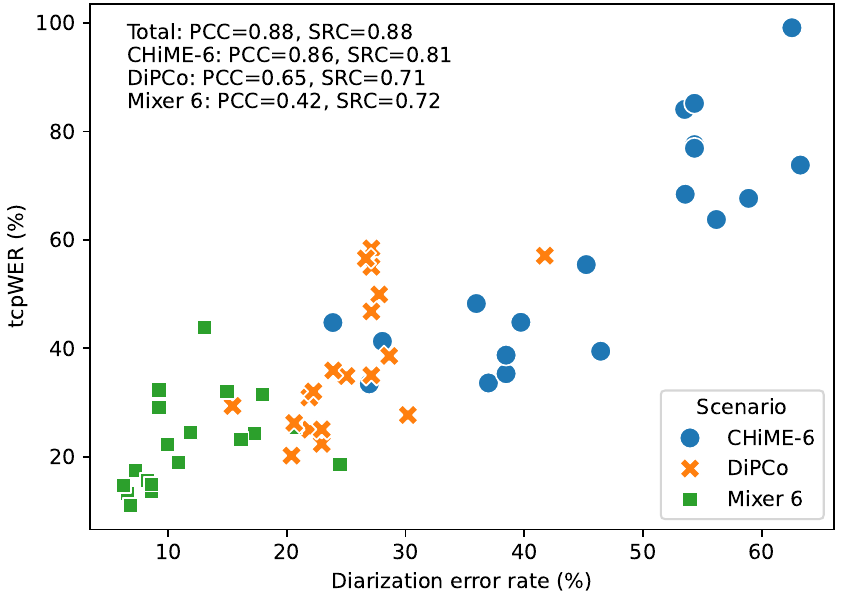}
    \caption{tcpWER (\%) vs. DER (\%) for C7DASR and C8DASR systems (all submissions are included for a total of 22 systems). Figures are computed for each core scenario separately (CH\-iME-6, DiPCo, MX6). Pearson correlation coefficient (PCC) and Spearman rank correlation (SRC) are also reported for all the scenarios jointly (Total) and also separately for each scenario.}
    \label{fig:c78_DER_tcpwer_correlation}
\end{figure}

In general, there is a clear correlation between JER and tcpWER ({PCC=0.93 and SRC=$0.89$}). This is largely expected. 
As mentioned in Section~\ref{sec:description}, all participants practically relied on GSS-based TSE, which requires accurate diarization to be effective. 
The best systems consistently have a macro JER around 20-30\%, while the worst over 40\%.
This is even more evident in Figure~\ref{fig:c78_jer_tcpwer_correlation} where we plot, respectively, JER versus tcpWER for each scenario (including the macro-average) and for all submitted systems (thus not only the best one for each team). The number of systems amounts to 22 (14 C7DASR systems plus 5 C8DASR and the three baseline systems). 
In Figure~\ref{fig:c78_DER_tcpwer_correlation} we instead report DER versus tcpWER in the same manner. 
% since the distribution is heteroschedastic we report PCC and SRC also for each scenario separately. 

\textcolor{black}{First, we observe that JER has slightly higher SRC and PCC than DER with respect to tcpWER. This is because JER better accounts for speaker counting errors. 
In a diarization $\rightarrow$ separation $\rightarrow$ ASR pipeline, speaker counting errors are particularly catastrophic as they compound through subsequent stages. If the number of speakers is estimated incorrectly in the first diarization pass, errors propagate through TS-VAD refinement (if employed), then GSS, and finally ASR. This compounding effect makes speaker counting errors more impactful than other types of diarization errors, even when those speakers have relatively short speech duration.
This phenomenon was also observed in our previous work~\citep{cornell2024chime} when analyzing the C8DASR baseline results. There may be instances where DER values are similar between systems, but tcpWER differs substantially due to differences in speaker counting accuracy.}

\begin{figure}
    \centering
    \includegraphics[width=0.7\linewidth]{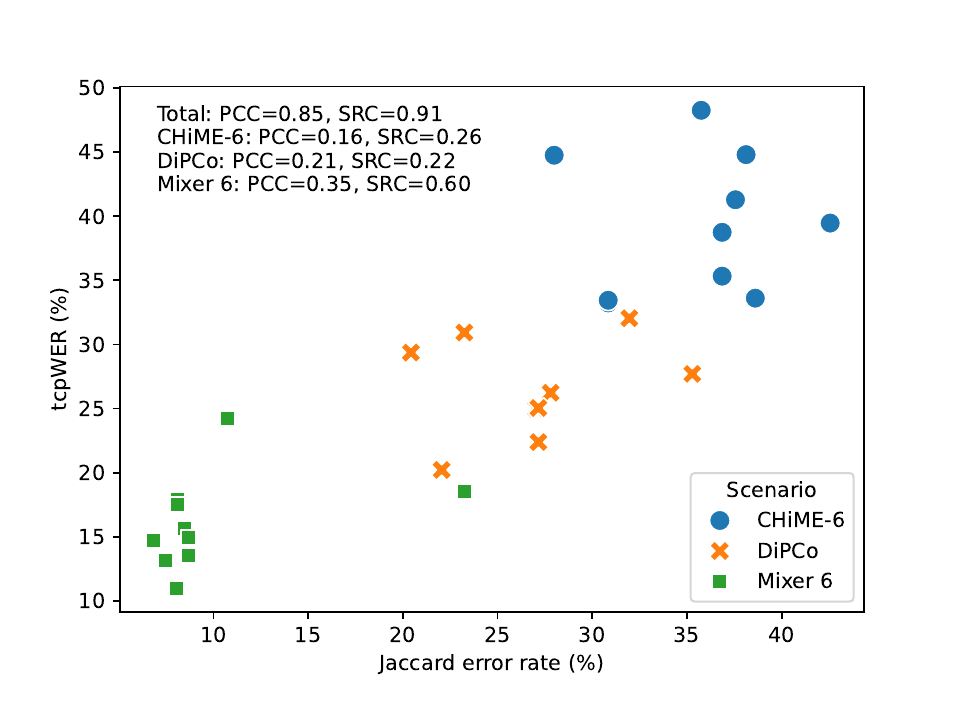}
    \caption{tcpWER (\%) vs. JER (\%) for C7DASR and C8DASR systems (all submissions but only from the top-6 ranking teams). Figures are computed for each core scenario separately (CH\-iME-6, DiPCo, MX6). Pearson correlation coefficient (PCC) and Spearman rank correlation (SRC) are also reported for all the scenarios jointly (Total) and also separately for each scenario.}
    \label{fig:c78_jer_tcpwer_correlation_top}
\end{figure}

%In a diarization $\rightarrow$\,separation$\rightarrow$\,ASR pipeline these are particularly critical, as an error in speaker counting has a compounding effect within the rest of the pipeline. }
%Note that the PCC and SRC are also in both instances lower when each different scenario is considered in isolation. This is because.
On the other hand, observing more closely Figure~\ref{fig:c78_jer_tcpwer_correlation}, it also appears that for the top-4 ranking systems (STCON (C8), NTT (C8), STCON, NTT, USTC-NERCSLIP, IACAS-Thinkit) a lower JER does not always guarantee a lower tcpWER. That is, the correlation seems weaker for such top performing systems. 
Figure~\ref{fig:c78_jer_tcpwer_correlation_top} reports tcpWER versus JER only when considering such top-performing systems, confirming that indeed the correlation is weaker when scenarios are considered independently. 
Especially for the CHiME-6 and DiPCo scenarios (PCC and SRC around 20\%). 

%This phenomenon for example observed in our previous work~\citep{} when analyzing C8DASR baseline system performance. 
% scenario  chime6 dipco mixer6
% systems all PCC SRC
% top-20%
% top-50%

%However, if only top systems are considered, such correlation seems weaker. 
This is, however, expected as such high performing systems do not make significant speaker counting errors and estimate correctly the number of speakers, especially on DiPCo and CHiME-6 scenarios. This is not true for Mixer 6 where even some high-performing systems overestimated the number of speakers (and, in fact, the SRC is lower than the one in Figure~\ref{fig:c78_jer_tcpwer_correlation} but still significant). 
On CHiME-6 and DiPCo most JER errors for these systems arise from segmentation and are thus not very reflective of the final ASR performance. 
This phenomenon was already observed in the previous CH\-iME-6 challenge by the top-ranking team (STCON) on the non-oracle diarization track~\citep{medennikov2020stc, medennikov2020target}. 
Among their system configuration, their best performing one was the one with worse DER as it had a higher false alarm rate.
In fact, for ASR, it is better to combine short utterances from the same speaker into one and have less tight segmentation, as ASR can handle short pauses and usually benefits from having longer context. This approach sacrifices precise utterance-wise segmentation accuracy in favor of improved recognition performance, which is often the preferred outcome. \textcolor{black}{Also recent work~\citep{boeddeker2023ts} on joint TS-VAD/TSE and ASR, found that segment boundaries overestimation is beneficial for ASR even if DER is increased and performed some analysis.}
%Similarly, also recent work on TSE~\citep{boeddeker2023ts} and ASR stresses the importance of diarization segmentation post-processing. 
Such important post-processing step is also performed (as explained in Section~\ref{sssec:diar_baseline}) in the two baseline systems and also by all participants. 
%Such output resegmentation step was is also employed in the ESPnet baseline system and by most participants. 
These observations raise important questions on the relevance of diarization metrics like JER and DER when evaluating diarization systems for ASR applications.
As shown, these metrics, past a certain threshold, may not always capture the impact of diarization quality on overall recognition performance and the resulting user experience.
Therefore, if the application of the diarization system involves ASR, it is crucial to complement DER and JER with WER-based metrics. 

%As such, it is always important to integrate DER and JER with WER-based metrics, again if the application at hand involves ASR. 
%after a certain ``threshold'', may not necessarily capture the impact of diarization quality on overall recognition performance and the resulting user experience.
%only metrics like JER or DER in ASR applications, which is one of the main applica. 

%This fact also raises important questions regarding the significance of diarization-only based metrics like JER or DER for ASR applications which are the most common as they may not fully reflect the impact of diarization quality on the final recognition and user experience. 

%This fact a will be discussed further in Section~\ref{sec:systems_description} during the analysis of participant techniques.
%These results will be further commented in Section~\ref{sec:systems_description} when analyzing participants techniques. 
%The IACAS-Thinkit C7DASR system even goes further by also rescoring the ASR model outputs using an inter-utterance LM. In general, it appears that, especially for conversational speech, long-term context is quite important, as also observed from a recent work which explored the use of LLMs for 
%For conversational speech having longer context seems to be particularly crucial: the IACAS-Thinkit system 

\begin{figure}
    \centering
    \includegraphics[width=\linewidth]{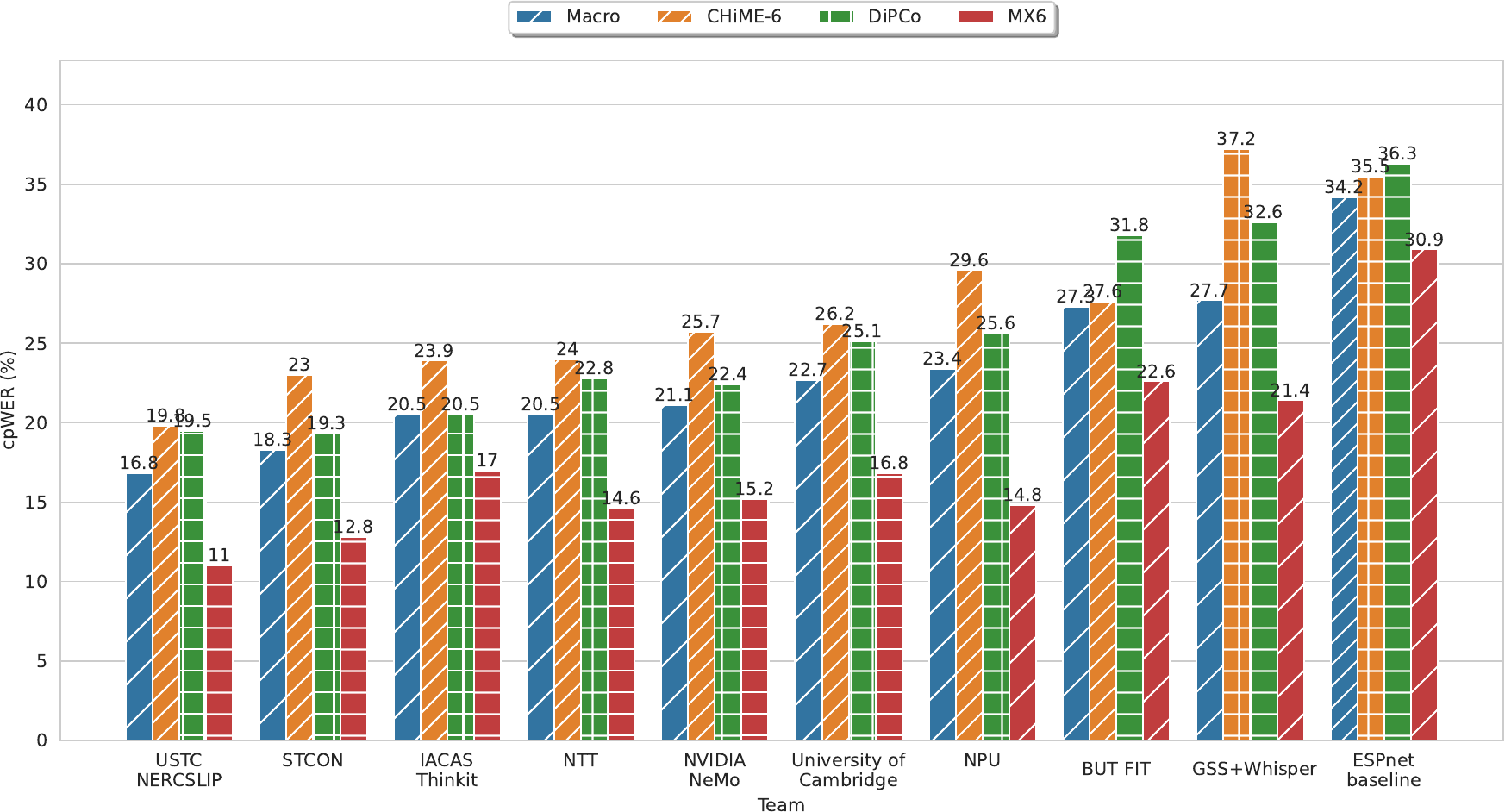}
    \caption{tcpWER (\%) for each C7DASR core scenario (CH\-iME-6, DiPCo, MX6) as well its macro average (Macro) for C7DASR acoustic robustness (oracle diarization) optional sub-track.}
    \label{fig:c8dasr_results_oracle_diarization}
\end{figure}

\begin{figure}
    \centering
    \includegraphics[width=\linewidth]{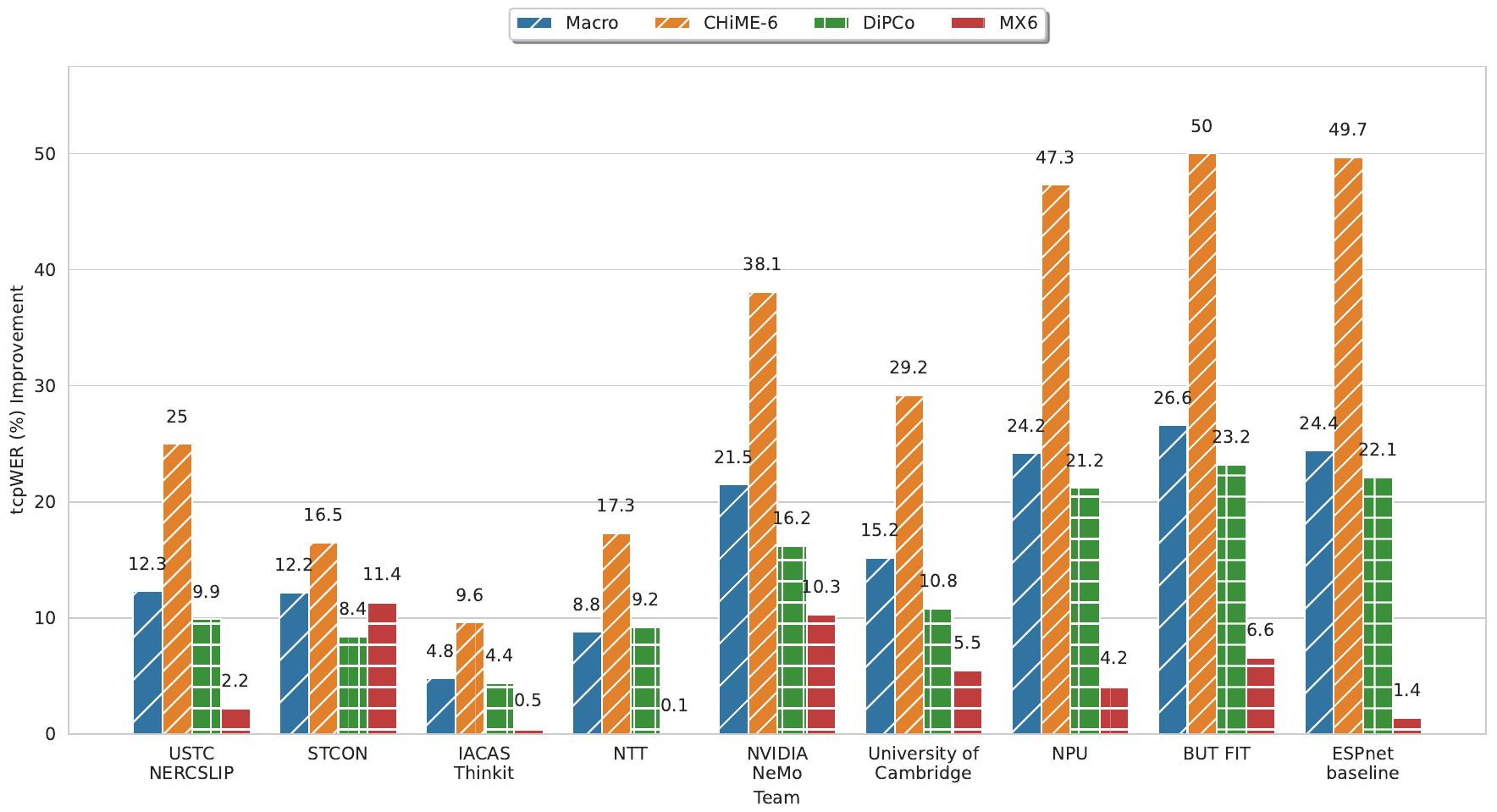}
    \caption{\textcolor{black}{tcpWER improvement (\%) comparing the oracle diarization sub-track to the main track (non-oracle diarization, Figure~\ref{fig:c78dasr_results}) across C7DASR core scenarios (CHiME-6, DiPCo, MX6) and their macro average.}}
    \label{fig:c8dasr_results_oracle_diarization_improvement}
\end{figure}

In Figure~\ref{fig:c8dasr_results_oracle_diarization} we report the results of the C7DASR acoustic robustness sub-track, in which participants could use oracle diarization information. \textcolor{black}{In Figure~\ref{fig:c8dasr_results_oracle_diarization_improvement} for the reader's convenience, we also report the absolute improvement vs. the main track (non oracle diarization, Figure~\ref{fig:c78dasr_results}).}
Compared to previous plots, here we lack C8DASR systems as this track was not included in the C8DASR challenge. Note that this track was also optional: for example, the Paderborn University team did not submit, as they mainly focused on improving the diarization component. 
In this figure, as an additional reference, we also include the results for a baseline system \textcolor{black}{(GSS+Whisper)} that uses Whisper \texttt{large-v3} instead of the baseline ASR WavLM system with all other things equal. 

In general, for all systems, \textcolor{black}{we can observe a significant tcpWER improvement when using oracle diarization especially for the CH\-iME-6 scenario (Figure~\ref{fig:c8dasr_results_oracle_diarization_improvement}). Instead, for DiPCo and MX6, at least for the best systems, the difference is much less pronounced except for STCON. Especially for MX6 which features an easier 2 speakers scenario. This suggests that much of the difficulty in the CH\-iME-6 scenario is still related to accurate speaker diarization and, in particular, speaker counting.
For example, the STCON submission here ranks second with oracle diarization when considering only C7DASR systems. However, in Figure~\ref{fig:c8dasr_results_oracle_diarization_improvement} we can observe a significant difference between oracle and non oracle in macro-averaged tcpWER due to poor performance in MX6. This in turn was due to the fact that their diarization clustering component over-estimated the number of speakers in MX6 as it was over-tuned towards the CH\-iME-6 scenario (it is the best C7DASR system on CH\-iME-6 scenario in Figure~\ref{fig:c78dasr_results}).}

%Overall, for all systems, tcpWER figures on the CH\-iME-6 scenario are noticeably smaller than what is observed in Figure~\ref{fig:c78dasr_results}. Instead, for DiPCo and MX6, at least for the best systems, the difference is much less pronounced. Especially for MX6 which features a much simpler 2-speaker scenario. 
%This suggests that much of the difficulty in the CH\-iME-6 scenario is still related to accurate speaker diarization and, in particular, speaker counting.
%For example, the STCON submission here ranks second with oracle diarization when considering only C7DASR systems. However, in Figure~\ref{fig:c78dasr_results} we can observe a significant degradation in the macro-averaged tcpWER due to poor performance in MX6. This in turn was due to the fact that their diarization clustering component over-estimated the number of speakers in MX6 as it was over-tuned towards the CH\-iME-6 scenario (it is the best C7DASR system on CH\-iME-6 scenario in Figure~\ref{fig:c78dasr_results}). 

\begin{landscape}
   \begin{figure}
    \centering
    \includegraphics[width=\linewidth]{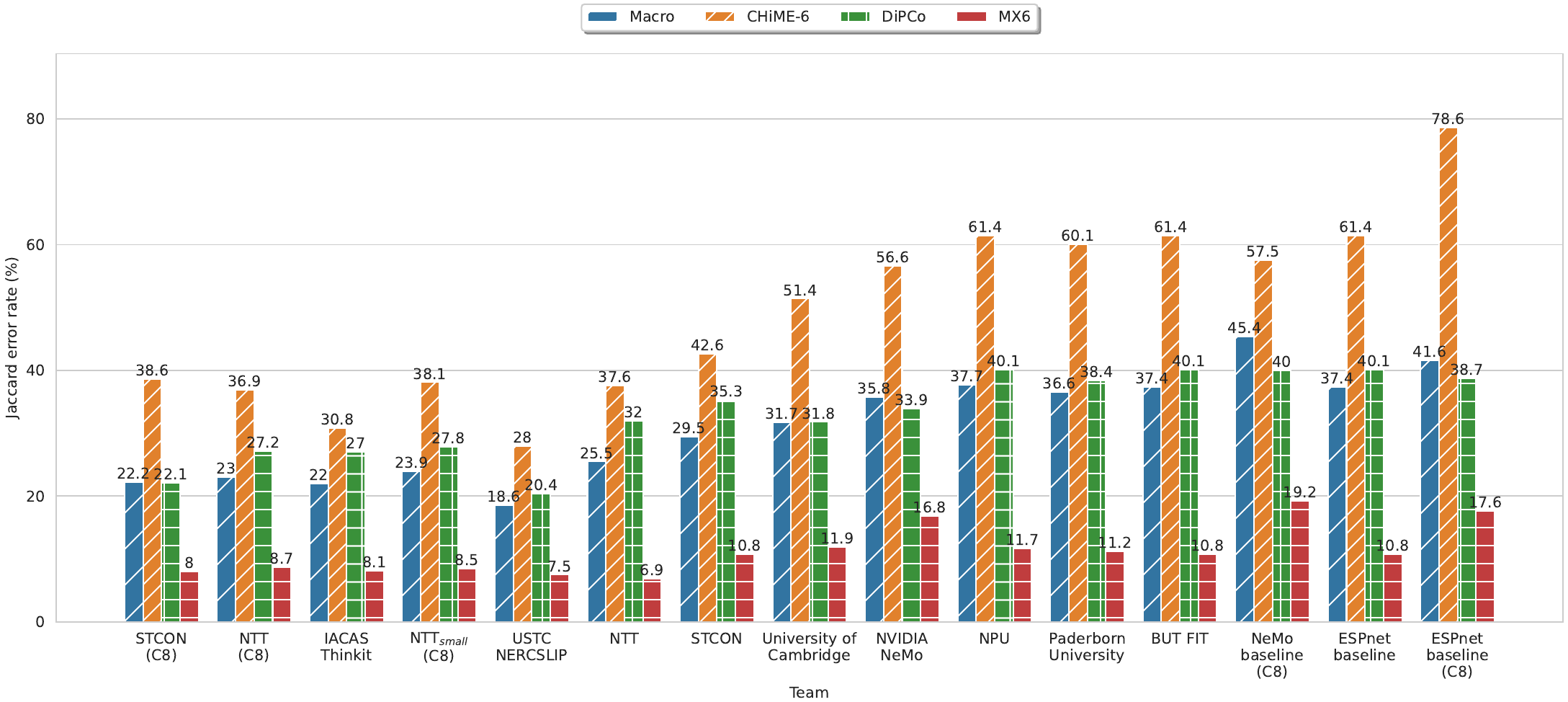}
    \caption{JER (\%) for each C7DASR core scenarios (CH\-iME-6, DiPCo, MX6) as well its macro average (Macro) for both C7DASR and C8DASR systems. C8DASR systems are denoted by (C8).}
    \label{fig:c78dasr_JER_results}
\end{figure} 
\end{landscape}
\noindent
In contrast, their C8DASR system does not suffer from this issue and demonstrates greater robustness, thanks to some improvements which were explained in Section~\ref{sec:systems_description} such as the wav2vec 2.0-based frame-level speaker embedding model and the improved clustering pipeline. 
These observations are confirmed by Figure~\ref{fig:c78dasr_JER_results} where we plot instead, for all C7-8DASR systems JER for each C7DASR core scenarios (CH\-iME-6, DiPCo, MX6) as well its macro average. 
It can be clearly seen that CHiME-6 is the scenario with the highest JER for all submissions followed by DiPCo. Also, C7DASR STCON system obtains higher JER on MX6 than the other top systems, and comparable to the C7DASR ESPnet baseline one, indicating sub-par diarization performance. 
It is evident the degradation in JER and thus in tcpWER (previous Figure~\ref{fig:c78dasr_results}) for the two C8DASR baselines. This is because their parameters have been re-tuned to handle also the NOTSOFAR-1 scenario, and thus make significant speaker counting errors on the other scenarios. Again, this fact underscores the difficulty of achieving a good performance balance across all four scenarios.  

%%%%%%%%%%%%%%%%
% ADD JER PLOT ? 
%%%%%%%%%%%%%%%%
%Their C8DASR system does not suffer from the same pitfall and it is more robust in this regard as it will be explained in Section~\ref{sec:systems_description}.
Similarly, USTC-NERCSLIP ranks first with oracle diarization, but its performance then degrades more compared to the IACAS-Thinkit submission. However, in this case, the degradation was mainly due to wrong word-level segmentation. In terms of DA-WER, the USTC-NERCSLIP team won and ranked first in the C7DASR challenge~\footnote{See leaderboard: \href{https://www.chimechallenge.org/challenges/chime7/task1/results}{https://www.chimechallenge.org/challenges/chime7/task1/results}}. However, in Figure~\ref{fig:c78dasr_results}, it is more penalized by tcpWER as it makes more utterance-wise timestamp errors. This appears to be the result of employing forced alignment during inference. 
In fact, this system uses a rather complex iterative inference procedure in which GSS is re-ran after 
a first inference pass, after using forced alignment with respect to ASR outputs in order to derive more precise word boundaries and thus speaker activity which could be re-used for GSS guidance.
While this approach aims to enhance accuracy, it may introduces vulnerabilities: if the ASR fails to recognize certain words correctly, the forced alignment may produce outlier word segments, leading to timestamp errors and affecting the overall tcpWER. JER as observed in Figure~\ref{fig:c78dasr_JER_results} is not affected, due to the fact that the collar mitigates such errors which consists mostly of single words. 
This phenomenon may be the reason why the CHiME-8 NOTSOFAR-1 USTC-NERCSLIP submission~\citep{niu2024ustc}, which is very similar to their C7DASR submission in terms of pipeline structure, no longer employs this approach.

\subsection{CH\-iME-8 DASR Results}

\begin{figure}
    \centering
    \includegraphics[width=0.9\linewidth]{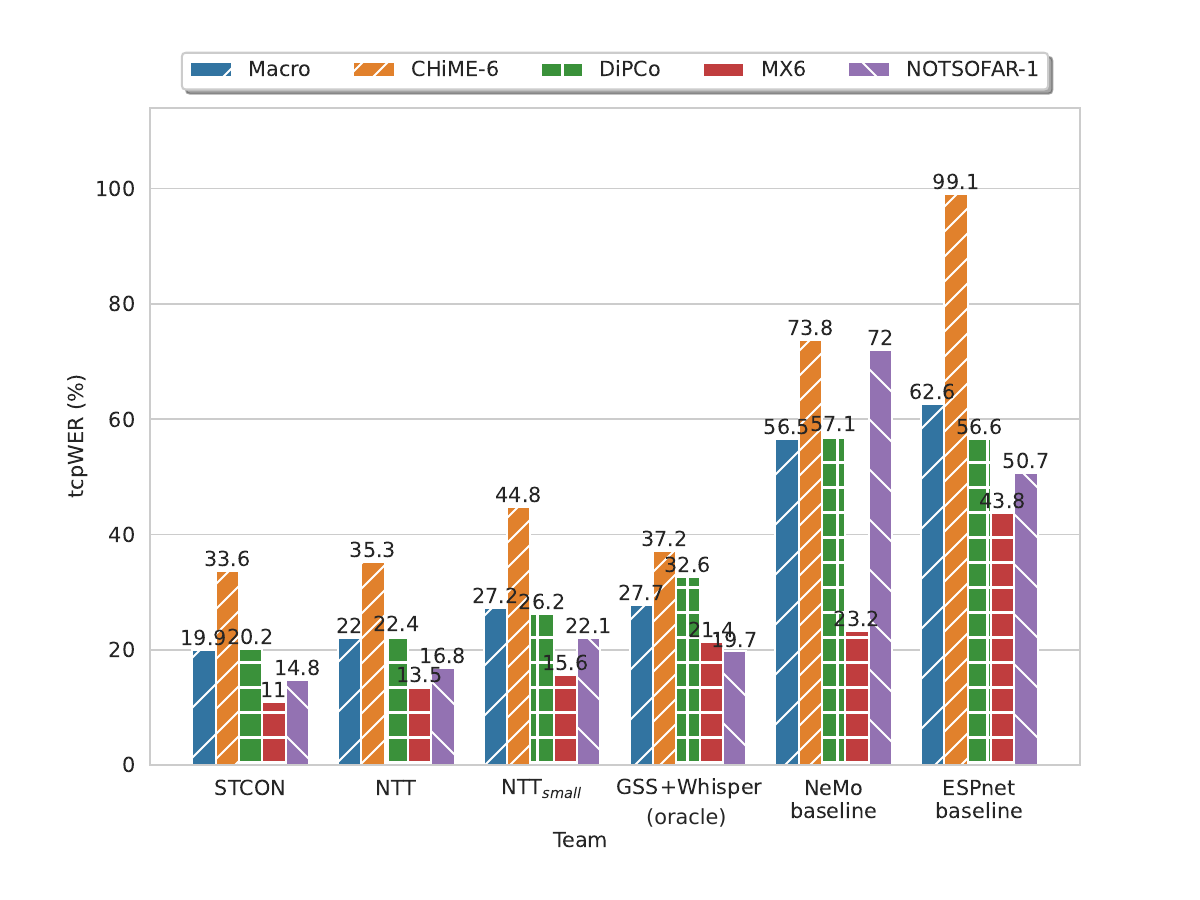}
    \caption{C8DASR results. We report tcpWER (\%) for each C8DASR core scenario (CH\-iME-6, DiPCo, MX6 and NOTSOFAR-1) as well as its macro average (Macro).}
    \label{fig:c8dasr_results}
\end{figure}

In Figure~\ref{fig:c8dasr_results} we plot the final results for the C8DASR challenge. As mentioned, the challenge saw limited participation from 3 teams, also due to the fact that many participants split between the two CHiME-8 DASR and NOTSOFAR-1 challenges. %One of the teams was disqualified and is here \textcolor{black}{anonymized} because it did not submit a system description paper. 
%Despite being disqualified, we have included this anonymized submission result to highlight the complexity and difficulty of the task, as it ended up performing worse than the baseline systems in the evaluation set. 
An oracle-based system, consisting of oracle diarization, GSS and Whisper \texttt{large-v3} \textcolor{black}{GSS+Whisper (oracle)} is also added in the figure as a reference.

%STCON system performs best overall, with the NTT system second but with close overall performance. Overall, both systems having very balanced performance across the four scenarios.
The STCON system achieved the best overall performance, with the NTT system closely following. Both systems have fairly well-balanced performance in all four evaluation scenarios.
Interestingly, overall, the tcpWER for NOTSOFAR-1 is, for these two systems, only slightly higher than the one obtained on MX6, indicating very high robustness in speaker counting capability, at least for relatively high SNR environments. 
The NTT$_{small}$ system overall does not exhibit much degraded performance despite not using any diarization refinement and only a single ASR model.
%The two baseline systems performs much worse, mostly due to wrong speaker counting as also explained in the previous section when presenting C7-8DASR JER results in Figure~\ref{fig:c78dasr_JER_results}.
%It is interesting to see how the two baselines have somewhat opposite behaviors regarding performance on NOTSOFAR-1. The NeMo one has very weak performance on NOTSOFAR-1, as it tends to under-estimate the number of speakers in such scenario~\citep{cornell2024chime}, while the ESPnet baseline has an opposite behavior and instead under-estimates on CHiME-6. As the two baseline relies on a too simple speaker counting strategy, they exhibit a ``zero-sum game'' behaviour where 

%The two baseline systems perform significantly worse, primarily due to incorrect speaker counting, as previously discussed in the context of C7-8DASR JER results shown in Figure~\ref{fig:c78dasr_JER_results}.
%It should be noted that the two baselines display contrasting behaviors on the NOTSOFAR-1 dataset. The NeMo baseline exhibits very weak performance in this scenario, as it tends to underestimate the number of speakers~\citep{cornell2024chime}. Similarly, the ESPnet baseline shows the opposite issue, underestimating the number of speakers in the CHiME-6 scenario.
%Since both baselines rely on classical speaker counting strategies, they exhibit a ``zero-sum game'' behavior when faced with such contrasting scenarios. Improvements in speaker counting for the CHiME-6 scenario often lead to a deterioration in performance for NOTSOFAR-1 and vice-versa, due to their almost opposite characteristics in duration and, for the latter, variability in the number of speakers. 
\textcolor{black}{The two baseline systems perform significantly worse, primarily due to incorrect speaker counting, as previously discussed in the context of C7-8DASR JER results shown in Figure~\ref{fig:c78dasr_JER_results}.
It should be noted that the two baselines display contrasting behaviors on the NOTSOFAR-1 dataset. The NeMo baseline exhibits very weak performance in this scenario, as it tends to underestimate the number of speakers~\citep{cornell2024chime}. The ESPnet baseline instead performs better on NOTSOFAR-1 but then underestimates the number of speakers in the CHiME-6 scenario.
Since both baselines rely on classical speaker counting strategies, they exhibit a ``zero-sum game'' behavior when faced with such diverse scenarios. Improvements in speaker counting for the CHiME-6 scenario often lead to a deterioration in performance for NOTSOFAR-1 and vice-versa, due to their contrasting session-level characteristics: CHiME-6 has 2+ hour sessions with 4 speakers, while NOTSOFAR-1 has $\sim$10 minute meetings with 4-8 speakers.}

As mentioned in Section~\ref{ssec:tracks}, C8DASR featured an additional optional track where several pre-trained LLM models were also allowed. 
As briefly mentioned in Section~\ref{sec:description}, only the STCON team participated in this sub-track with an approach based on Llama-2 and similar to~\citep{ogawa2024applying}. However, as mentioned, this resulted in very limited performance gains: a reduction of only 0.5\% in absolute macro-tcpWER compared to their main track submission. 

In general, the best-performing systems (STCON and NTT) are remarkable in being superior even to the \textcolor{black}{GSS+Whisper (oracle)} system (which uses oracle diarization). This includes even the NTT$_{small}$ system, despite the fact that it does not use any ASR ensembling technique and no diarization refinement. 

%\texttt{7B} fine-tuned on the whole C8DASR \texttt{train} text data augmented using LibriSpeech text, for ASR hypothesis LM rescoring with an approach similar to. 
% mention the nttsmall is quite good. 
% mention the fact that the STCON system with LLM refinement has basically the same performance. 

\subsection{Joint CH\-iME-8 DASR and NOTSOFAR-1 Challenges Results}\label{ssec:c8dasr_notsofar1_results}

Some of the C7DASR participants did not participate in C8DASR but only in the ``twin'' time-concurrent CH\-iME-8 NOTSOFAR-1 challenge that, as mentioned, focused exclusively on the NOTSOFAR-1 scenario. In C8DASR, this latter is instead one of the four scenarios participants should address (together with CH\-iME-6, DiPCo and MX6). As mentioned in Section~\ref{sec:intro}, the main motivation for having both challenges was to compare across the two. The idea was to assess how ``generalist'', recording setup agnostic solutions can measure up with respect to domain-specific ones, considering that \textcolor{black}{the latter} can use a-priori information about the array configuration and the deployment domain. 

\begin{figure}
    \centering
    \includegraphics[width=0.98\linewidth]{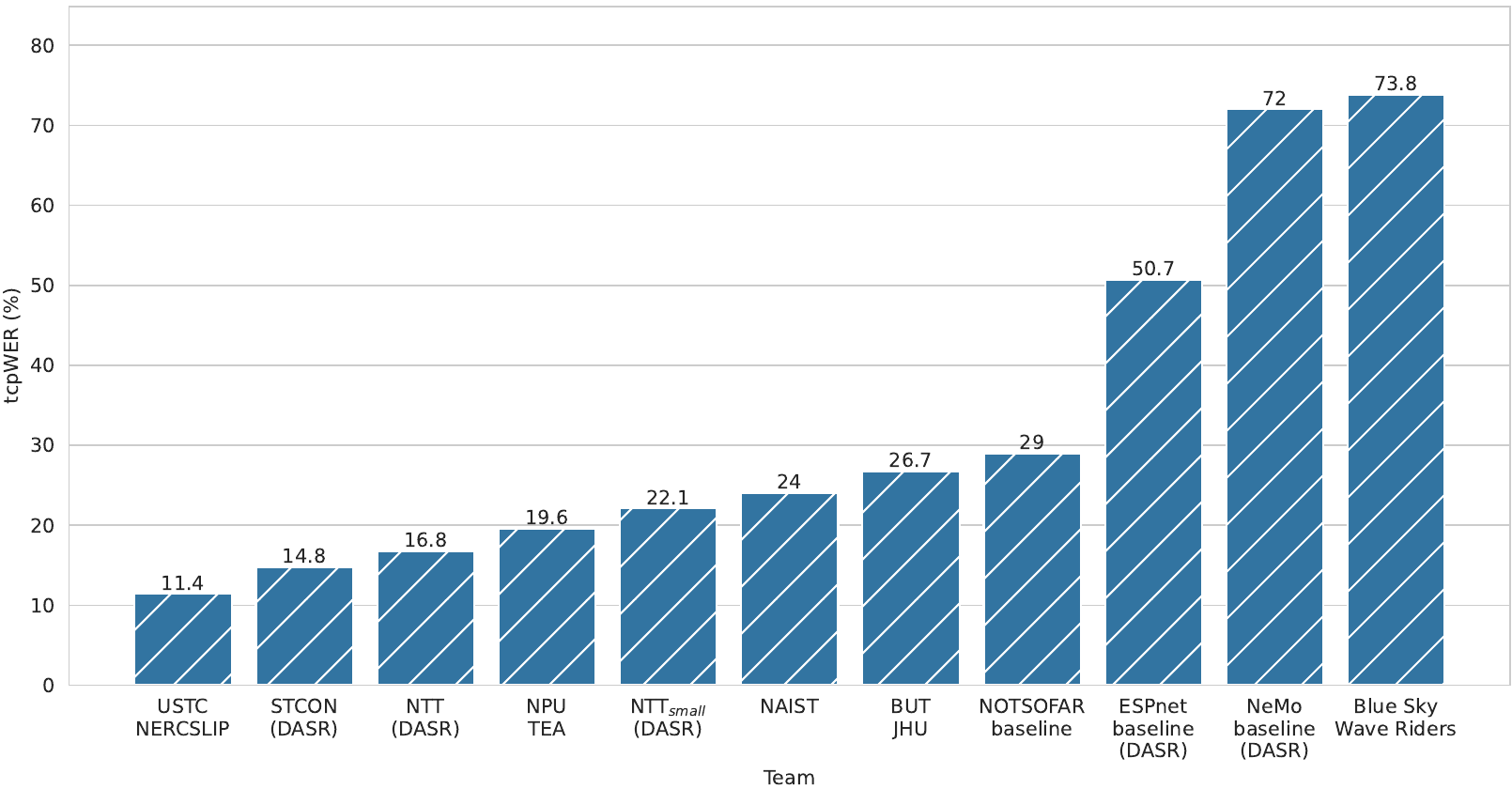}
    \caption{tcpWER (\%) for CH\-iME-8 NOTSOFAR-1 (multi-channel track) and DASR challenges systems on the NOTSOFAR-1 scenario.}
    \label{fig:c8_notsofar1_tcpwer}
\end{figure}

CH\-iME-8 DASR and CH\-iME-8 NOTSOFAR-1 joint results are in Figure~\ref{fig:c8_notsofar1_tcpwer} where we plot, for each system, the tcpWER on the NOTSOFAR-1 scenario.
For NOTSOFAR-1, we report only the best systems from the multi-channel track and not those from the single-channel one, which, as expected, have slightly worse performance. 
Note that the values are different from those reported in the official NOTSOFAR-1 challenge results, as we accumulate error statistics rather than average tcpWER across the sessions as done in the CHiME-8 NOTSOFAR-1 challenge.
The reason is that averaging tcpWER per session biases the overall figure on the easier sessions, which have fewer words and also fewer participants.
\textcolor{black}{However, it is worth mentioning that the overall system rankings remain consistent between both computation methods, with differences in absolute tcpWER values of only 1-2\%\footnote{Results with NOTSOFAR-1 tcpWER averaging: \url{https://www.chimechallenge.org/challenges/chime8/task2/results\#notsofar-and-dasr-results-supplementary}}.} 
The CHiME-8 NOTSOFAR-1 challenge had 5 teams participating, in alphabetical order: Blue Sky Wave Riders, BUT JHU~\citep{maciejewskibut}, NAIST~\citep{hirano2024naist}, NPU~\citep{huang2024npu} and
USTC-NERCSLIP~\citep{niu2024ustc}.

In Table~\ref{tab:sys_summary_frontend_notsofar1}, we summarize the front-end and diarization techniques employed by participants to aid in the analysis of results, following a similar approach to that used in Table~\ref{tab:dasr_sys_summary_frontend} for the C7-8DASR challenges.
For the reader's convenience, the characteristics of the C8DASR STCON and NTT submissions are also included here, marked with a (C8) tag. C8DASR baselines are omitted, as they have already been discussed in ample detail.

%also regarding the CHiME-8 NOTSOFAR-1 challenge, most 
% we also report the solutions for the front end 
% we should note that NOTSOFAR-1 baseline uses CSS, performs better than baselines but also 
% needs not to address the other scenarios. 
% plot of CH\-iME-7 DASR + 8 results. 
% plot systems of C8DASR only 
% plot each scenario in terms of tcpWER as done for the website.
%\subsubsection{Joint CH\-iME-8 DASR and CH\-iME-8 NOTSOFAR-1 Results}
% only one scenario here plot. 
% note that the tcpWER is different from one in the website as we accumulate statistics

%\begin{figure}
%    \centering
%    \includegraphics[width=0.8\linewidth]{images/c8notsofar_jer_tcpwer.pdf}
    %\input{}
%    \caption{tcpWER (\%) vs. JER (\%) for CH\-iME-8 NOTSOFAR-1 and DASR systems on the NOTSOFAR-1 scenario.}
%    \label{fig:c8_notsofar1_jer_tcpwer}
%\end{figure}

\begin{landscape}
\begin{table}[htbp]{
\centering
\scriptsize
\caption{Summary of diarization and front-end components for CH\-iME-8 NOTSOFAR-1 challenge multi-channel track. Top panel (\colorbox{neutralOne}{\vphantom{a}\smash{grey}}): baseline system; bottom panel: participant submissions. Within each panel, the participants' systems are ordered according to macro-averaged tcpWER. C7-8DASR baselines are omitted here. STCON and NTT systems are the same across NOTSOFAR-1 and CH\-iME-8 challenges. C8DASR submissions are in \colorbox{ai2offwhite}{\vphantom{a}\smash{yellow}}.}
\label{tab:sys_summary_frontend_notsofar1}
\setlength{\tabcolsep}{0.2em}  % Increased spacing between columns
\begin{tabular}{lccccccc}  % Simplified column alignment
\toprule   
Team & \multicolumn{4}{c}{Diarization} & \multicolumn{2}{c}{Front-End} \\
&  &  &  &  &  \\ 
\cmidrule(l{2pt}r{2pt}){2-5}  \cmidrule(l{2pt}r{2pt}){6-7} 
 & Segmentation & Spk-id Extr. & Refinement & Multi-Channel & Channel  & Separation  \\
 &     & \& Clustering  &      & Mechanism  &  Selection  &   \\ 
 
 \midrule
 \rowcolor{neutralOne} NOTSOFAR-1  & Whisper & TitaNet &  &  \multirow{2}{*}{-}   &     &  Conformer CSS +  \\ 
\rowcolor{neutralOne}  Baseline & word timestamps  & + NME-SC  & \multirow{-2}{*}{-}  &   & \multirow{-2}{*}{-} & MVDR  \\ 
 
\midrule
 \midrule
 \rowcolor{ai2lightgreen}
   & Conformer CSS/OSD + JDS & ResNet-221  & NSD-MS2S &   &    &  WPE + Conformer \\ 
\rowcolor{ai2lightgreen}  & + Enhanced Whisper & + NME-SC  & + JDS &   &  &   CSS/OSD + GSS  \\ 
\rowcolor{ai2lightgreen} \multirow{-3}{*}{USTC} &  word timestamps  &  &  & \multirow{-3}{*}{-}   & \multirow{-3}{*}{-} &   + JDS refinement  \\ 

\midrule

 \rowcolor{ai2offwhite} & TDNN  &  ECAPA-TDNN \& &\multirow{3}{*}{NSD-MS2S}  & DOVER-Lap and   &  &  GSS with \\ 
\rowcolor{ai2offwhite}&  stats-based  & wav2vec 2.0 AED multi-speaker OSD  &  &  TS-VAD &  & neural  \\ 
\rowcolor{ai2offwhite} \multirow{-3}{*}{STCON (C8)} &   &  with UMAP+DBSCAN+GMM clustering&   & posterior averaging  &  \multirow{-3}{*}{EV} & refinement (G-TSE) \\ 

\midrule

\rowcolor{ai2offwhite} &   & EEND-VC \&  & &  DOVER-Lap \&  & EV \& &  \\ 
\rowcolor{ai2offwhite} &   & ECAPA-TDNN &  &  channel clustering  & Brouhaha C$_{50}$ &   \\ 
\rowcolor{ai2offwhite} \multirow{-3}{*}{NTT (C8)} &  \multirow{-3}{*}{EEND-VC}  &  with NME-SC  & \multirow{-3}{*}{NSD-MS2S}  & for speaker counting  &   &  \multirow{-3}{*}{GSS with SP-MWF} \\

\midrule

% NOTE that they perform rover with no separation too 
% the performance is similar. 

  \rowcolor{ai2lightgreen} NPU  & Silero + Whisper & ResNet293 + NME-SC  &  &     &    &  WavLM Conformer \\ 
 \rowcolor{ai2lightgreen}  & word timestamps  &  + speaker merging & \multirow{-2}{*}{-} & \multirow{-2}{*}{-}  & \multirow{-2}{*}{-}  &   CSS + MVDR  \\ 
\midrule

 \rowcolor{ai2lightgreen} NAIST  & Whisper & TitaNet &   &     &     &   WPE + Conformer \\ 
 \rowcolor{ai2lightgreen}  & word timestamps  & + NME-SC  & \multirow{-2}{*}{-} & \multirow{-2}{*}{-}  & \multirow{-2}{*}{-} &   CSS + MVDR  \\  
\midrule

 \rowcolor{ai2lightgreen} & WavLM multi-channel & ECAPA-TDNN +  & -  &   & -  & - \\
\rowcolor{ai2lightgreen} \multirow{-2}{*}{BUT/JHU} & local EEND & NME-SC  &  &   \multirow{-2}{*}{TAC-layers} &  &   \\ 

%STCON   &   &   & NSD-MA-MSE  &   &  EV  &  GSS+neural TSep refinement\\ 
%NTT   &   &   &   &  NSD-MA-MSE  &  EV  &  GSS (rank-1 Wiener filter) \\ 

\bottomrule
\end{tabular}
}
\end{table}
\end{landscape}

Notably, compared to C7-8DASR, the approaches in this context are more varied. They are divided among three main pipelines: diarization~\textrightarrow~GSS~\textrightarrow~ASR (adopted by USTC-NERCSLIP, STCON, and NTT), separation~\textrightarrow~ASR~\textrightarrow~diarization (NOTSOFAR-1 baseline, NAIST and NPU), and a simpler diarization~\textrightarrow~ASR pipeline, as used by the BUT JHU submission.
The former, as said, has been the prevalent approach for CHiME-6 and CHiME-7 DASR challenges; the second is similar to \citep{yoshioka2019advances, raj2021integration, kanda2022vararray} and here is adopted mainly because it is the approach that the NOTSOFAR-1 challenge baseline uses~\citep{vinnikov2024notsofar}.
\textcolor{black}{The latter} uses a Conformer-based CSS model, which is applied without VAD on the entire session. 
%This CSS model is based on \citep{chen2020continuous}, and estimates an short-time Fourier transform (STFT) magnitude mask for up to three local speakers given a context of 3\,s and with an hop-size of 1.5\,s. This magnitude mask is then used to drive an MVDR beamformer, following~\citep{souden2009optimal}. 
%This model takes as input from all 7 channels magnitude and phase STFT 
%Thus, it is sub-optimal for ad-hoc array network processing, as these require more refined approaches that can allow generalization to unknown configurations~\citep{luo2020end}.
%and is trained on a 1000\,h synthetic multi-channel conversational dataset which has been created by using real-world measured array transfer functions. 
After CSS, Whisper \texttt{large-v2} is employed for recognition and word-level segmentation, as it can predict word-level timestamps. Such segmentation is then used to extract TitaNet embeddings which are then clustered with NME-SC to obtain also speaker-attribution. 
NPU~\citep{huang2024npu} and NAIST~\citep{hirano2024naist} teams adopt the NOTSOFAR-1 baseline CSS model. NPU modifies it by using WavLM as a feature extractor, while NAIST uses MIMO-WPE before CSS to enhance dereverberation capabilities.
For the backend, NPU used Whisper \texttt{large-v2} but fine-tuned it using AdaLoRA~\citep{zhangadaptive}. NAIST instead uses K2 Zipformer with WavLM \texttt{large} as the frontend, reporting a great speed up in inference time over the baseline and better performance. 

The USTC-NERCSLIP submission~\citep{niu2024ustc}, also uses the same CSS component as the NOTSOFAR-1 baseline, but only within its diarization sub-module and for diarization purposes as they extend it to perform also overlapped speech detection jointly with separation. 
The system has some similarities with their C7DASR submission and in fact heavily relies on NSD-MS2S TS-VAD diarization refinement followed by GSS. 
GSS is initialized by a neural TSE model, similarly to the C7DASR IACAS-Thinkit system but with the difference that here this neural TSE model is integrated with the NSD-MS2S TS-VAD in a single joint diarization-separation model (JDS). 
%An initial diarization hypothesis is then extracted using an in-house ResNet-221~\citep{he2016deep} speaker-id discriminative embedding model after CSS. CSS+MVDR is performed only on segments that contain overlapped speech, otherwise GSS is used. 
%This initial diarization is then refined using the and after diarization refinement, a neural TSE model is employed to initialize GSS. As such GSS is used to perform separation. In this sense, this system is clearly inspired by their previous C7DASR system, with the neural TSE being a clear novelty. 
%However, the use of a neural TSE model for initialization is very interesting, and similar to what IACAS-Thinkit did in the C7DASR challenge. STCON instead in C8DASR used neural TSE for GSS refinement. 
%The USTC-NERCSLIP approach differs from these as the TSE network is jointly optimized and combined with the NSD-MS2S, in a . The TSE network is based on Conformer and estimates the target speaker mask, it is conditioned by NSD-MS2S frame-level target speaker activities. 
The back-end is \textcolor{black}{a heavily} modified version of Whisper (both \texttt{large-v2} and \texttt{large-v3} are used, as the final system is an ensemble), which is augmented/modified with RoPE positional encodings~\citep{heo2025rotary}, 
a mixture of expert component from~\citep{you2021speechmoe} and even WavLM \texttt{large} features. 

%Its output is used only on overlapped speech segments in order to extract R 

The BUT-JHU~\citep{maciejewskibut} system instead significantly departs from the baseline system by not relying at all on explicit separation. This system uses a target speaker augmented Whisper model~\citep{polok2024target}, which can perform TSE implicitly, without relying on techniques such as GSS.
For diarization, they use a WavLM-based local EEND model~\citep{han2024leveraging} similar to the one employed in the Pyannote diarization pipeline 2.1 (and thus the ESPnet baseline). 
%This model is extended to handle multiple channels via a technique similar to transform-average-concatenate (TAC)~\citep{luo2020end} following the approach of~\citep{movsner2024multi}.

Regarding the three C8DASR systems, we observe that they compare favorably with the CHiME-8 NOTSOFAR-1 submissions, despite also addressing three additional scenarios, two of which are arguably more challenging, as it is evident from tcpWER figures in Figure~\ref{fig:c8dasr_results}. 
More broadly, all top-three systems utilize GSS and diarization refinement through the NSD-MS2S TS-VAD model, reaffirming the effectiveness and versatility of these techniques. Furthermore, this highlights, again, the superior robustness of GSS compared to current CSS or purely neural TSE methods, even, crucially, when the domain and array configuration is known and it is not ad-hoc. 
As such, there is a clear need for further research in this area, particularly by focusing on self/semi-supervised learning techniques~\citep{wisdom2020unsupervised, bando2024neural}. 

%Regarding the three C8DASR systems, we can see that they compare favorably with CHiME-8 NOTSOFAR-1 submissions, despite having also to tackle other three scenarios, two of which are arguably more challenging as seen in Figure~\ref{fig:c8dasr_results}. 
%More in general, all top three systems employ GSS and diarization refinement through the NSD-MS2S TS-VAD model, confirming the efficacy of these techniques and their wide applicability, and, more, in general, their superior robustness to CSS or pure neural TSE, as noted by the C8DASR STCON system description paper. 

%Again, as mentioned, also when examing these results it is still clear that there using neural TSE or neural separation (as in CSS) in real-world meeting is challenging. The only system that is using it directly is the STCON one, but, again because e2e fine-tuning with the ASR criterion was done. 

Among the C8DASR systems, NTT$_{small}$ is particularly noteworthy, as it does not employ any ensembling techniques (unlike most other submissions, except BUT JHU) or diarization refinement. % however it must be mentioned that its RTF as reported in it is still high at 
The BUT JHU system also shows promise due to its streamlined pipeline. However, it falls short of the performance achieved by other submissions. One limitation is that its separation is handled implicitly within the ASR system, limiting its ability to fully leverage multi-channel information. %And in fact their performance gap between multi-channel and single channel 

%USTC-NERCSLIP system is an improved version of their C7DASR submission. 
%It must be noted however that the CSS component is only applied in the multi-channel track. 
%In fact, during baseline development. 

% mention that STCON found CSS not to work on C8DASR it is too brittle in dipco and chime 6 scenarios. Plus challenging CSS for array agnostic generalization. 
% ask whether MixIT can be leveraged in future ? or other SSL tehcniques. 

% general approaches compare favourably in general. 
% NTT small does not use ensembles actually  but only whisper m

% report also the standard deviation here ? mmmh maybe not necessary ? 
% report it later after the 

%\section{Further Analysis and Discussion}\label{sec:results}
%\subsection{Joint C7-8DASR and CH\-iME-6 Challenge Results}
%Several teams have been participating to the . 
%It is thus possible to track 

%Progress seems dictated mainly by advancements in the diarization part. With TS-VAD. 
% after TS-VAD returns are lower but the task has gotten more complicated. 
% C8 can improve further. NTT_small especially robustness due to whisper and large scale training is playing a role here. Aside from diarization. 

%\begin{figure}
 %   \centering
%    \includegraphics[width=0.8\linewidth]{images/c678_jer_tcpwer_improved.pdf}
   % \input{}
 %   \caption{tcpWER (\%) vs. JER (\%) for CH\-iME-6 (Track 2 unconstrained LM), C7DASR and C8DASR challenges systems on the CH\-iME-6 scenario.}
%    \label{fig:c678_jer_tcpwer}
%\end{figure}

%\subsection{Joint CH\-iME-8 DASR and CH\-iME-8 NOTSOFAR-1 Challenge Results}
% should we put into appendix a table with NOTSOFAR-1 systems characteristics ?

\subsubsection{System evaluation on meeting summarization for the NOTSOFAR-1 scenario}\label{ssec:summarization_results}

%The NOTSOFAR-1 dataset, as described in Section~\ref{sssec:notsofar1}, differs from the other scenarios by featuring an office meeting centered around a specific topic. This makes it possible to perform downstream evaluation on meeting summarization tasks, unlike the other three C8DASR scenarios, which represent purely conversational, speech-in-the-wild settings.
%The NOTSOFAR-1 dataset, described in Section~\ref{sssec:notsofar1}, differs from the other scenarios by focusing on an office meeting centered on a specific topic. This feature makes it suitable to perform meeting summarization tasks. The other three C8DASR scenarios, instead, represent purely conversational, speech-in-the-wild settings and thus are unsuitable for it. 
The NOTSOFAR-1 dataset, described in Section~\ref{sssec:notsofar1}, differs from the other scenarios by focusing on structured office meetings centered on specific topics. This feature makes it well-suited for meeting summarization tasks. In contrast, the other three C8DASR scenarios represent purely conversational, speech-in-the-wild settings without particular structure, making them less appropriate for \textcolor{black}{gauging downstream summarization performance}.
\textcolor{black}{As mentioned in the introduction, summarization} is an important downstream task for joint diarization and ASR which has recently gained even more traction due to the rise of LLMs. It is thus interesting to evaluate how much the quality of meeting summaries can depend on accurate transcription and correct speaker attribution. 
In numerous practical settings, such as routine office meetings, perfect verbatim transcription may be unnecessary, while concise, accurate meeting summaries remain highly desirable.
For these applications, traditional ASR and diarization metrics derived from WER (e.g. tcpWER, cpWER) or DER/JER may inadequately reflect actual user experience and over-estimate the amounts of errors. This is especially true for spontaneous conversational speech where it is common to find contraptions, hesitations, false starts, filler words, and repetitions that may be counted as errors but have minimal impact on the overall meaning~\citep{kim2021semantic}. % For example, WER-derived metrics are high dependent on text normalization
This is particularly relevant considering that the LLM summarization process can correct or compensate for certain inaccuracies in transcription and speaker attribution, producing useful summaries despite imperfect input. 
On the other hand, text summarization has other issues such as the fact that the evaluation is difficult due to the inherent subjectivity of what constitutes a ``good'' summary~\citep{lee2024unisumeval}. 

%Especially for meeting and dialog summaries 
%Furthermore, especthere's a trade-off between factual accuracy and conciseness in meeting summaries. While all key decisions and action items should be preserved, there's significant flexibility in how much contextual discussion to include. This makes it difficult to establish a single gold standard for evaluation.

%Moreover, this kind of downstream evaluation is valuable because it measures what can ultimately matter to users in a particular application scenario. 
%In many settings, such as non-critical office meetings, verbatim transcription is not really needed while a generally accurate and useful meeting summary may be desired. 
%In these instances objective, low-level metrics derived from WER (e.g. tcpWER) or DER may be sub-optimal. 
%getting accurate, useful meeting summaries - rather than just optimizing for technical metrics like WER (Word Error Rate) or DER (Diarization Error Rate) in isolation.

%For this evaluation, we used the Gemini 2.0 flash~\citep{team2024gemini} LLM model to generate 10 summaries for each system submission transcript, aiming for approximately 200 words per summary. We set for a 200 words limit because the average number of words in grount-truth transcriptions is $\sim 1600$ words as NOTSOFAR-1 meetings are rather short ($\sim 6 mins$, Table~\ref{tab:dset_overview}, Sectio~\ref{sec:core_scenarios}). 
For this evaluation, we used the Gemini 2.0 Flash~\citep{team2024gemini} LLM to generate summaries for each NOTSOFAR-1 session starting from \textcolor{black}{each submitted transcript}. We prompt the LLM to generate summaries of  approximately 200 words. We selected this 200 words limit because NOTSOFAR-1 meetings are relatively short (averaging 6 minutes as shown in Table\ref{tab:dset_overview}, Section~\ref{sec:core_scenarios}) with ground-truth transcriptions averaging $\sim 1600$ words. This summary length was considered a reasonable trade-off between compression and content preservation.

The following prompt was used, where \texttt{\{transcription\}} was replaced with the actual system transcription: 
\footnotesize
\begin{gbox}[red!02]
You are provided with the transcription of a meeting
involving a minimum of two to maximum eight participants.

Please create a comprehensive summary (approximately 200 words) 
of the provided meeting transcription. Your summary should capture 
all key points, decisions, action items, and significant exchanges. 
The transcription has speakers labeled as [speaker 1], 
[speaker 2] and so on. 
Your summary must maintain these speaker attributions,
clearly indicating who said what, proposed ideas, or took on action items.

Meeting transcription: {transcription}
Summary (write only the requested summary hereafter):
\end{gbox}

\noindent
\normalsize
This prompt encourages the LLM to preserve speaker attributions in the summaries, which is a highly desired feature in most real-world applications.  
Each system transcript was pre-processed using the C8DASR scoring text normalization procedure (Section~\ref{ssec:normalization}). Then, a further processing step was performed to add speaker-id tags at each utterance in addition to the recognized words, i.e.: 

\footnotesize
\begin{gbox}[red!02]
[speaker 1] ok [speaker 2] but yeah [speaker 1] yes what i was
suggesting that we [speaker 3] it is fine so if you think that we can 
move forward. 
\end{gbox}
\normalsize
\noindent
We used the convention of \texttt{[speaker 1]} being the first speaker to speak in the whole meeting, \texttt{[speaker 2]} the second, and so on. 
We did not include start and end timestamps for each utterance as it is not critical for meeting summarization. %Instead, it makes the token length significantly longer, potentially leading to performance issues due to the increased context. 
Note that these speaker-id tags might be misaligned between reference and hypothesis transcriptions due to the fact that the hypothesis diarization usually contains errors. 
To address this issue, we used tcpWER to derive optimal reference vs. hypothesis speaker-id assignments for each speaker and re-assign the hypothesis speaker-id tags based on this optimal matching with respect to the ground truth transcript. 
This step ensures that the ground truth derived summaries and the hypothesis transcripts summaries have aligned speaker-id labels and thus our summarization evaluation is well defined. 
Note that this step still accounts for missed and false alarm speakers. False alarm speakers would still be present in the hypothesis transcript and would be assigned tags incrementally as \texttt{[speaker N]} with $N > S$ where $S$ is the number of speakers in both the reference and hypothesis transcriptions. 
\textcolor{black}{We acknowledge that this speaker alignment approach has potential limitations. The optimal permutation is computed at the transcript level using tcpWER, rather than at the summary level across all possible permutations. While the latter would be theoretically preferable, it is computationally prohibitive: for a meeting with 6 speakers, this would require 720 (6!) metric evaluations per summary. The proposed approach represents a practical trade-off between theoretical optimality and computational feasibility. 
Moreover, we also experimented with using alternative labeling schemes such as sequential letters A, B, C (in randomized order) and even the pseudonyms that participants used in the NOTSOFAR-1 challenge (e.g. Walter, Melissa etc.).
However, we did not observe an appreciable difference in the results, suggesting that the LLM relies primarily on content rather than speaker label formatting for summarization.}

%The outlined procedure is general enough it can be 
%To ensure that reference and hypothesis transcriptions speaker id tags are aligned, we assign 

To measure downstream meeting summarization performance, we employ complementary metrics to minimize potential evaluation biases. These include general LLM prompt-based approaches, specialized LLM-based methods, and n-gram based techniques, which are detailed below.
\begin{itemize}
\item G-Eval~\citep{liu2023g}: a SotA automatic meeting summarization technique that leverages an LLM (OpenAI GPT-4o~\citep{openai-2024-gpt4o}) and prompt engineering in order to evaluate the summarization performance. G-Eval is summary reference free and uses only the system output transcript and the original ground truth transcription.
It measures:
\begin{itemize}
    \item \emph{Coherence}: assesses how well-structured and logically connected the summary is, evaluating whether information flows naturally and maintains clear relationships between ideas.
    \item \emph{Consistency}: measures if the summary contains only factual information entailed in the original transcript. 
    \item \emph{Relevance}: evaluates how well the summary captures the most important and salient information from the original meeting transcript.
    \item \emph{Fluency}: assesses the linguistic quality of the summary, including grammatical correctness, natural language use, and readability.
\end{itemize}
G-Eval scores are natural numbers in the range $[1, 5]$ ($[1, 3]$ for \emph{fluency}), with higher values indicating better performance. 
%All these metrics are natural numbers (the higher the better) $\in [1, 5]$, except for \emph{fluency}, whose maximum score is instead 3. 
% make here another bullet
\item UniEval~\citep{zhong2022towards}: A unified evaluation framework that fine-tunes pre-trained LM models (which are T5-based~\citep{raffel2020exploring}) and evaluates text generation across multiple dimensions. Among other tasks, it is also specifically designed for summarization and, as G-Eval, also evaluates \emph{coherence}, \emph{conciseness}, \emph{relevance}, and \emph{fluency} of the system summary. 
Unlike G-Eval, UniEval also uses the reference summary on top of the original meeting transcription to calculate these scores. UniEval scores are normalized real numbers in the range $[0, 1]$, with higher values indicating better performance.
\item ROUGE-1, ROUGE-2, and ROUGE-L~\citep{lin2004rouge}: standard lexical overlap metrics that measure unigram, bigram, and longest common subsequence matches between the generated and reference summaries, respectively. For every ROUGE metric, here we report only the F1 scores. 
\end{itemize}

\noindent
As mentioned, meeting summarization suffers from the fact that the evaluation procedure is inherently ``noisy'' due to its intrinsic subjective nature.
To mitigate this issue and ensure more reliable evaluation, we employed multiple reference summaries and multiple system hypothesis in order to be able to compute standard error bounds.
As reference summaries, for each NOTSOFAR-1 session, we generated 8 different summaries obtained by using ground-truth transcription by varying the seed at each Gemini 2.0 flash inference run. These are needed for UniEval and ROUGE metrics. 
Similarly, from each submitted system transcript, we generate 8 different summaries using same process.
This allows us to consider, for UniEval and ROUGE metrics, 64 hypothesis and reference combinations per meeting summary. Instead, since G-Eval is reference free, we only consider 8 different evaluation scores per meeting summary. This is also reasonable, as, in fact, this latter is also more expensive to run. 

\begin{landscape}
\begin{figure}
    \centering
    \includegraphics[width=\linewidth]{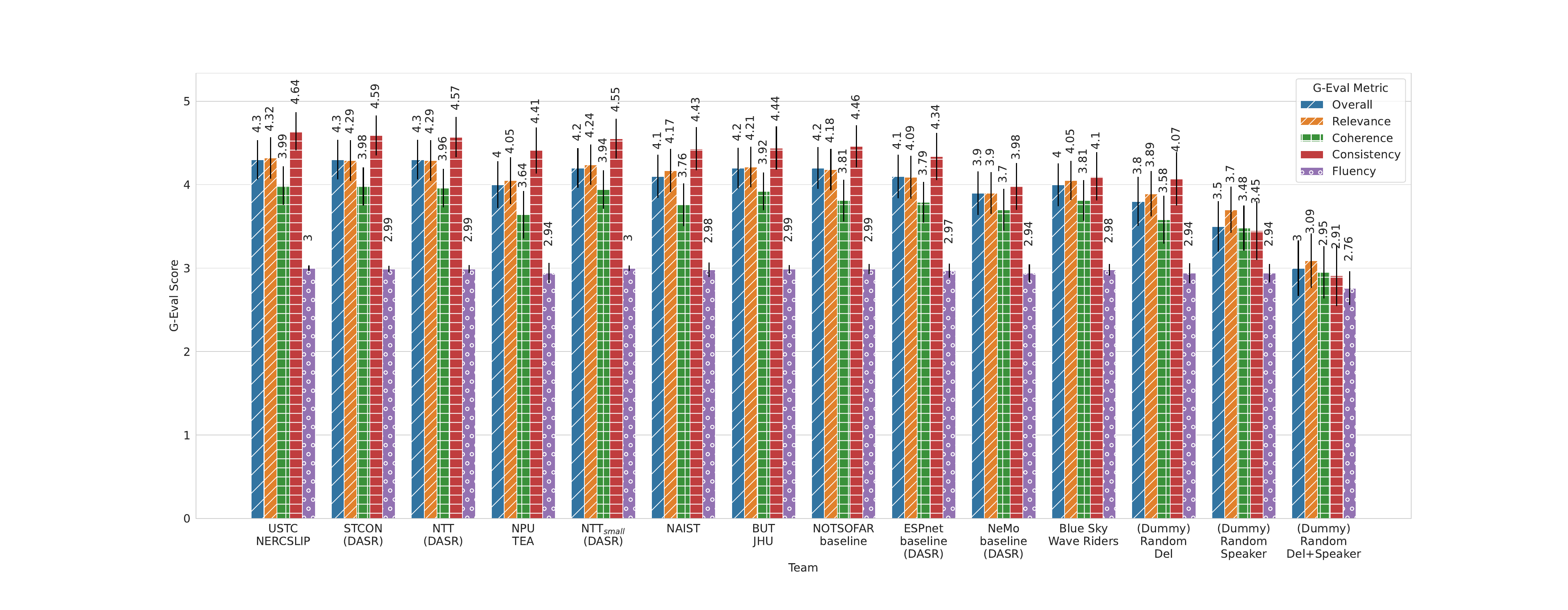}
    \caption{G-Eval scores for summarization downstream task on the NOTSOFAR-1 scenario for each team best performing system in terms of tcpWER. The \emph{overall} score here is the average of \emph{coherence}, \emph{relevance}, and \emph{consistency} metrics. %Note that \emph{fluency} is omitted as this metric showed negligible variation across all submissions. 
    Error bars indicate standard error for each metric.} %For reference, we include results for several ``dummy'' submissions with artificially introduced errors in the ground-truth transcription.}
    \label{fig:c8notsofar1_gevalplus_results}
\end{figure}
\end{landscape}

%\begin{landscape}
%\begin{figure}
%    \centering
%    \includegraphics[width=\linewidth]{images/unieval_notsofar1.pdf}
   % \input{}
%    \caption{UniEval scores for summarization downstream task on the NOTSOFAR-1 scenario for each team best performing system in terms of tcpWER. Error bars indicate standard error for each metric.}
%    \label{fig:c8notsofar1_unieval_results}
%\end{figure}
%\end{landscape}

In Figure~\ref{fig:c8notsofar1_gevalplus_results} we report the G-Eval metrics for each NOTSOFAR-1 submitted system, including C8DASR and those submitted in the CHiME-8 NOTSOFAR-1 challenge. We only included the best system in terms of tcpWER for each team and order the teams in the plot based on their tcpWER ranking to be consistent with previous Figure~\ref{fig:c8_notsofar1_tcpwer}. 
For reference, we include results for several ``dummy'' submissions with artificially introduced errors in the ground-truth transcription: 
\begin{itemize}
    \item \emph{Random Del}: each word in the ground-truth transcription is randomly removed with 50\% probability. The resulting tcpWER is $49.83\%$.
    \item \emph{Random Speaker}: speaker IDs in the ground-truth transcription are randomly reassigned for each utterance. The resulting tcpWER is $114\%$.
    \item \emph{Random Del+Speaker}: a combination of both error types: speaker IDs are randomly reassigned for each utterance, and each word is randomly removed with 50\% probability. The resulting tcpWER is $101\%$. This value is lower than the Random Speaker condition because fewer words are incorrectly attributed to different speakers, as many are now counted only as deletions. In tcpWER, speaker-word misattributions count as two errors: once as a deletion and once as an insertion.
\end{itemize}

%We did not include \emph{fluency} because we found that almost all systems had always the maximum score of $3$ for this metric. 
%This can stem  

\begin{figure}
    \centering
    \includegraphics[width=\linewidth]{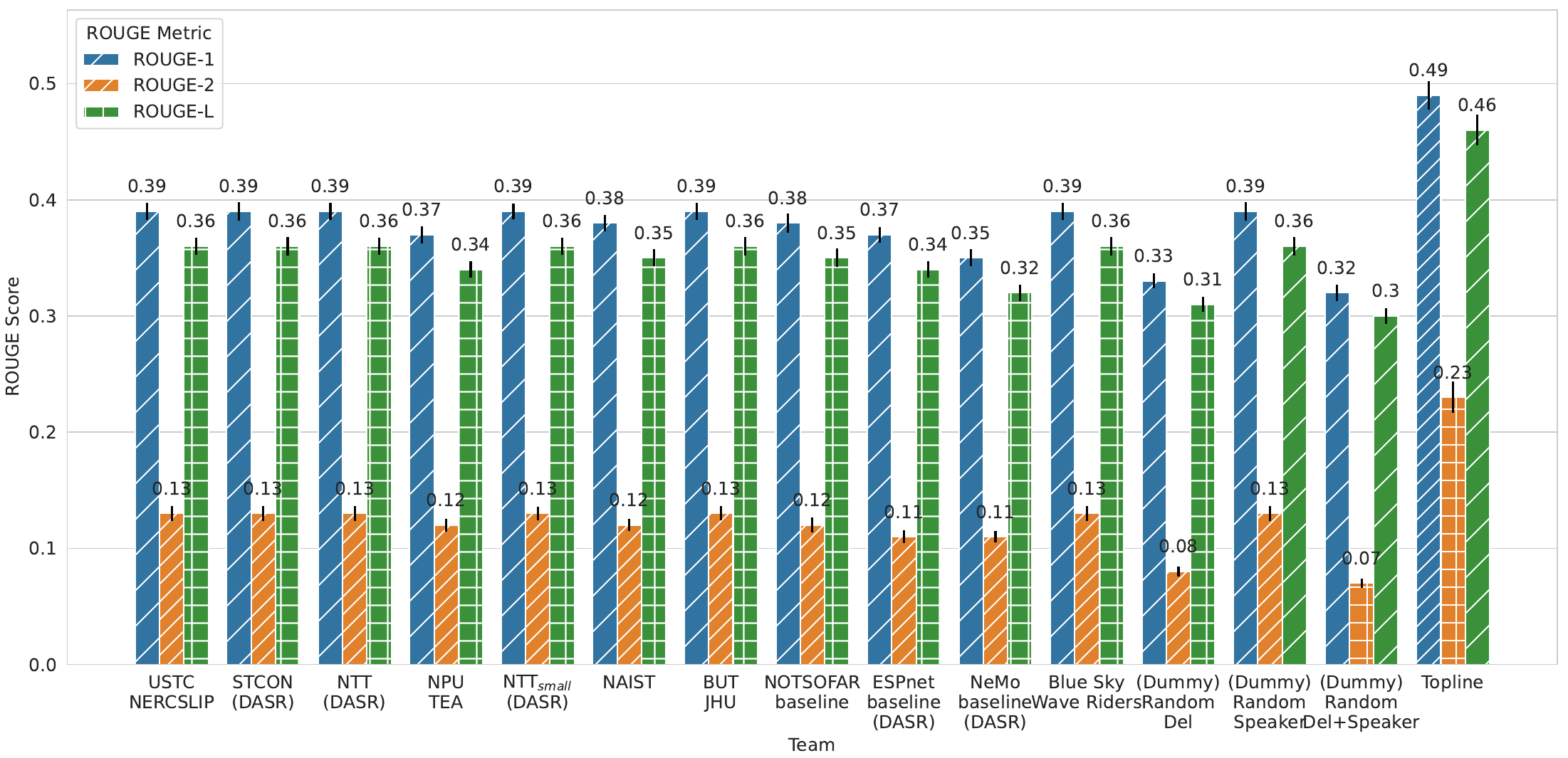}
    \caption{ROUGE-1, ROUGE-2 and ROUGE-L F1-scores for summarization downstream task on the NOTSOFAR-1 scenario for each team best performing system in terms of tcpWER. Error bars indicate standard error for each metric. 
    \textcolor{black}{The ``topline'' system on the right ("Reference vs Reference") shows ROUGE scores between reference summaries generated from ground-truth transcriptions with different seeds serves as an approximate upper bound for evaluation.}}
    \label{fig:c8notsofar1_rouge_results}
\end{figure}

\begin{figure}
    \centering
    \includegraphics[width=\linewidth]{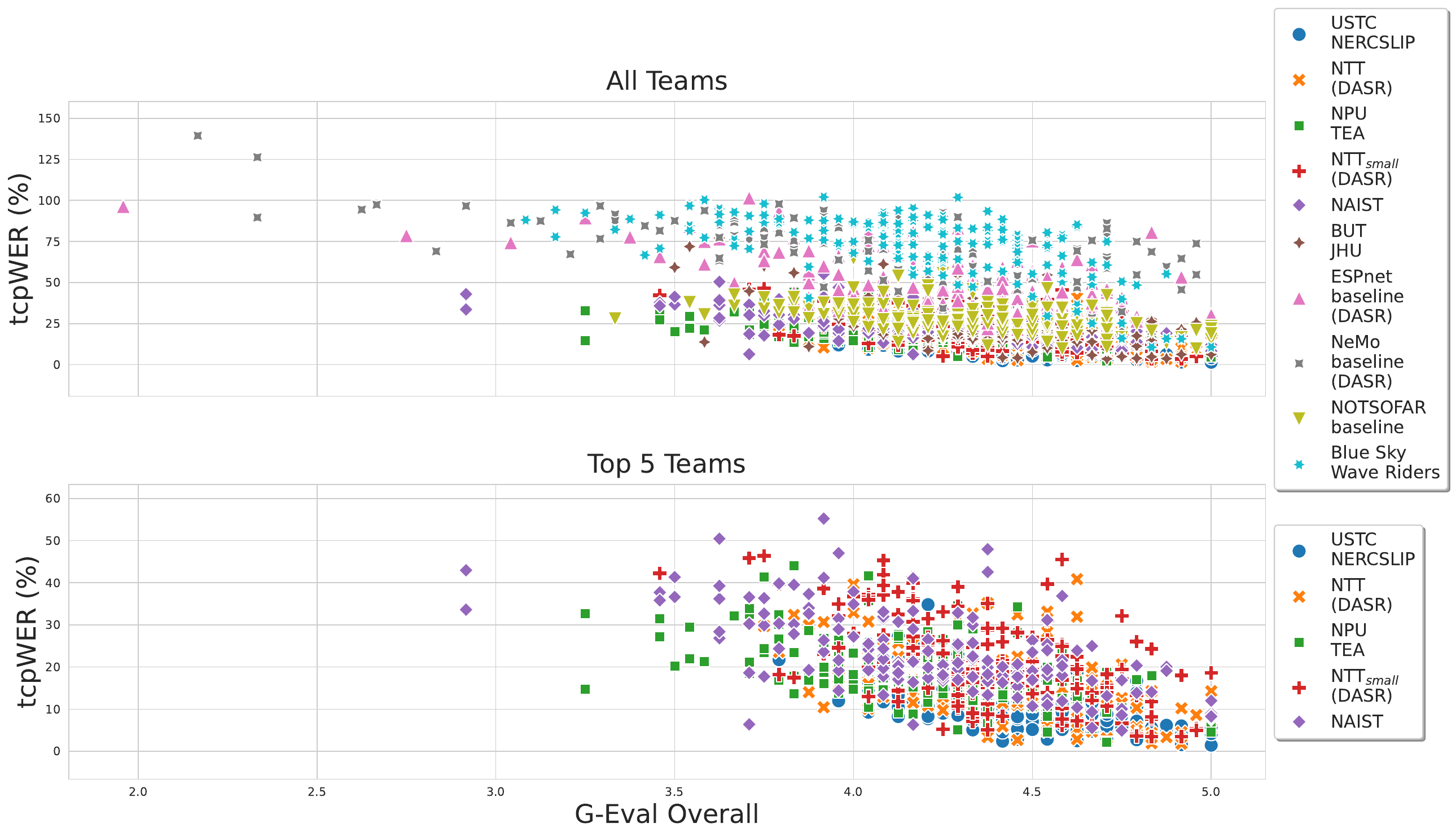}
    \caption{\textcolor{black}{Scatter plot of tcpWER (\%) vs. G-Eval Overall scores for all submitted systems on the NOTSOFAR-1 scenario. Top: all systems; Bottom: top-5 systems by tcpWER. Each point represents one NOTSOFAR-1 evaluation session.}}
    \label{fig:c8notsofar1_gevalvstcpwer}
\end{figure}

We can observe that, despite the evident tcpWER difference observed in Figure~\ref{fig:c8_notsofar1_tcpwer}, for which, the best system (USTC-NERCSLIP) had a tcpWER $\sim 11\%$ while the worst $\sim 70\%$, the G-Eval metrics show reduced variation between submissions. That is, even systems for which roughly only one out of two words is correctly recognized and speaker-attributed (e.g. ESPnet baseline) appear to generate summaries which are roughly on-par with systems for which $\sim$80\% of the words are correctly recognized and attributed (e.g. NPU TEA). 
In fact, most differences between systems in Figure~\ref{fig:c8_notsofar1_tcpwer} can be considered to be not significant enough since they fall within one standard error range. 
Only when the amount of errors are extremely severe as in the Random Del and Random Del+Speaker systems an appreciable degradation in summarization metrics can be observed.  
Among the G-Eval metrics, \emph{fluency} is the one that is most consistent across all submissions, including the ``dummy'' systems and virtually shows no variation. 
\textcolor{black}{This same trend is also observed when computing ROUGE scores which are reported in Figure~\ref{fig:c8notsofar1_rouge_results}. %Except for systems for which the error rate is significant, 
Crucially G-Eval consistently assigns a high \emph{fluency} score to all systems even for the Random Del and Random Del+Speaker ``dummy'' systems.}
This is due to the fact that all our summaries were generated using the same Gemini Flash 2.0 model, and modern SotA LLMs have become increasingly good at generating linguistically well-formed output, thus saturating these fluency summarization metrics.
It is also possible that there is a potential systemic bias where LLM evaluators tend to rate LLM-generated text highly on fluency metrics~\citep{liu2023g} generated via Gemini Flash 2.0. 
However, upon inspection of the generated summaries, we found no noticeable issues and all inspected summaries appeared to be well written.
%However, upon inspection we found no significant issues 
%However, upon inspection we found the resulting summaries to be 
%Instead, it is likely due to the fact that our summaries are LLM-generated, and modern SotA LLMs have become increasingly good at generating linguistically well-formed outputs, thus saturating these evaluation metrics for summarization. 
Table~\ref{tab:pcc_src_summ} reports PCC and SRC of tcpWER vs. G-Eval, ROUGE and also UniEval scores.
These correlation coefficients are computed considering each submitted system and each session independently for a total of 1760 samples. For this study we did not consider including the ``dummy'' systems with artificially generated errors. 
%For both PCC and SRC we also report the p-value indicating the probability of an uncorrelated and normally distributed dataset producing the same observed correlation coefficient value~\citep{kowalski1972effects}. 
These figures indicate that only G-Eval has a moderate negative correlation with tcpWER with \emph{consistency} being the most correlated. This can be expected since it quantifies if the summary contains only factually correct information that was contained in the original transcript, thus penalizing hallucinated content and speaker mis-attributions due to transcription errors.
\textcolor{black}{
For ROUGE and even more so for UniEval metrics, the correlation is much weaker. 
To give a reader a sense of the weakness of the correlation between tcpWER and G-Eval, in Figure~\ref{fig:c8notsofar1_gevalvstcpwer} we report also a scatter plot for all submitted systems (top: all systems, bottom: top 5 systems) of tcpWER vs. G-Eval \textit{overall} metric. Each data-point corresponds to one NOTSOFAR-1 session in the evaluation set.}

\begin{table}[htbp]
\centering
\scriptsize
\caption{Pearson correlation coefficients (PCC) and Spearmank rank correlation (SRC) between tcpWER and G-Eval, UniEval and ROUGE F1 scores metrics.}
\label{tab:pcc_src_summ}
\setlength{\tabcolsep}{0.5em}  % Increased spacing between columns
\begin{tabular}{lcc}  % Simplified column alignment
\toprule   
 \textbf{Summarization Metric} & \multicolumn{1}{c}{\textbf{PCC}} & \multicolumn{1}{c}{\textbf{SRC}}   \\
\midrule  % Added midrule for better visual separation
\rowcolor{ai2offwhite}  G-Eval Overall & -0.51 & -0.5  \\ 
 G-Eval Relevance & -0.46 &  -0.45  \\ 
\rowcolor{ai2offwhite}  G-Eval Consistency & -0.54 &  -0.55   \\ 
  G-Eval Coherence & -0.27 & -0.28   \\ 
\rowcolor{ai2offwhite}  G-Eval Fluency & -0.22 & -0.13  \\ 
\midrule
UniEval Overall  & -0.15 &  -0.18  \\ 
\rowcolor{ai2offwhite} UniEval Relevance  & -0.11 &  -0.16   \\ 
UniEval Consistency  & -0.07  &  -0.10   \\ 
\rowcolor{ai2offwhite} UniEval Coherence  &  -0.12 & -0.20   \\ 
UniEval Fluency  &  0.01 & -0.01   \\ 
 \midrule
\rowcolor{ai2offwhite} ROUGE-1 F1 score  & -0.34 & -0.36  \\ 
 ROUGE-2 F1 score  & -0.28 & -0.32  \\ 
\rowcolor{ai2offwhite} ROUGE-L F1 score  & -0.33 & -0.35 \\ 

\bottomrule
\end{tabular}
\end{table}

%\begin{table}[htbp]
%\centering
%\scriptsize
%\caption{Pearson correlation coefficients (PCC) and Spearmank rank correlation (SRC) between tcpWER and G-Eval, UniEval and ROUGE F1 scores metrics.}
%\label{tab:pcc_src_summ}
%\setlength{\tabcolsep}{0.5em}  % Increased spacing between columns
%\begin{tabular}{lcc}  % Simplified column alignment
%\toprule   
% \textbf{Summarization Metric} & \multicolumn{1}{c}{\textbf{PCC}} & \multicolumn{1}{c}{\textbf{SRC}}   \\
%\midrule  % Added midrule for better visual separation
%\rowcolor{ai2offwhite} G-Eval Overall & -0.48 & -0465  \\ 
% G-Eval Relevance & -0.46 &  -0.45  \\ 
%\rowcolor{ai2offwhite} G-Eval Consistency & -0.57 &  -0.55   \\ 
%  G-Eval Coherence & -0.29 & -0.32   \\ 
%\rowcolor{ai2offwhite} G-Eval Fluency & -0.24 & -0.18  \\
%\bottomrule
%\end{tabular}
%\end{table}

These results underscore the challenges of evaluating transcription systems through meeting summarization, as the use of LLMs introduces an additional black-box component that can significantly bias and influence the results.
When an application requires reliable or even acceptable transcription quality, meeting summarization may not be a good proxy evaluation task due to its high robustness to transcription errors. 
Among the metrics we have analyzed, only G-Eval showed moderate correlation with transcription errors. Thus, despite promising advantages in handling text normalization and common linguistic artifacts in spontaneous speech, this same robustness makes summarization currently sub-optimal as a proxy evaluation task for transcription quality. At least until better summarization metrics or ad-hoc evaluation frameworks are developed for dialogue-oriented summarization, though G-Eval already shows improved correlation compared to legacy metrics like UniEval and ROUGE.
%Further research in this direction remains essential, although G-Eval already shows improved correlation compared to more established metrics like UniEval and ROUGE.
For example, in our preliminary experiments, we found it crucial to include explicit speaker attribution instructions in the LLM summarization prompt; without these, the produced summaries were too generic, and no appreciable correlation was observed across all summarization metrics, including G-Eval. We hope that this observation and, more generally, the summarization evaluation framework adopted in this work can be a valuable contribution towards improving dialogue summarization evaluation, but further research in this direction remains essential.

From a more optimistic perspective, these findings can also be interpreted in another way. The near-consistent performance across all systems and metrics demonstrates that current LLM-based summarization approaches are remarkably powerful in recovering from transcription errors and generating accurate summaries.
This suggests that some degree of decoupling between transcription and effective summarization exists, challenging traditional assumptions about the dependency chain in speech processing pipelines.
As such, if the application at hand does not strictly require accurate transcription, these findings could also motivate research toward fully end-to-end spoken meeting summarization. 
\textcolor{black}{An ad-hoc end-to-end approach could perhaps} even offer more contextually aware summarization capabilities as the raw audio is much richer in content and may offer additional cues that are unavailable in transcripts. An example is speech prosody indicating surprise, emphasis, or hesitation.  %even boffering more efficient and 
This promising research direction has been enabled by recent advances in multimodal audio and text LLM models (SpeechLMs)~\citep{slm_survey1, slm_survey2} and is already supported by models such as Gemini, GPT-4o, and Llama 3~\citep{fang2024llama}. However, practical challenges remain, particularly regarding reliable long-context modeling to effectively capture extended meeting interactions and keyfacts. These problems are amplified for SpeechLMs as the sequence length increases dramatically. As a consequence, computational demands (and thus cost) also increase substantially, presenting a critical barrier to widespread adoption of these technologies in many practical settings.

\section{Conclusions}
\label{sec:conclusion}
%%%%%%%%%%%%%%%%%%%%%%

\subsection{Summary of challenges outcomes and findings}

In this work, we described the motivations, design, and outcomes of the recent CHiME-7 and CHiME-8 distant automatic speech recognition (DASR) challenges. The core objective was to promote the development of joint ASR and diarization transcription systems capable of generalizing across diverse scenarios by testing their performance on multiple datasets: CHiME-6, DiPCo, Mixer 6, and NOTSOFAR-1 (with NOTSOFAR-1 added in the CHiME-8 challenge). These datasets encompass significant variations in acoustic environments, recording setups, meeting durations, participant numbers, and linguistic and paralinguistic speaking styles.
A total of 9 teams participated in the two challenges, submitting 33 different transcription systems. Our analysis of the techniques employed by participants spans both front-end components (including diarization and speech separation) and back-end ASR components, revealing several key trends.

First, the transition towards e2e ASR systems, facilitated by the availability of large-scale pretrained models, represents a significant evolution from the previous CHiME-6 challenge.
Second, guided source separation (GSS) remains a crucial component in most high-performing systems, as evidenced by submissions across both the CHiME-7/8 DASR and CHiME-8 NOTSOFAR-1 multi-channel track challenges. Few participants attempted to integrate neural-based speech separation methods, as challenges persist in reliably handling the complexities of real-world conversational environments.

Third, cross-comparison of results between CHiME-8 DASR and CHiME-8 NOTSOFAR-1 challenges revealed that generalizable array-agnostic systems can achieve comparable or competitive results versus ad-hoc systems developed for specific narrow scenarios or known microphone array configurations. This is a significant finding, as more flexible systems are highly desirable for real-world applications.

Fourth, with few exceptions, most high-performing systems heavily relied on diarization refinement via target-speaker voice activity detection (TS-VAD) techniques. These approaches, together with the availability of large-scale pre-trained models, were the primary drivers of performance improvement. The improved TS-VAD techniques such as the NSD-MA-MSE model from USTC-NERCSLIP can be considered a significant outcome of the CHiME-7 DASR challenge.
Finally, accurate speaker counting remains crucial, as errors propagate catastrophically through the rest of the pipeline. In this regard, novel methods were developed during the CHiME-8 DASR challenge by NTT and STCON teams, featuring improved clustering techniques and overlapped speech handling in the diarization component.
\textcolor{black}{However, developing systems more robust to speaker counting errors remains an important and practical research direction. As discussed in Section~\ref{ssec:joint_c78dasr_results}, such errors have catastrophic downstream effects in typical pipelined approaches. Few works have addressed this issue, for example by making TS-VAD robust to such speaker counting errors~\citep{wang2024profile}.}

As an additional contribution, we explored the suitability of meeting summarization as a downstream evaluation task for meeting transcription using NOTSOFAR-1 data. We outlined an evaluation framework where we used an LLM to generate summaries from system transcriptions while ensuring and measuring correct speaker attribution. 
Our results indicate that recent summarization metrics, such as G-Eval, show moderate correlation with transcription accuracy. However, we also found that contemporary LLMs demonstrate remarkable ability in handling transcription errors and inferring missing information. This resilience suggests that summarization is not a suitable proxy evaluation task for applications where precise transcription accuracy is crucial and, instead, justify the research direction of direct e2e meeting summarization, when verbatim transcription is not needed.

\subsection{Limitations and future research directions}

Despite the contributions of the CHiME-7 and CHiME-8 DASR challenges, there are several limitations that warrant further discussion. These limitations provide valuable insights for future research and the design of more effective evaluation campaigns.

First, the datasets used in the challenges, while diverse, may not fully capture the complexities of all real-world conversational environments. 
The scenarios considered are still limited in scope, and moreover, they only consider the English language. 
More effort should be done in the future to also consider the possibility of multilingual meeting transcription that can also be robust to code switching so that this technology can benefit a larger amount of people.

Second, most importantly, to truly gauge system generalization, future benchmarks should consider having at least some fully blind evaluation scenarios for which the domain is not known a-priori by the participants. In the C7-8DASR challenges, all the evaluation scenarios were already known to the participants and this could have biased the development of submitted systems. On the other hand, in real-world applications, some deployment conditions/domains \textcolor{black}{are not known a-priori}. 
Future challenges should include evaluation scenarios that are completely hidden from participants until submission time to better simulate real-world deployment conditions.

\textcolor{black}{
Third, while our downstream evaluation via meeting summarization provided some valuable insights, several limitations remain. Our speaker alignment approach (using tcpWER-based optimal permutation before summary generation) represents a practical trade-off given computational constraints. Moreover, developing more robust summarization evaluation metrics that better capture the nuances of multi-speaker dialogue remains an open challenge. However, this will require specialized data collection efforts with human-annotated reference summaries that explicitly ground key facts, decisions, and action items.}
%Third, 
Finally, while the participants' submissions show promise, as e.g. best submissions on CHiME-8 DASR challenge were even able to surpass Whisper \texttt{large-v3} applied to oracle diarization and GSS, most rely on rather impractical ensembling techniques. While some mechanisms were introduced in the CHiME-8 DASR challenge to spur research towards more efficient and practical approaches, the resulting systems are still far from being practical. For example, among all submissions in both C7-8DASR challenges, NTT$_{small}$ is the one that achieved the best compromise between performance and efficiency, by avoiding ensembling and diarization refinement techniques. However, as reported by the authors, the RTF was more than $2$, even if this could be brought down by ad-hoc optimizations. Thus, future evaluation campaigns should also try to devise ways to better encourage the practicality of developed techniques.

\section{Acknowledgements}
%%%%%%%%%%%%%%%%%%%%%%%%%%
We would like to thank the CH\-iME Steering Committee members Marc Delcroix, Michael Mandel, and Jon Barker for their invaluable help, support, and guidance in organizing these challenges. 
S. Cornell was supported by the IC Postdoctoral
Research Fellowship Program via Oak Ridge Institute for Science and Education through an agreement between the U.S. Department of Energy and the Office of the Director of National Intelligence.

% \appendix
% \section{Example Appendix Section}


%%       \citet{<label>} ==> Lamport (1994)
%%       \citep{<label>} ==> (Lamport, 1994)
%%
% Example citation, See \citet{lamport94}.

\bibliographystyle{elsarticle-harv} 
\bibliography{references}

\begin{thebibliography}{180}
\expandafter\ifx\csname natexlab\endcsname\relax\def\natexlab#1{#1}\fi
\providecommand{\url}[1]{\texttt{#1}}
\providecommand{\href}[2]{#2}
\providecommand{\path}[1]{#1}
\providecommand{\DOIprefix}{doi:}
\providecommand{\ArXivprefix}{arXiv:}
\providecommand{\URLprefix}{URL: }
\providecommand{\Pubmedprefix}{pmid:}
\providecommand{\doi}[1]{\href{http://dx.doi.org/#1}{\path{#1}}}
\providecommand{\Pubmed}[1]{\href{pmid:#1}{\path{#1}}}
\providecommand{\bibinfo}[2]{#2}
\ifx\xfnm\relax \def\xfnm[#1]{\unskip,\space#1}\fi
%Type = Inproceedings
\bibitem[{Akiba et~al.(2019)Akiba, Sano, Yanase, Ohta and Koyama}]{akiba2019optuna}
\bibinfo{author}{Akiba, T.}, \bibinfo{author}{Sano, S.}, \bibinfo{author}{Yanase, T.}, \bibinfo{author}{Ohta, T.}, \bibinfo{author}{Koyama, M.}, \bibinfo{year}{2019}.
\newblock \bibinfo{title}{Optuna: A next-generation hyperparameter optimization framework}, in: \bibinfo{booktitle}{Proceedings of the 25th ACM SIGKDD international conference on knowledge discovery \& data mining}, pp. \bibinfo{pages}{2623--2631}.
%Type = Inproceedings
\bibitem[{Ardila et~al.(2020)Ardila, Branson, Davis, Kohler, Meyer, Henretty, Morais, Saunders, Tyers and Weber}]{ardila2020common}
\bibinfo{author}{Ardila, R.}, \bibinfo{author}{Branson, M.}, \bibinfo{author}{Davis, K.}, \bibinfo{author}{Kohler, M.}, \bibinfo{author}{Meyer, J.}, \bibinfo{author}{Henretty, M.}, \bibinfo{author}{Morais, R.}, \bibinfo{author}{Saunders, L.}, \bibinfo{author}{Tyers, F.}, \bibinfo{author}{Weber, G.}, \bibinfo{year}{2020}.
\newblock \bibinfo{title}{Common voice: A massively-multilingual speech corpus}, in: \bibinfo{booktitle}{Proceedings of the Twelfth Language Resources and Evaluation Conference}, pp. \bibinfo{pages}{4218--4222}.
%Type = Article
\bibitem[{Arora et~al.(2020)Arora, Raj, Subramanian, Li, Ben-Yair, Maciejewski, {\.Z}elasko, Garcia, Watanabe and Khudanpur}]{arora2020jhu}
\bibinfo{author}{Arora, A.}, \bibinfo{author}{Raj, D.}, \bibinfo{author}{Subramanian, A.S.}, \bibinfo{author}{Li, K.}, \bibinfo{author}{Ben-Yair, B.}, \bibinfo{author}{Maciejewski, M.}, \bibinfo{author}{{\.Z}elasko, P.}, \bibinfo{author}{Garcia, P.}, \bibinfo{author}{Watanabe, S.}, \bibinfo{author}{Khudanpur, S.}, \bibinfo{year}{2020}.
\newblock \bibinfo{title}{The {JHU} multi-microphone multi-speaker asr system for the chime-6 challenge}.
\newblock \bibinfo{journal}{CHiME Workshop} .
%Type = Inproceedings
\bibitem[{Baevski et~al.(2020)Baevski, Zhou, Mohamed and Auli}]{baevski2020wav2vec}
\bibinfo{author}{Baevski, A.}, \bibinfo{author}{Zhou, Y.}, \bibinfo{author}{Mohamed, A.}, \bibinfo{author}{Auli, M.}, \bibinfo{year}{2020}.
\newblock \bibinfo{title}{{Wav2Vec} 2.0: A framework for self-supervised learning of speech representations}, in: \bibinfo{booktitle}{NeurIPS}.
%Type = Article
\bibitem[{Bando et~al.(2024)Bando, Nakamura and Watanabe}]{bando2024neural}
\bibinfo{author}{Bando, Y.}, \bibinfo{author}{Nakamura, T.}, \bibinfo{author}{Watanabe, S.}, \bibinfo{year}{2024}.
\newblock \bibinfo{title}{Neural blind source separation and diarization for distant speech recognition}.
\newblock \bibinfo{journal}{Interspeech} .
%Type = Inproceedings
\bibitem[{Barker et~al.(2015)Barker, Marxer, Vincent and Watanabe}]{barker2015third}
\bibinfo{author}{Barker, J.}, \bibinfo{author}{Marxer, R.}, \bibinfo{author}{Vincent, E.}, \bibinfo{author}{Watanabe, S.}, \bibinfo{year}{2015}.
\newblock \bibinfo{title}{The third {CH\-iME} speech separation and recognition challenge: Dataset, task and baselines}, in: \bibinfo{booktitle}{IEEE ASRU}.
%Type = Article
\bibitem[{Barker et~al.(2017)Barker, Marxer, Vincent and Watanabe}]{barker2017third}
\bibinfo{author}{Barker, J.}, \bibinfo{author}{Marxer, R.}, \bibinfo{author}{Vincent, E.}, \bibinfo{author}{Watanabe, S.}, \bibinfo{year}{2017}.
\newblock \bibinfo{title}{The third ‘chime’speech separation and recognition challenge: Analysis and outcomes}.
\newblock \bibinfo{journal}{Computer Speech \& Language} \bibinfo{volume}{46}, \bibinfo{pages}{605--626}.
%Type = Article
\bibitem[{Barker et~al.(2013)Barker, Vincent, Ma, Christensen and Green}]{barker2013pascal}
\bibinfo{author}{Barker, J.}, \bibinfo{author}{Vincent, E.}, \bibinfo{author}{Ma, N.}, \bibinfo{author}{Christensen, H.}, \bibinfo{author}{Green, P.}, \bibinfo{year}{2013}.
\newblock \bibinfo{title}{The {PASCAL} {CHiME} speech separation and recognition challenge}.
\newblock \bibinfo{journal}{Computer Speech \& Language} \bibinfo{volume}{27}.
%Type = Inproceedings
\bibitem[{Barker et~al.(2018)Barker, Watanabe, Vincent and Trmal}]{barker2018fifth}
\bibinfo{author}{Barker, J.}, \bibinfo{author}{Watanabe, S.}, \bibinfo{author}{Vincent, E.}, \bibinfo{author}{Trmal, J.}, \bibinfo{year}{2018}.
\newblock \bibinfo{title}{The fifth {CHiME} speech separation and recognition challenge: Dataset, task and baselines}, in: \bibinfo{booktitle}{Interspeech}.
%Type = Inproceedings
\bibitem[{Boeddeker et~al.(2018)Boeddeker, Heitkaemper, Schmalenstroeer, Drude, Heymann and Haeb-Umbach}]{boeddeker2018front}
\bibinfo{author}{Boeddeker, C.}, \bibinfo{author}{Heitkaemper, J.}, \bibinfo{author}{Schmalenstroeer, J.}, \bibinfo{author}{Drude, L.}, \bibinfo{author}{Heymann, J.}, \bibinfo{author}{Haeb-Umbach, R.}, \bibinfo{year}{2018}.
\newblock \bibinfo{title}{Front-end processing for the {CHiME-5} dinner party scenario}, in: \bibinfo{booktitle}{CHiME5 Workshop}.
%Type = Article
\bibitem[{Boeddeker et~al.(2024)Boeddeker, Subramanian, Wichern, Haeb-Umbach and Le~Roux}]{boeddeker2023ts}
\bibinfo{author}{Boeddeker, C.}, \bibinfo{author}{Subramanian, A.S.}, \bibinfo{author}{Wichern, G.}, \bibinfo{author}{Haeb-Umbach, R.}, \bibinfo{author}{Le~Roux, J.}, \bibinfo{year}{2024}.
\newblock \bibinfo{title}{{TS-SEP}: Joint diarization and separation conditioned on estimated speaker embeddings}.
\newblock \bibinfo{journal}{IEEE/ACM Transactions on Audio, Speech, and Language Processing} \bibinfo{volume}{32}, \bibinfo{pages}{1185--1197}.
%Type = Inproceedings
\bibitem[{Boeddeker et~al.(2023)Boeddeker, Cord-Landwehr, Neumann and Haeb-Umbach}]{boeddeker2023multi}
\bibinfo{author}{Boeddeker, C.B.}, \bibinfo{author}{Cord-Landwehr, T.}, \bibinfo{author}{Neumann, T.v.}, \bibinfo{author}{Haeb-Umbach, R.}, \bibinfo{year}{2023}.
\newblock \bibinfo{title}{Multi-stage diarization refinement for the chime-7 dasr scenario}, in: \bibinfo{booktitle}{CHiME Workshop}, pp. \bibinfo{pages}{51--56}.
%Type = Inproceedings
\bibitem[{Brandschain et~al.(2010)Brandschain, Graff, Cieri, Walker, Caruso and Neely}]{brandschain2010mixer}
\bibinfo{author}{Brandschain, L.}, \bibinfo{author}{Graff, D.}, \bibinfo{author}{Cieri, C.}, \bibinfo{author}{Walker, K.}, \bibinfo{author}{Caruso, C.}, \bibinfo{author}{Neely, A.}, \bibinfo{year}{2010}.
\newblock \bibinfo{title}{{The Mixer 6 corpus}: Resources for cross-channel and text independent speaker recognition}, in: \bibinfo{booktitle}{LREC}.
%Type = Inproceedings
\bibitem[{Bredin(2023)}]{bredin2023pyannote}
\bibinfo{author}{Bredin, H.}, \bibinfo{year}{2023}.
\newblock \bibinfo{title}{{Pyannote}. audio 2.1 speaker diarization pipeline: principle, benchmark, and recipe}, in: \bibinfo{booktitle}{Interspeech}, \bibinfo{organization}{ISCA}. pp. \bibinfo{pages}{1983--1987}.
%Type = Inproceedings
\bibitem[{Bredin and Laurent(2021)}]{bredin2021end}
\bibinfo{author}{Bredin, H.}, \bibinfo{author}{Laurent, A.}, \bibinfo{year}{2021}.
\newblock \bibinfo{title}{End-to-end speaker segmentation for overlap-aware resegmentation}, in: \bibinfo{booktitle}{Interspeech}.
%Type = Inproceedings
\bibitem[{Bredin et~al.(2020)Bredin, Yin, Coria, Gelly, Korshunov, Lavechin, Fustes, Titeux, Bouaziz and Gill}]{bredin2020pyannote}
\bibinfo{author}{Bredin, H.}, \bibinfo{author}{Yin, R.}, \bibinfo{author}{Coria, J.M.}, \bibinfo{author}{Gelly, G.}, \bibinfo{author}{Korshunov, P.}, \bibinfo{author}{Lavechin, M.}, \bibinfo{author}{Fustes, D.}, \bibinfo{author}{Titeux, H.}, \bibinfo{author}{Bouaziz, W.}, \bibinfo{author}{Gill, M.P.}, \bibinfo{year}{2020}.
\newblock \bibinfo{title}{Pyannote. audio: neural building blocks for speaker diarization}, in: \bibinfo{booktitle}{Proc. of ICASSP}.
%Type = Article
\bibitem[{Capon(1969)}]{capon_mvdr}
\bibinfo{author}{Capon, J.}, \bibinfo{year}{1969}.
\newblock \bibinfo{title}{High-resolution frequency-wavenumber spectrum analysis}.
\newblock \bibinfo{journal}{IEEE} \bibinfo{volume}{57}.
%Type = Inproceedings
\bibitem[{Carletta et~al.(2005)Carletta, Ashby, Bourban, Flynn, Guillemot, Hain, Kadlec, Karaiskos, Kraaij, Kronenthal et~al.}]{carletta2005ami}
\bibinfo{author}{Carletta, J.}, \bibinfo{author}{Ashby, S.}, \bibinfo{author}{Bourban, S.}, \bibinfo{author}{Flynn, M.}, \bibinfo{author}{Guillemot, M.}, \bibinfo{author}{Hain, T.}, \bibinfo{author}{Kadlec, J.}, \bibinfo{author}{Karaiskos, V.}, \bibinfo{author}{Kraaij, W.}, \bibinfo{author}{Kronenthal, M.}, et~al., \bibinfo{year}{2005}.
\newblock \bibinfo{title}{The {AMI} meeting corpus: A pre-announcement}, in: \bibinfo{booktitle}{International workshop on machine learning for multimodal interaction}.
%Type = Inproceedings
\bibitem[{Chang et~al.(2022)Chang, Maekaku, Fujita and Watanabe}]{chang2022end}
\bibinfo{author}{Chang, X.}, \bibinfo{author}{Maekaku, T.}, \bibinfo{author}{Fujita, Y.}, \bibinfo{author}{Watanabe, S.}, \bibinfo{year}{2022}.
\newblock \bibinfo{title}{End-to-end integration of speech recognition, speech enhancement, and self-supervised learning representation}, in: \bibinfo{booktitle}{Interspeech}.
%Type = Inproceedings
\bibitem[{Chen et~al.(2021)Chen, Chai, Wang, Du, Zhang, Weng, Su, Povey, Trmal, Zhang et~al.}]{chen2021gigaspeech}
\bibinfo{author}{Chen, G.}, \bibinfo{author}{Chai, S.}, \bibinfo{author}{Wang, G.}, \bibinfo{author}{Du, J.}, \bibinfo{author}{Zhang, W.Q.}, \bibinfo{author}{Weng, C.}, \bibinfo{author}{Su, D.}, \bibinfo{author}{Povey, D.}, \bibinfo{author}{Trmal, J.}, \bibinfo{author}{Zhang, J.}, et~al., \bibinfo{year}{2021}.
\newblock \bibinfo{title}{Gigaspeech: An evolving, multi-domain asr corpus with 10,000 hours of transcribed audio}, in: \bibinfo{booktitle}{Interspeech}.
%Type = Inproceedings
\bibitem[{Chen et~al.(2020a)Chen, Zhang, Shi and Liu}]{chen2020improved}
\bibinfo{author}{Chen, H.}, \bibinfo{author}{Zhang, P.}, \bibinfo{author}{Shi, Q.}, \bibinfo{author}{Liu, Z.}, \bibinfo{year}{2020}a.
\newblock \bibinfo{title}{Improved guided source separation integrated with a strong back-end for the chime-6 dinner party scenario.}, in: \bibinfo{booktitle}{Interspeech}, pp. \bibinfo{pages}{334--338}.
%Type = Article
\bibitem[{Chen et~al.(2022)Chen, Wang, Chen, Wu, Liu, Chen, Li, Kanda, Yoshioka, Xiao et~al.}]{chen2022wavlm}
\bibinfo{author}{Chen, S.}, \bibinfo{author}{Wang, C.}, \bibinfo{author}{Chen, Z.}, \bibinfo{author}{Wu, Y.}, \bibinfo{author}{Liu, S.}, \bibinfo{author}{Chen, Z.}, \bibinfo{author}{Li, J.}, \bibinfo{author}{Kanda, N.}, \bibinfo{author}{Yoshioka, T.}, \bibinfo{author}{Xiao, X.}, et~al., \bibinfo{year}{2022}.
\newblock \bibinfo{title}{{WavLM}: Large-scale self-supervised pre-training for full stack speech processing}.
\newblock \bibinfo{journal}{IEEE Journal of Selected Topics in Signal Processing} \bibinfo{volume}{16}.
%Type = Inproceedings
\bibitem[{Chen et~al.(2020b)Chen, Yoshioka, Lu, Zhou, Meng, Luo, Wu, Xiao and Li}]{chen2020continuous}
\bibinfo{author}{Chen, Z.}, \bibinfo{author}{Yoshioka, T.}, \bibinfo{author}{Lu, L.}, \bibinfo{author}{Zhou, T.}, \bibinfo{author}{Meng, Z.}, \bibinfo{author}{Luo, Y.}, \bibinfo{author}{Wu, J.}, \bibinfo{author}{Xiao, X.}, \bibinfo{author}{Li, J.}, \bibinfo{year}{2020}b.
\newblock \bibinfo{title}{Continuous speech separation: Dataset and analysis}, in: \bibinfo{booktitle}{Proc. of ICASSP}.
%Type = Inproceedings
\bibitem[{Chorowski et~al.(2015)Chorowski, Bahdanau, Serdyuk, Cho and Bengio}]{Chorowski2015}
\bibinfo{author}{Chorowski, J.K.}, \bibinfo{author}{Bahdanau, D.}, \bibinfo{author}{Serdyuk, D.}, \bibinfo{author}{Cho, K.}, \bibinfo{author}{Bengio, Y.}, \bibinfo{year}{2015}.
\newblock \bibinfo{title}{Attention-based models for speech recognition}, in: \bibinfo{booktitle}{Proc. NeurIPS}.
%Type = Inproceedings
\bibitem[{Christensen et~al.(2010)Christensen, Barker, Ma and Green}]{christensen2010chime}
\bibinfo{author}{Christensen, H.}, \bibinfo{author}{Barker, J.}, \bibinfo{author}{Ma, N.}, \bibinfo{author}{Green, P.D.}, \bibinfo{year}{2010}.
\newblock \bibinfo{title}{The {CHiME corpus}: a resource and a challenge for computational hearing in multisource environments}, in: \bibinfo{booktitle}{Eleventh annual conference of the international speech communication association}.
%Type = Inproceedings
\bibitem[{Cieri et~al.(2004)Cieri, Miller and Walker}]{cieri2004fisher}
\bibinfo{author}{Cieri, C.}, \bibinfo{author}{Miller, D.}, \bibinfo{author}{Walker, K.}, \bibinfo{year}{2004}.
\newblock \bibinfo{title}{The {Fisher} {Corpus}: A resource for the next generations of speech-to-text.}, in: \bibinfo{booktitle}{LREC}, pp. \bibinfo{pages}{69--71}.
%Type = Article
\bibitem[{Cooke et~al.(2010)Cooke, Hershey and Rennie}]{cooke2010monaural}
\bibinfo{author}{Cooke, M.}, \bibinfo{author}{Hershey, J.R.}, \bibinfo{author}{Rennie, S.J.}, \bibinfo{year}{2010}.
\newblock \bibinfo{title}{Monaural speech separation and recognition challenge}.
\newblock \bibinfo{journal}{Computer Speech \& Language} \bibinfo{volume}{24}, \bibinfo{pages}{1--15}.
%Type = Article
\bibitem[{Cornell et~al.(2021)Cornell, Brutti, Matassoni and Squartini}]{cornell2021learning}
\bibinfo{author}{Cornell, S.}, \bibinfo{author}{Brutti, A.}, \bibinfo{author}{Matassoni, M.}, \bibinfo{author}{Squartini, S.}, \bibinfo{year}{2021}.
\newblock \bibinfo{title}{Learning to rank microphones for distant speech recognition}.
\newblock \bibinfo{journal}{arXiv preprint arXiv:2104.02819} .
%Type = Article
\bibitem[{Cornell et~al.(2024a)Cornell, Darefsky, Duan and Watanabe}]{cornell2024generating}
\bibinfo{author}{Cornell, S.}, \bibinfo{author}{Darefsky, J.}, \bibinfo{author}{Duan, Z.}, \bibinfo{author}{Watanabe, S.}, \bibinfo{year}{2024}a.
\newblock \bibinfo{title}{Generating data with text-to-speech and large-language models for conversational speech recognition}.
\newblock \bibinfo{journal}{SynData4GenAI Workshop} .
%Type = Inproceedings
\bibitem[{Cornell et~al.(2024b)Cornell, Jung, Watanabe and Squartini}]{Cornell2024icassp}
\bibinfo{author}{Cornell, S.}, \bibinfo{author}{Jung, J.w.}, \bibinfo{author}{Watanabe, S.}, \bibinfo{author}{Squartini, S.}, \bibinfo{year}{2024}b.
\newblock \bibinfo{title}{One model to rule them all? towards end-to-end joint speaker diarization and speech recognition}, in: \bibinfo{booktitle}{Proc. of ICASSP}, pp. \bibinfo{pages}{11856--11860}.
%Type = Article
\bibitem[{Cornell et~al.(2024c)Cornell, Park, Huang, Boeddeker, Chang, Maciejewski, Wiesner, Garcia and Watanabe}]{cornell2024chime}
\bibinfo{author}{Cornell, S.}, \bibinfo{author}{Park, T.}, \bibinfo{author}{Huang, S.}, \bibinfo{author}{Boeddeker, C.}, \bibinfo{author}{Chang, X.}, \bibinfo{author}{Maciejewski, M.}, \bibinfo{author}{Wiesner, M.}, \bibinfo{author}{Garcia, P.}, \bibinfo{author}{Watanabe, S.}, \bibinfo{year}{2024}c.
\newblock \bibinfo{title}{The {CHiME-8 DASR} challenge for generalizable and array agnostic distant automatic speech recognition and diarization}.
\newblock \bibinfo{journal}{CHiME Workshop} .
%Type = Article
\bibitem[{Cornell et~al.(2023)Cornell, Wiesner, Watanabe, Raj, Chang, Garcia, Masuyama, Wang, Squartini and Khudanpur}]{cornell2023chime}
\bibinfo{author}{Cornell, S.}, \bibinfo{author}{Wiesner, M.}, \bibinfo{author}{Watanabe, S.}, \bibinfo{author}{Raj, D.}, \bibinfo{author}{Chang, X.}, \bibinfo{author}{Garcia, P.}, \bibinfo{author}{Masuyama, Y.}, \bibinfo{author}{Wang, Z.Q.}, \bibinfo{author}{Squartini, S.}, \bibinfo{author}{Khudanpur, S.}, \bibinfo{year}{2023}.
\newblock \bibinfo{title}{{The CHiME-7 DASR Challenge}: Distant meeting transcription with multiple devices in diverse scenarios}.
\newblock \bibinfo{journal}{CHiME Workshop} .
%Type = Inproceedings
\bibitem[{Cristoforetti et~al.(2014)Cristoforetti, Ravanelli, Omologo, Sosi, Abad, Hagm{\"u}ller and Maragos}]{cristoforetti2014dirha}
\bibinfo{author}{Cristoforetti, L.}, \bibinfo{author}{Ravanelli, M.}, \bibinfo{author}{Omologo, M.}, \bibinfo{author}{Sosi, A.}, \bibinfo{author}{Abad, A.}, \bibinfo{author}{Hagm{\"u}ller, M.}, \bibinfo{author}{Maragos, P.}, \bibinfo{year}{2014}.
\newblock \bibinfo{title}{The {DIRHA} simulated corpus.}, in: \bibinfo{booktitle}{LREC}.
%Type = Article
\bibitem[{Cui et~al.(2024)}]{slm_survey1}
\bibinfo{author}{Cui, W.}, et~al., \bibinfo{year}{2024}.
\newblock \bibinfo{title}{Recent advances in speech language models: A survey}.
\newblock \bibinfo{journal}{arXiv preprint arXiv:2410.03751} .
%Type = Article
\bibitem[{D{\'e}fossez et~al.(2024)D{\'e}fossez, Mazar{\'e}, Orsini, Royer, P{\'e}rez, J{\'e}gou, Grave and Zeghidour}]{defossez2024moshi}
\bibinfo{author}{D{\'e}fossez, A.}, \bibinfo{author}{Mazar{\'e}, L.}, \bibinfo{author}{Orsini, M.}, \bibinfo{author}{Royer, A.}, \bibinfo{author}{P{\'e}rez, P.}, \bibinfo{author}{J{\'e}gou, H.}, \bibinfo{author}{Grave, E.}, \bibinfo{author}{Zeghidour, N.}, \bibinfo{year}{2024}.
\newblock \bibinfo{title}{Moshi: a speech-text foundation model for real-time dialogue}.
\newblock \bibinfo{journal}{arXiv preprint arXiv:2410.00037} .
%Type = Article
\bibitem[{Deng and Woodland(2024)}]{deng2024label}
\bibinfo{author}{Deng, K.}, \bibinfo{author}{Woodland, P.C.}, \bibinfo{year}{2024}.
\newblock \bibinfo{title}{Label-synchronous neural transducer for adaptable online e2e speech recognition}.
\newblock \bibinfo{journal}{IEEE/ACM Transactions on Audio, Speech, and Language Processing} .
%Type = Inproceedings
\bibitem[{Deng et~al.(2023)Deng, Zheng and Woodland}]{deng2023university}
\bibinfo{author}{Deng, K.}, \bibinfo{author}{Zheng, X.}, \bibinfo{author}{Woodland, P.C.}, \bibinfo{year}{2023}.
\newblock \bibinfo{title}{The {University} of {Cambridge} system for the {CHiME-7 DASR Task}}, in: \bibinfo{booktitle}{Proc. CHiME 2023}, pp. \bibinfo{pages}{73--76}.
%Type = Article
\bibitem[{Desplanques et~al.(2020)Desplanques, Thienpondt and Demuynck}]{desplanques2020ecapa}
\bibinfo{author}{Desplanques, B.}, \bibinfo{author}{Thienpondt, J.}, \bibinfo{author}{Demuynck, K.}, \bibinfo{year}{2020}.
\newblock \bibinfo{title}{{ECAPA-TDNN}: Emphasized channel attention, propagation and aggregation in tdnn based speaker verification}.
\newblock \bibinfo{journal}{Interspeech} .
%Type = Article
\bibitem[{Du~Bois et~al.(2000)Du~Bois, Chafe, Meyer, Thompson and Martey}]{du2000santa}
\bibinfo{author}{Du~Bois, J.W.}, \bibinfo{author}{Chafe, W.L.}, \bibinfo{author}{Meyer, C.}, \bibinfo{author}{Thompson, S.A.}, \bibinfo{author}{Martey, N.}, \bibinfo{year}{2000}.
\newblock \bibinfo{title}{Santa {Barbara} corpus of spoken {American} {English}}.
\newblock \bibinfo{journal}{CD-ROM. Philadelphia: Linguistic Data Consortium} .
%Type = Inproceedings
\bibitem[{Erdogan et~al.(2016)Erdogan, Hershey, Watanabe, Mandel and Le~Roux}]{erdogan2016improved}
\bibinfo{author}{Erdogan, H.}, \bibinfo{author}{Hershey, J.R.}, \bibinfo{author}{Watanabe, S.}, \bibinfo{author}{Mandel, M.I.}, \bibinfo{author}{Le~Roux, J.}, \bibinfo{year}{2016}.
\newblock \bibinfo{title}{Improved {MVDR} beamforming using single-channel mask prediction networks}, in: \bibinfo{booktitle}{Interspeech}, pp. \bibinfo{pages}{1981--1985}.
%Type = Article
\bibitem[{Fang et~al.(2024)Fang, Guo, Zhou, Ma, Zhang and Feng}]{fang2024llama}
\bibinfo{author}{Fang, Q.}, \bibinfo{author}{Guo, S.}, \bibinfo{author}{Zhou, Y.}, \bibinfo{author}{Ma, Z.}, \bibinfo{author}{Zhang, S.}, \bibinfo{author}{Feng, Y.}, \bibinfo{year}{2024}.
\newblock \bibinfo{title}{Llama-omni: Seamless speech interaction with large language models}.
\newblock \bibinfo{journal}{arXiv preprint arXiv:2409.06666} .
%Type = Inproceedings
\bibitem[{Fiscus(1997)}]{fiscus1997post}
\bibinfo{author}{Fiscus, J.G.}, \bibinfo{year}{1997}.
\newblock \bibinfo{title}{A post-processing system to yield reduced word error rates: Recognizer output voting error reduction (rover)}, in: \bibinfo{booktitle}{Proc. ASRU}.
%Type = Inproceedings
\bibitem[{Fiscus et~al.(2007a)Fiscus, Ajot and Garofolo}]{Fiscus2007}
\bibinfo{author}{Fiscus, J.G.}, \bibinfo{author}{Ajot, J.}, \bibinfo{author}{Garofolo, J.S.}, \bibinfo{year}{2007}a.
\newblock \bibinfo{title}{The rich transcription 2007 meeting recognition evaluation}, in: \bibinfo{booktitle}{Proc. Multimodal Tech. Percept. Hum.}, pp. \bibinfo{pages}{373--389}.
%Type = Inproceedings
\bibitem[{Fiscus et~al.(2007b)Fiscus, Ajot and Garofolo}]{fiscus2007rich}
\bibinfo{author}{Fiscus, J.G.}, \bibinfo{author}{Ajot, J.}, \bibinfo{author}{Garofolo, J.S.}, \bibinfo{year}{2007}b.
\newblock \bibinfo{title}{The {Rich Transcription} 2007 meeting recognition evaluation}, in: \bibinfo{booktitle}{International Evaluation Workshop on Rich Transcription}, \bibinfo{organization}{Springer}. pp. \bibinfo{pages}{373--389}.
%Type = Inproceedings
\bibitem[{Fiscus et~al.(2006a)Fiscus, Ajot, Radde, Laprun et~al.}]{fiscus2006multiple}
\bibinfo{author}{Fiscus, J.G.}, \bibinfo{author}{Ajot, J.}, \bibinfo{author}{Radde, N.}, \bibinfo{author}{Laprun, C.}, et~al., \bibinfo{year}{2006}a.
\newblock \bibinfo{title}{Multiple dimension {Levenshtein} edit distance calculations for evaluating automatic speech recognition systems during simultaneous speech.}, in: \bibinfo{booktitle}{LREC}, \bibinfo{organization}{Citeseer}. pp. \bibinfo{pages}{803--808}.
%Type = Article
\bibitem[{Fiscus et~al.(2007c)Fiscus, Doddington, Le, Sanders, Przybocki and Pallett}]{fiscus20072003}
\bibinfo{author}{Fiscus, J.G.}, \bibinfo{author}{Doddington, G.}, \bibinfo{author}{Le, A.}, \bibinfo{author}{Sanders, G.}, \bibinfo{author}{Przybocki, M.}, \bibinfo{author}{Pallett, D.}, \bibinfo{year}{2007}c.
\newblock \bibinfo{title}{2003 nist rich transcription evaluation data ldc2007s10}.
\newblock \bibinfo{journal}{Web Download. Philadelphia: Linguistic Data Consortium} .
%Type = Inproceedings
\bibitem[{Fiscus et~al.(2006b)Fiscus, Radde, Garofolo, Le, Ajot and Laprun}]{fiscus2006rich}
\bibinfo{author}{Fiscus, J.G.}, \bibinfo{author}{Radde, N.}, \bibinfo{author}{Garofolo, J.S.}, \bibinfo{author}{Le, A.}, \bibinfo{author}{Ajot, J.}, \bibinfo{author}{Laprun, C.}, \bibinfo{year}{2006}b.
\newblock \bibinfo{title}{The rich transcription 2005 spring meeting recognition evaluation}, in: \bibinfo{booktitle}{Machine Learning for Multimodal Interaction: Second International Workshop, MLMI 2005, Edinburgh, UK, July 11-13, 2005, Revised Selected Papers 2}, \bibinfo{organization}{Springer}. pp. \bibinfo{pages}{369--389}.
%Type = Article
\bibitem[{Fonseca et~al.(2021)Fonseca, Favory, Pons, Font and Serra}]{fonseca2021fsd50k}
\bibinfo{author}{Fonseca, E.}, \bibinfo{author}{Favory, X.}, \bibinfo{author}{Pons, J.}, \bibinfo{author}{Font, F.}, \bibinfo{author}{Serra, X.}, \bibinfo{year}{2021}.
\newblock \bibinfo{title}{{FSD50k}: An open dataset of human-labeled sound events}.
\newblock \bibinfo{journal}{IEEE/ACM Trans. Audio, Speech, Lang. Process.} \bibinfo{volume}{30}.
%Type = Inproceedings
\bibitem[{Fox et~al.(2013)Fox, Liu, Zwyssig and Hain}]{fox2013sheffield}
\bibinfo{author}{Fox, C.}, \bibinfo{author}{Liu, Y.}, \bibinfo{author}{Zwyssig, E.}, \bibinfo{author}{Hain, T.}, \bibinfo{year}{2013}.
\newblock \bibinfo{title}{The {Sheffield} wargames corpus}, in: \bibinfo{booktitle}{Interspeech}.
%Type = Article
\bibitem[{Fu et~al.(2021)Fu, Cheng, Lv, Jv, Kong, Chen, Hu, Xie, Wu, Bu et~al.}]{fu2021aishell}
\bibinfo{author}{Fu, Y.}, \bibinfo{author}{Cheng, L.}, \bibinfo{author}{Lv, S.}, \bibinfo{author}{Jv, Y.}, \bibinfo{author}{Kong, Y.}, \bibinfo{author}{Chen, Z.}, \bibinfo{author}{Hu, Y.}, \bibinfo{author}{Xie, L.}, \bibinfo{author}{Wu, J.}, \bibinfo{author}{Bu, H.}, et~al., \bibinfo{year}{2021}.
\newblock \bibinfo{title}{{AISHELL-4}: An open source dataset for speech enhancement, separation, recognition and speaker diarization in conference scenario} .
%Type = Inproceedings
\bibitem[{Fujita et~al.(2019a)Fujita, Kanda, Horiguchi, Nagamatsu and Watanabe}]{fujita2019end_lstm}
\bibinfo{author}{Fujita, Y.}, \bibinfo{author}{Kanda, N.}, \bibinfo{author}{Horiguchi, S.}, \bibinfo{author}{Nagamatsu, K.}, \bibinfo{author}{Watanabe, S.}, \bibinfo{year}{2019}a.
\newblock \bibinfo{title}{End-to-end neural speaker diarization with permutation-free objectives}, in: \bibinfo{booktitle}{Interspeech 2019}, pp. \bibinfo{pages}{4300--4304}.
%Type = Inproceedings
\bibitem[{Fujita et~al.(2019b)Fujita, Kanda, Horiguchi, Xue, Nagamatsu and Watanabe}]{fujita2019end}
\bibinfo{author}{Fujita, Y.}, \bibinfo{author}{Kanda, N.}, \bibinfo{author}{Horiguchi, S.}, \bibinfo{author}{Xue, Y.}, \bibinfo{author}{Nagamatsu, K.}, \bibinfo{author}{Watanabe, S.}, \bibinfo{year}{2019}b.
\newblock \bibinfo{title}{End-to-end neural speaker diarization with self-attention}, in: \bibinfo{booktitle}{Proc. of ASRU}, \bibinfo{organization}{IEEE}. pp. \bibinfo{pages}{296--303}.
%Type = Article
\bibitem[{Gales et~al.(2008)Gales, Young et~al.}]{gales2008application}
\bibinfo{author}{Gales, M.}, \bibinfo{author}{Young, S.}, et~al., \bibinfo{year}{2008}.
\newblock \bibinfo{title}{The application of hidden markov models in speech recognition}.
\newblock \bibinfo{journal}{Foundations and Trends{\textregistered} in Signal Processing} \bibinfo{volume}{1}, \bibinfo{pages}{195--304}.
%Type = Book
\bibitem[{Garofolo et~al.(2004)Garofolo, Fiscus and Laprun}]{garofolo2004rich}
\bibinfo{author}{Garofolo, J.S.}, \bibinfo{author}{Fiscus, J.G.}, \bibinfo{author}{Laprun, C.D.}, \bibinfo{year}{2004}.
\newblock \bibinfo{title}{The rich transcription 2004 spring meeting recognition evaluation}.
\newblock \bibinfo{publisher}{US Department of Commerce, National Institute of Standards and Technology}.
%Type = Inproceedings
\bibitem[{Garofolo et~al.(2002)Garofolo, Fiscus, Martin, Pallett and Przybocki}]{garofolo2002nist}
\bibinfo{author}{Garofolo, J.S.}, \bibinfo{author}{Fiscus, J.G.}, \bibinfo{author}{Martin, A.F.}, \bibinfo{author}{Pallett, D.S.}, \bibinfo{author}{Przybocki, M.A.}, \bibinfo{year}{2002}.
\newblock \bibinfo{title}{Nist rich transcription 2002 evaluation: A preview.}, in: \bibinfo{booktitle}{LREC}.
%Type = Article
\bibitem[{Georgiev et~al.(2024)Georgiev, Lei, Burnell, Bai, Gulati, Tanzer, Vincent, Pan, Wang et~al.}]{team2024gemini}
\bibinfo{author}{Georgiev, P.}, \bibinfo{author}{Lei, V.I.}, \bibinfo{author}{Burnell, R.}, \bibinfo{author}{Bai, L.}, \bibinfo{author}{Gulati, A.}, \bibinfo{author}{Tanzer, G.}, \bibinfo{author}{Vincent, D.}, \bibinfo{author}{Pan, Z.}, \bibinfo{author}{Wang, S.}, et~al., \bibinfo{year}{2024}.
\newblock \bibinfo{title}{Gemini 1.5: Unlocking multimodal understanding across millions of tokens of context}.
\newblock \bibinfo{journal}{arXiv preprint arXiv:2403.05530} .
%Type = Inproceedings
\bibitem[{Godfrey et~al.(1992)Godfrey, Holliman and McDaniel}]{godfrey1992switchboard}
\bibinfo{author}{Godfrey, J.J.}, \bibinfo{author}{Holliman, E.C.}, \bibinfo{author}{McDaniel, J.}, \bibinfo{year}{1992}.
\newblock \bibinfo{title}{{SWITCHBOARD}: Telephone speech corpus for research and development}, in: \bibinfo{booktitle}{Proc. of ICASSP}.
%Type = Inproceedings
\bibitem[{Grauman et~al.(2022)Grauman, Westbury, Byrne, Chavis, Furnari, Girdhar, Hamburger, Jiang, Liu, Liu et~al.}]{grauman2022ego4d}
\bibinfo{author}{Grauman, K.}, \bibinfo{author}{Westbury, A.}, \bibinfo{author}{Byrne, E.}, \bibinfo{author}{Chavis, Z.}, \bibinfo{author}{Furnari, A.}, \bibinfo{author}{Girdhar, R.}, \bibinfo{author}{Hamburger, J.}, \bibinfo{author}{Jiang, H.}, \bibinfo{author}{Liu, M.}, \bibinfo{author}{Liu, X.}, et~al., \bibinfo{year}{2022}.
\newblock \bibinfo{title}{{Ego4d}: Around the world in 3,000 hours of egocentric video}, in: \bibinfo{booktitle}{CVPR}.
%Type = Inproceedings
\bibitem[{Graves(2012)}]{gravessequence}
\bibinfo{author}{Graves, A.}, \bibinfo{year}{2012}.
\newblock \bibinfo{title}{Sequence transduction with recurrent neural networks}, in: \bibinfo{booktitle}{ICML}.
%Type = Inproceedings
\bibitem[{Graves et~al.(2006)Graves, Fern{\'a}ndez, Gomez and Schmidhuber}]{graves2006connectionist}
\bibinfo{author}{Graves, A.}, \bibinfo{author}{Fern{\'a}ndez, S.}, \bibinfo{author}{Gomez, F.}, \bibinfo{author}{Schmidhuber, J.}, \bibinfo{year}{2006}.
\newblock \bibinfo{title}{Connectionist temporal classification: labelling unsegmented sequence data with recurrent neural networks}, in: \bibinfo{booktitle}{ICML}.
%Type = Article
\bibitem[{Gu et~al.(2021)Gu, Goel and R{\'e}}]{gu2021efficiently}
\bibinfo{author}{Gu, A.}, \bibinfo{author}{Goel, K.}, \bibinfo{author}{R{\'e}, C.}, \bibinfo{year}{2021}.
\newblock \bibinfo{title}{Efficiently modeling long sequences with structured state spaces}.
\newblock \bibinfo{journal}{arXiv preprint arXiv:2111.00396} .
%Type = Article
\bibitem[{Gulati et~al.(2020)Gulati, Qin, Chiu, Parmar, Zhang, Yu, Han, Wang, Zhang, Wu et~al.}]{gulati2020conformer}
\bibinfo{author}{Gulati, A.}, \bibinfo{author}{Qin, J.}, \bibinfo{author}{Chiu, C.C.}, \bibinfo{author}{Parmar, N.}, \bibinfo{author}{Zhang, Y.}, \bibinfo{author}{Yu, J.}, \bibinfo{author}{Han, W.}, \bibinfo{author}{Wang, S.}, \bibinfo{author}{Zhang, Z.}, \bibinfo{author}{Wu, Y.}, et~al., \bibinfo{year}{2020}.
\newblock \bibinfo{title}{Conformer: Convolution-augmented transformer for speech recognition}.
\newblock \bibinfo{journal}{Interspeech} .
%Type = Article
\bibitem[{Haeb-Umbach et~al.(2019)Haeb-Umbach, Watanabe, Nakatani, Bacchiani, Hoffmeister, Seltzer, Zen and Souden}]{haeb2019speech}
\bibinfo{author}{Haeb-Umbach, R.}, \bibinfo{author}{Watanabe, S.}, \bibinfo{author}{Nakatani, T.}, \bibinfo{author}{Bacchiani, M.}, \bibinfo{author}{Hoffmeister, B.}, \bibinfo{author}{Seltzer, M.L.}, \bibinfo{author}{Zen, H.}, \bibinfo{author}{Souden, M.}, \bibinfo{year}{2019}.
\newblock \bibinfo{title}{Speech processing for digital home assistants: Combining signal processing with deep-learning techniques}.
\newblock \bibinfo{journal}{IEEE Signal processing magazine} \bibinfo{volume}{36}, \bibinfo{pages}{111--124}.
%Type = Article
\bibitem[{Han et~al.(2024)Han, Landini, Rohdin, Silnova, Diez and Burget}]{han2024leveraging}
\bibinfo{author}{Han, J.}, \bibinfo{author}{Landini, F.}, \bibinfo{author}{Rohdin, J.}, \bibinfo{author}{Silnova, A.}, \bibinfo{author}{Diez, M.}, \bibinfo{author}{Burget, L.}, \bibinfo{year}{2024}.
\newblock \bibinfo{title}{Leveraging self-supervised learning for speaker diarization}.
\newblock \bibinfo{journal}{Submitted to ICASSP} .
%Type = Inproceedings
\bibitem[{Han et~al.(2019)Han, Prieto and Ma}]{han2019state}
\bibinfo{author}{Han, K.J.}, \bibinfo{author}{Prieto, R.}, \bibinfo{author}{Ma, T.}, \bibinfo{year}{2019}.
\newblock \bibinfo{title}{State-of-the-art speech recognition using multi-stream self-attention with dilated 1d convolutions}, in: \bibinfo{booktitle}{Proc. of ASRU}, \bibinfo{organization}{IEEE}. pp. \bibinfo{pages}{54--61}.
%Type = Inproceedings
\bibitem[{Harper(2015)}]{harper2015automatic}
\bibinfo{author}{Harper, M.}, \bibinfo{year}{2015}.
\newblock \bibinfo{title}{The automatic speech recognition in reverberant environments (aspire) challenge}, in: \bibinfo{booktitle}{IEEE ASRU}.
%Type = Article
\bibitem[{He et~al.(2023)He, Du, Liu and Lee}]{he2023ansd}
\bibinfo{author}{He, M.K.}, \bibinfo{author}{Du, J.}, \bibinfo{author}{Liu, Q.F.}, \bibinfo{author}{Lee, C.H.}, \bibinfo{year}{2023}.
\newblock \bibinfo{title}{Ansd-ma-mse: Adaptive neural speaker diarization using memory-aware multi-speaker embedding}.
\newblock \bibinfo{journal}{IEEE/ACM Transactions on Audio, Speech, and Language Processing} \bibinfo{volume}{31}, \bibinfo{pages}{1561--1573}.
%Type = Inproceedings
\bibitem[{Heafield(2011)}]{heafield2011kenlm}
\bibinfo{author}{Heafield, K.}, \bibinfo{year}{2011}.
\newblock \bibinfo{title}{Kenlm: Faster and smaller language model queries}, in: \bibinfo{booktitle}{Proceedings of the sixth workshop on statistical machine translation}, pp. \bibinfo{pages}{187--197}.
%Type = Inproceedings
\bibitem[{Heo et~al.(2025)Heo, Park, Han and Yun}]{heo2025rotary}
\bibinfo{author}{Heo, B.}, \bibinfo{author}{Park, S.}, \bibinfo{author}{Han, D.}, \bibinfo{author}{Yun, S.}, \bibinfo{year}{2025}.
\newblock \bibinfo{title}{Rotary position embedding for vision transformer}, in: \bibinfo{booktitle}{European Conference on Computer Vision}, \bibinfo{organization}{Springer}. pp. \bibinfo{pages}{289--305}.
%Type = Inproceedings
\bibitem[{Hernandez et~al.(2018)Hernandez, Nguyen, Ghannay, Tomashenko and Esteve}]{hernandez2018ted}
\bibinfo{author}{Hernandez, F.}, \bibinfo{author}{Nguyen, V.}, \bibinfo{author}{Ghannay, S.}, \bibinfo{author}{Tomashenko, N.}, \bibinfo{author}{Esteve, Y.}, \bibinfo{year}{2018}.
\newblock \bibinfo{title}{{TED-LIUM 3}: Twice as much data and corpus repartition for experiments on speaker adaptation}, in: \bibinfo{booktitle}{Speech and Computer: 20th International Conference, SPECOM 2018, Leipzig, Germany, September 18--22, 2018, Proceedings 20}, \bibinfo{organization}{Springer}. pp. \bibinfo{pages}{198--208}.
%Type = Inproceedings
\bibitem[{Hirano et~al.(2024)Hirano, Nguyen, Azuma, Saragih and Sakti}]{hirano2024naist}
\bibinfo{author}{Hirano, Y.}, \bibinfo{author}{Nguyen, M.}, \bibinfo{author}{Azuma, K.}, \bibinfo{author}{Saragih, J.M.}, \bibinfo{author}{Sakti, S.}, \bibinfo{year}{2024}.
\newblock \bibinfo{title}{The naist system for the chime-8 notsofar-1 task}, in: \bibinfo{booktitle}{CHiME Workshop}, pp. \bibinfo{pages}{59--63}.
%Type = Inproceedings
\bibitem[{Hirsch and Pearce(2000)}]{hirsch2000aurora}
\bibinfo{author}{Hirsch, H.G.}, \bibinfo{author}{Pearce, D.}, \bibinfo{year}{2000}.
\newblock \bibinfo{title}{The {AURORA} experimental framework for the performance evaluation of speech recognition systems under noisy conditions}, in: \bibinfo{booktitle}{ASR2000-Automatic speech recognition: challenges for the new Millenium ISCA tutorial and research workshop (ITRW)}.
%Type = Inproceedings
\bibitem[{Horiguchi et~al.(2022)Horiguchi, Takashima, Garcia, Watanabe and Kawaguchi}]{horiguchi2022multi}
\bibinfo{author}{Horiguchi, S.}, \bibinfo{author}{Takashima, Y.}, \bibinfo{author}{Garcia, P.}, \bibinfo{author}{Watanabe, S.}, \bibinfo{author}{Kawaguchi, Y.}, \bibinfo{year}{2022}.
\newblock \bibinfo{title}{Multi-channel end-to-end neural diarization with distributed microphones}, in: \bibinfo{booktitle}{Proc. of ICASSP}, pp. \bibinfo{pages}{7332--7336}.
%Type = Article
\bibitem[{Hsu et~al.(2021)Hsu, Bolte, Tsai, Lakhotia, Salakhutdinov and Mohamed}]{hsu2021hubert}
\bibinfo{author}{Hsu, W.N.}, \bibinfo{author}{Bolte, B.}, \bibinfo{author}{Tsai, Y.H.H.}, \bibinfo{author}{Lakhotia, K.}, \bibinfo{author}{Salakhutdinov, R.}, \bibinfo{author}{Mohamed, A.}, \bibinfo{year}{2021}.
\newblock \bibinfo{title}{{HuBERT}: Self-supervised speech representation learning by masked prediction of hidden units}.
\newblock \bibinfo{journal}{IEEE/ACM Trans. Audio, Speech, Lang. Process.} \bibinfo{volume}{29}.
%Type = Article
\bibitem[{Hu et~al.(2024)Hu, Chen, Yang, Qin, Chen, Chng and Zhang}]{hu2024self}
\bibinfo{author}{Hu, Y.}, \bibinfo{author}{Chen, C.}, \bibinfo{author}{Yang, C.H.H.}, \bibinfo{author}{Qin, C.}, \bibinfo{author}{Chen, P.Y.}, \bibinfo{author}{Chng, E.S.}, \bibinfo{author}{Zhang, C.}, \bibinfo{year}{2024}.
\newblock \bibinfo{title}{Self-taught recognizer: Toward unsupervised adaptation for speech foundation models}.
\newblock \bibinfo{journal}{arXiv preprint arXiv:2405.14161} .
%Type = Inproceedings
\bibitem[{Huang et~al.(2024)Huang, Li, Wang, Wang, Rao, Sun, Tang, Huang, Wang, Yu et~al.}]{huang2024npu}
\bibinfo{author}{Huang, K.}, \bibinfo{author}{Li, Y.}, \bibinfo{author}{Wang, Z.}, \bibinfo{author}{Wang, H.}, \bibinfo{author}{Rao, W.}, \bibinfo{author}{Sun, Z.}, \bibinfo{author}{Tang, Z.}, \bibinfo{author}{Huang, S.}, \bibinfo{author}{Wang, Y.}, \bibinfo{author}{Yu, T.}, et~al., \bibinfo{year}{2024}.
\newblock \bibinfo{title}{The npu-tea system for the chime-8 notsofar-1 challenge}, in: \bibinfo{booktitle}{CHiME Workshop}, pp. \bibinfo{pages}{45--48}.
%Type = Inproceedings
\bibitem[{Ito et~al.(2016)Ito, Araki and Nakatani}]{ito2016cACGMM}
\bibinfo{author}{Ito, N.}, \bibinfo{author}{Araki, S.}, \bibinfo{author}{Nakatani, T.}, \bibinfo{year}{2016}.
\newblock \bibinfo{title}{Complex angular central gaussian mixture model for directional statistics in mask-based microphone array signal processing}, in: \bibinfo{booktitle}{2016 24th European Signal Processing Conference (EUSIPCO)}, \bibinfo{organization}{IEEE}. pp. \bibinfo{pages}{1153--1157}.
%Type = Inproceedings
\bibitem[{Janin et~al.(2003)Janin, Baron, Edwards, Ellis, Gelbart, Morgan, Peskin, Pfau, Shriberg, Stolcke et~al.}]{janin2003icsi}
\bibinfo{author}{Janin, A.}, \bibinfo{author}{Baron, D.}, \bibinfo{author}{Edwards, J.}, \bibinfo{author}{Ellis, D.}, \bibinfo{author}{Gelbart, D.}, \bibinfo{author}{Morgan, N.}, \bibinfo{author}{Peskin, B.}, \bibinfo{author}{Pfau, T.}, \bibinfo{author}{Shriberg, E.}, \bibinfo{author}{Stolcke, A.}, et~al., \bibinfo{year}{2003}.
\newblock \bibinfo{title}{The {ICSI} meeting corpus}, in: \bibinfo{booktitle}{Proc. of ICASSP}.
%Type = Inproceedings
\bibitem[{Jeub et~al.(2009)Jeub, Schafer and Vary}]{jeub2009binaural}
\bibinfo{author}{Jeub, M.}, \bibinfo{author}{Schafer, M.}, \bibinfo{author}{Vary, P.}, \bibinfo{year}{2009}.
\newblock \bibinfo{title}{A binaural room impulse response database for the evaluation of dereverberation algorithms}, in: \bibinfo{booktitle}{2009 16th International Conference on Digital Signal Processing}, \bibinfo{organization}{IEEE}. pp. \bibinfo{pages}{1--5}.
%Type = Article
\bibitem[{Kamo et~al.(2023)Kamo, Tawara, Ando, Kano, Sato, Ikeshita, Moriya, Horiguchi, Matsuura, Ogawa et~al.}]{kamo2023ntt}
\bibinfo{author}{Kamo, N.}, \bibinfo{author}{Tawara, N.}, \bibinfo{author}{Ando, A.}, \bibinfo{author}{Kano, T.}, \bibinfo{author}{Sato, H.}, \bibinfo{author}{Ikeshita, R.}, \bibinfo{author}{Moriya, T.}, \bibinfo{author}{Horiguchi, S.}, \bibinfo{author}{Matsuura, K.}, \bibinfo{author}{Ogawa, A.}, et~al., \bibinfo{year}{2023}.
\newblock \bibinfo{title}{Ntt multi-speaker asr system for the dasr task of chime-7 challenge}.
\newblock \bibinfo{journal}{CHiME Workshop} .
%Type = Inproceedings
\bibitem[{Kamo et~al.(2024)Kamo, Tawara, Ando, Kano, Sato, Ikeshita, Moriya, Horiguchi, Matsuura, Ogawa et~al.}]{kamo2024ntt}
\bibinfo{author}{Kamo, N.}, \bibinfo{author}{Tawara, N.}, \bibinfo{author}{Ando, A.}, \bibinfo{author}{Kano, T.}, \bibinfo{author}{Sato, H.}, \bibinfo{author}{Ikeshita, R.}, \bibinfo{author}{Moriya, T.}, \bibinfo{author}{Horiguchi, S.}, \bibinfo{author}{Matsuura, K.}, \bibinfo{author}{Ogawa, A.}, et~al., \bibinfo{year}{2024}.
\newblock \bibinfo{title}{Ntt multi-speaker asr system for the dasr task of chime-8 challenge}, in: \bibinfo{booktitle}{CHiME Workshop}.
%Type = Article
\bibitem[{Kanda et~al.(2022a)Kanda, Wu, Wang, Chen, Li and Yoshioka}]{kanda2022vararray}
\bibinfo{author}{Kanda, N.}, \bibinfo{author}{Wu, J.}, \bibinfo{author}{Wang, X.}, \bibinfo{author}{Chen, Z.}, \bibinfo{author}{Li, J.}, \bibinfo{author}{Yoshioka, T.}, \bibinfo{year}{2022}a.
\newblock \bibinfo{title}{{VarArray} meets {t-SOT}: Advancing the state of the art of streaming distant conversational speech recognition}.
\newblock \bibinfo{journal}{ArXiv} .
%Type = Inproceedings
\bibitem[{Kanda et~al.(2022b)Kanda, Xiao, Gaur, Wang, Meng, Chen and Yoshioka}]{kanda2022transcribe}
\bibinfo{author}{Kanda, N.}, \bibinfo{author}{Xiao, X.}, \bibinfo{author}{Gaur, Y.}, \bibinfo{author}{Wang, X.}, \bibinfo{author}{Meng, Z.}, \bibinfo{author}{Chen, Z.}, \bibinfo{author}{Yoshioka, T.}, \bibinfo{year}{2022}b.
\newblock \bibinfo{title}{Transcribe-to-diarize: Neural speaker diarization for unlimited number of speakers using end-to-end speaker-attributed {ASR}}, in: \bibinfo{booktitle}{Proc. of ICASSP}, \bibinfo{organization}{IEEE}. pp. \bibinfo{pages}{8082--8086}.
%Type = Inproceedings
\bibitem[{Karafi{\'a}t et~al.(2023)Karafi{\'a}t, Vesel{\`y}, Sz{\"o}ke, Mo{\v{s}}ner, Bene{\v{s}}, Witkowski, Barchi and Pepino}]{karafiat2023but}
\bibinfo{author}{Karafi{\'a}t, M.}, \bibinfo{author}{Vesel{\`y}, K.}, \bibinfo{author}{Sz{\"o}ke, I.}, \bibinfo{author}{Mo{\v{s}}ner, L.}, \bibinfo{author}{Bene{\v{s}}, K.}, \bibinfo{author}{Witkowski, M.}, \bibinfo{author}{Barchi, G.}, \bibinfo{author}{Pepino, L.}, \bibinfo{year}{2023}.
\newblock \bibinfo{title}{{BUT CHiME-7} system description}, in: \bibinfo{booktitle}{CHiME Workshop}.
%Type = Article
\bibitem[{Kim et~al.(2020)Kim, Lee and Kim}]{kim2020accelerating}
\bibinfo{author}{Kim, J.}, \bibinfo{author}{Lee, Y.}, \bibinfo{author}{Kim, E.}, \bibinfo{year}{2020}.
\newblock \bibinfo{title}{Accelerating rnn transducer inference via adaptive expansion search}.
\newblock \bibinfo{journal}{IEEE Signal Processing Letters} \bibinfo{volume}{27}, \bibinfo{pages}{2019--2023}.
%Type = Inproceedings
\bibitem[{Kim et~al.(2021)Kim, Arora, Le, Yeh, Fuegen, Kalinli and Seltzer}]{kim2021semantic}
\bibinfo{author}{Kim, S.}, \bibinfo{author}{Arora, A.}, \bibinfo{author}{Le, D.}, \bibinfo{author}{Yeh, C.F.}, \bibinfo{author}{Fuegen, C.}, \bibinfo{author}{Kalinli, O.}, \bibinfo{author}{Seltzer, M.L.}, \bibinfo{year}{2021}.
\newblock \bibinfo{title}{Semantic distance: A new metric for asr performance analysis towards spoken language understanding}, in: \bibinfo{booktitle}{Interspeech}.
%Type = Article
\bibitem[{Kinoshita et~al.(2021a)Kinoshita, Delcroix and Tawara}]{kinoshita2021advances}
\bibinfo{author}{Kinoshita, K.}, \bibinfo{author}{Delcroix, M.}, \bibinfo{author}{Tawara, N.}, \bibinfo{year}{2021}a.
\newblock \bibinfo{title}{Advances in integration of end-to-end neural and clustering-based diarization for real conversational speech}.
\newblock \bibinfo{journal}{Interspeech} .
%Type = Inproceedings
\bibitem[{Kinoshita et~al.(2021b)Kinoshita, Delcroix and Tawara}]{kinoshita2021integrating}
\bibinfo{author}{Kinoshita, K.}, \bibinfo{author}{Delcroix, M.}, \bibinfo{author}{Tawara, N.}, \bibinfo{year}{2021}b.
\newblock \bibinfo{title}{Integrating end-to-end neural and clustering-based diarization: Getting the best of both worlds}, in: \bibinfo{booktitle}{Proc. of ICASSP}.
%Type = Inproceedings
\bibitem[{Kinoshita et~al.(2013)Kinoshita, Delcroix, Yoshioka, Nakatani, Habets, Haeb-Umbach, Leutnant, Sehr, Kellermann, Maas et~al.}]{kinoshita2013reverb}
\bibinfo{author}{Kinoshita, K.}, \bibinfo{author}{Delcroix, M.}, \bibinfo{author}{Yoshioka, T.}, \bibinfo{author}{Nakatani, T.}, \bibinfo{author}{Habets, E.}, \bibinfo{author}{Haeb-Umbach, R.}, \bibinfo{author}{Leutnant, V.}, \bibinfo{author}{Sehr, A.}, \bibinfo{author}{Kellermann, W.}, \bibinfo{author}{Maas, R.}, et~al., \bibinfo{year}{2013}.
\newblock \bibinfo{title}{The reverb challenge: A common evaluation framework for dereverberation and recognition of reverberant speech}, in: \bibinfo{booktitle}{Proc. of WASPAA}, \bibinfo{organization}{IEEE}. pp. \bibinfo{pages}{1--4}.
%Type = Inproceedings
\bibitem[{Ko et~al.(2017)Ko, Peddinti, Povey, Seltzer and Khudanpur}]{ko2017study}
\bibinfo{author}{Ko, T.}, \bibinfo{author}{Peddinti, V.}, \bibinfo{author}{Povey, D.}, \bibinfo{author}{Seltzer, M.L.}, \bibinfo{author}{Khudanpur, S.}, \bibinfo{year}{2017}.
\newblock \bibinfo{title}{A study on data augmentation of reverberant speech for robust speech recognition}, in: \bibinfo{booktitle}{Proc. of ICASSP}, \bibinfo{organization}{IEEE}. pp. \bibinfo{pages}{5220--5224}.
%Type = Inproceedings
\bibitem[{Koluguri et~al.(2022)Koluguri, Park and Ginsburg}]{koluguri2022titanet}
\bibinfo{author}{Koluguri, N.R.}, \bibinfo{author}{Park, T.}, \bibinfo{author}{Ginsburg, B.}, \bibinfo{year}{2022}.
\newblock \bibinfo{title}{Titanet: Neural model for speaker representation with 1d depth-wise separable convolutions and global context}, in: \bibinfo{booktitle}{Proc. of ICASSP}, \bibinfo{organization}{IEEE}. pp. \bibinfo{pages}{8102--8106}.
%Type = Inproceedings
\bibitem[{Kuchaiev et~al.(2019)}]{nemo_2019}
\bibinfo{author}{Kuchaiev, O.}, et~al., \bibinfo{year}{2019}.
\newblock \bibinfo{title}{{NeMo: a toolkit for building AI applications using neural modules}}, in: \bibinfo{booktitle}{Proc. Systems for ML Worshop, NeurIPS}.
%Type = Inproceedings
\bibitem[{Kudo and Richardson(2018)}]{kudo2018sentencepiece}
\bibinfo{author}{Kudo, T.}, \bibinfo{author}{Richardson, J.}, \bibinfo{year}{2018}.
\newblock \bibinfo{title}{Sentencepiece: A simple and language independent subword tokenizer and detokenizer for neural text processing}, in: \bibinfo{booktitle}{Proceedings of the 2018 Conference on Empirical Methods in Natural Language Processing: System Demonstrations}, pp. \bibinfo{pages}{66--71}.
%Type = Inproceedings
\bibitem[{Kumar and Byrne(2004)}]{kumar2004minimum}
\bibinfo{author}{Kumar, S.}, \bibinfo{author}{Byrne, B.}, \bibinfo{year}{2004}.
\newblock \bibinfo{title}{Minimum bayes-risk decoding for statistical machine translation}, in: \bibinfo{booktitle}{Proceedings of the Human Language Technology Conference of the North American Chapter of the Association for Computational Linguistics: HLT-NAACL 2004}, pp. \bibinfo{pages}{169--176}.
%Type = Inproceedings
\bibitem[{Lavechin et~al.(2023)Lavechin, M{\'e}tais, Titeux, Boissonnet, Copet, Rivi{\`e}re, Bergelson, Cristia, Dupoux and Bredin}]{lavechin2023brouhaha}
\bibinfo{author}{Lavechin, M.}, \bibinfo{author}{M{\'e}tais, M.}, \bibinfo{author}{Titeux, H.}, \bibinfo{author}{Boissonnet, A.}, \bibinfo{author}{Copet, J.}, \bibinfo{author}{Rivi{\`e}re, M.}, \bibinfo{author}{Bergelson, E.}, \bibinfo{author}{Cristia, A.}, \bibinfo{author}{Dupoux, E.}, \bibinfo{author}{Bredin, H.}, \bibinfo{year}{2023}.
\newblock \bibinfo{title}{Brouhaha: multi-task training for voice activity detection, speech-to-noise ratio, and c50 room acoustics estimation}, in: \bibinfo{booktitle}{Proc. of ASRU}, \bibinfo{organization}{IEEE}. pp. \bibinfo{pages}{1--7}.
%Type = Article
\bibitem[{Lee et~al.(1990)Lee, Hon and Reddy}]{lee1990overview}
\bibinfo{author}{Lee, K.F.}, \bibinfo{author}{Hon, H.W.}, \bibinfo{author}{Reddy, R.}, \bibinfo{year}{1990}.
\newblock \bibinfo{title}{An overview of the sphinx speech recognition system}.
\newblock \bibinfo{journal}{IEEE Transactions on Acoustics, Speech, and Signal Processing} \bibinfo{volume}{38}, \bibinfo{pages}{35--45}.
%Type = Inproceedings
\bibitem[{Lee et~al.(2024)Lee, Yun, Cai, Su and Song}]{lee2024unisumeval}
\bibinfo{author}{Lee, Y.}, \bibinfo{author}{Yun, T.}, \bibinfo{author}{Cai, J.}, \bibinfo{author}{Su, H.}, \bibinfo{author}{Song, H.}, \bibinfo{year}{2024}.
\newblock \bibinfo{title}{Unisumeval: Towards unified, fine-grained, multi-dimensional summarization evaluation for llms}, in: \bibinfo{booktitle}{Findings of the Association for Computational Linguistics: EMNLP 2024}, pp. \bibinfo{pages}{3941--3960}.
%Type = Article
\bibitem[{Li et~al.(2014)Li, Deng, Gong and Haeb-Umbach}]{li2014overview}
\bibinfo{author}{Li, J.}, \bibinfo{author}{Deng, L.}, \bibinfo{author}{Gong, Y.}, \bibinfo{author}{Haeb-Umbach, R.}, \bibinfo{year}{2014}.
\newblock \bibinfo{title}{An overview of noise-robust automatic speech recognition}.
\newblock \bibinfo{journal}{IEEE/ACM Transactions on Audio, Speech, and Language Processing} \bibinfo{volume}{22}, \bibinfo{pages}{745--777}.
%Type = Inproceedings
\bibitem[{Lin(2004)}]{lin2004rouge}
\bibinfo{author}{Lin, C.Y.}, \bibinfo{year}{2004}.
\newblock \bibinfo{title}{Rouge: A package for automatic evaluation of summaries}, in: \bibinfo{booktitle}{Text summarization branches out}, pp. \bibinfo{pages}{74--81}.
%Type = Inproceedings
\bibitem[{Lin et~al.(2022)Lin, Li and Lee}]{lin2022listen}
\bibinfo{author}{Lin, G.T.}, \bibinfo{author}{Li, S.W.}, \bibinfo{author}{Lee, H.y.}, \bibinfo{year}{2022}.
\newblock \bibinfo{title}{Listen, adapt, better wer: Source-free single-utterance test-time adaptation for automatic speech recognition}, in: \bibinfo{booktitle}{Interspeech}.
%Type = Article
\bibitem[{Liu et~al.(2023)Liu, Iter, Xu, Wang, Xu and Zhu}]{liu2023g}
\bibinfo{author}{Liu, Y.}, \bibinfo{author}{Iter, D.}, \bibinfo{author}{Xu, Y.}, \bibinfo{author}{Wang, S.}, \bibinfo{author}{Xu, R.}, \bibinfo{author}{Zhu, C.}, \bibinfo{year}{2023}.
\newblock \bibinfo{title}{G-eval: Nlg evaluation using gpt-4 with better human alignment}.
\newblock \bibinfo{journal}{arXiv preprint arXiv:2303.16634} .
%Type = Article
\bibitem[{Luo and Mesgarani(2019)}]{luo2018convtasnet}
\bibinfo{author}{Luo, Y.}, \bibinfo{author}{Mesgarani, N.}, \bibinfo{year}{2019}.
\newblock \bibinfo{title}{{Conv-TasNet}: Surpassing ideal time–frequency magnitude masking for speech separation} \bibinfo{volume}{27}, \bibinfo{pages}{1256–1266}.
%Type = Inproceedings
\bibitem[{Masuyama et~al.(2022)Masuyama, Chang, Cornell, Watanabe and Ono}]{masuyama2022end}
\bibinfo{author}{Masuyama, Y.}, \bibinfo{author}{Chang, X.}, \bibinfo{author}{Cornell, S.}, \bibinfo{author}{Watanabe, S.}, \bibinfo{author}{Ono, N.}, \bibinfo{year}{2022}.
\newblock \bibinfo{title}{End-to-end integration of speech recognition, dereverberation, beamforming, and self-supervised learning representation}.
%Type = Inproceedings
\bibitem[{Medennikov et~al.(2020a)Medennikov, Korenevsky, Prisyach, Khokhlov, Korenevskaya, Sorokin, Timofeeva, Mitrofanov, Andrusenko, Podluzhny et~al.}]{medennikov2020stc}
\bibinfo{author}{Medennikov, I.}, \bibinfo{author}{Korenevsky, M.}, \bibinfo{author}{Prisyach, T.}, \bibinfo{author}{Khokhlov, Y.}, \bibinfo{author}{Korenevskaya, M.}, \bibinfo{author}{Sorokin, I.}, \bibinfo{author}{Timofeeva, T.}, \bibinfo{author}{Mitrofanov, A.}, \bibinfo{author}{Andrusenko, A.}, \bibinfo{author}{Podluzhny, I.}, et~al., \bibinfo{year}{2020}a.
\newblock \bibinfo{title}{The stc system for the chime-6 challenge}, in: \bibinfo{booktitle}{CHiME Workshop}.
%Type = Inproceedings
\bibitem[{Medennikov et~al.(2020b)Medennikov, Korenevsky, Prisyach, Khokhlov, Korenevskaya, Sorokin, Timofeeva, Mitrofanov, Podluzhny, Romanenko et~al.}]{medennikov2020target}
\bibinfo{author}{Medennikov, I.}, \bibinfo{author}{Korenevsky, M.}, \bibinfo{author}{Prisyach, T.}, \bibinfo{author}{Khokhlov, Y.}, \bibinfo{author}{Korenevskaya, M.}, \bibinfo{author}{Sorokin, I.}, \bibinfo{author}{Timofeeva, T.}, \bibinfo{author}{Mitrofanov, A.}, \bibinfo{author}{Podluzhny, I.}, \bibinfo{author}{Romanenko, A.}, et~al., \bibinfo{year}{2020}b.
\newblock \bibinfo{title}{Target-speaker voice activity detection: A novel approach for multi-speaker diarization in a dinner party scenario}, in: \bibinfo{booktitle}{Interspeech}, pp. \bibinfo{pages}{274--278}.
%Type = Inproceedings
\bibitem[{Merity et~al.(2018)Merity, Keskar and Socher}]{merity2018regularizing}
\bibinfo{author}{Merity, S.}, \bibinfo{author}{Keskar, N.S.}, \bibinfo{author}{Socher, R.}, \bibinfo{year}{2018}.
\newblock \bibinfo{title}{Regularizing and optimizing lstm language models}, in: \bibinfo{booktitle}{International Conference on Learning Representations}.
%Type = Inproceedings
\bibitem[{Mitrofanov et~al.(2024)Mitrofanov, Prisyach, Timofeeva, Novoselov, Korenevsky, Khokhlov, Akulov, Anikin, Khalili, Lezhenin et~al.}]{mitrofanov2024stcon}
\bibinfo{author}{Mitrofanov, A.}, \bibinfo{author}{Prisyach, T.}, \bibinfo{author}{Timofeeva, T.}, \bibinfo{author}{Novoselov, S.}, \bibinfo{author}{Korenevsky, M.}, \bibinfo{author}{Khokhlov, Y.}, \bibinfo{author}{Akulov, A.}, \bibinfo{author}{Anikin, A.}, \bibinfo{author}{Khalili, R.}, \bibinfo{author}{Lezhenin, I.}, et~al., \bibinfo{year}{2024}.
\newblock \bibinfo{title}{Stcon system for the chime-8 challenge}, in: \bibinfo{booktitle}{CHiME Workshop}.
%Type = Article
\bibitem[{Mostefa et~al.(2007)Mostefa, Moreau, Choukri, Potamianos, Chu, Tyagi, Casas, Turmo, Cristoforetti, Tobia et~al.}]{mostefa2007chil}
\bibinfo{author}{Mostefa, D.}, \bibinfo{author}{Moreau, N.}, \bibinfo{author}{Choukri, K.}, \bibinfo{author}{Potamianos, G.}, \bibinfo{author}{Chu, S.M.}, \bibinfo{author}{Tyagi, A.}, \bibinfo{author}{Casas, J.R.}, \bibinfo{author}{Turmo, J.}, \bibinfo{author}{Cristoforetti, L.}, \bibinfo{author}{Tobia, F.}, et~al., \bibinfo{year}{2007}.
\newblock \bibinfo{title}{The chil audiovisual corpus for lecture and meeting analysis inside smart rooms}.
\newblock \bibinfo{journal}{Language resources and evaluation} \bibinfo{volume}{41}, \bibinfo{pages}{389--407}.
%Type = Inproceedings
\bibitem[{Mu et~al.(2023)Mu, Guo, Wang, Li, Li, Zhou, Chen and Xie}]{mu2023npu}
\bibinfo{author}{Mu, B.}, \bibinfo{author}{Guo, P.}, \bibinfo{author}{Wang, H.}, \bibinfo{author}{Li, Y.}, \bibinfo{author}{Li, Y.}, \bibinfo{author}{Zhou, P.}, \bibinfo{author}{Chen, W.}, \bibinfo{author}{Xie, L.}, \bibinfo{year}{2023}.
\newblock \bibinfo{title}{The npu system for dasr task of chime-7 challenge}, in: \bibinfo{booktitle}{CHiME Workshop}, pp. \bibinfo{pages}{63--66}.
%Type = Article
\bibitem[{von Neumann et~al.(2023)von Neumann, Boeddeker, Delcroix and Haeb-Umbach}]{von2023meeteval}
\bibinfo{author}{von Neumann, T.}, \bibinfo{author}{Boeddeker, C.}, \bibinfo{author}{Delcroix, M.}, \bibinfo{author}{Haeb-Umbach, R.}, \bibinfo{year}{2023}.
\newblock \bibinfo{title}{Meeteval: A toolkit for computation of word error rates for meeting transcription systems}.
\newblock \bibinfo{journal}{CHiME Workshop} .
%Type = Article
\bibitem[{Nguyen et~al.(2023)Nguyen, Kharitonov, Copet, Adi, Hsu, Elkahky et~al.}]{nguyen2023generative}
\bibinfo{author}{Nguyen, T.A.}, \bibinfo{author}{Kharitonov, E.}, \bibinfo{author}{Copet, J.}, \bibinfo{author}{Adi, Y.}, \bibinfo{author}{Hsu, W.N.}, \bibinfo{author}{Elkahky, A.}, et~al., \bibinfo{year}{2023}.
\newblock \bibinfo{title}{{Generative Spoken Dialogue Language Modeling}}.
\newblock \bibinfo{journal}{Transactions of the Association for Computational Linguistics} .
%Type = Inproceedings
\bibitem[{Niu et~al.(2024)Niu, Wang, Du, Yang, Tu, Wu, Qian, Wu, Xu, Zhang et~al.}]{niu2024ustc}
\bibinfo{author}{Niu, S.}, \bibinfo{author}{Wang, R.}, \bibinfo{author}{Du, J.}, \bibinfo{author}{Yang, G.}, \bibinfo{author}{Tu, Y.}, \bibinfo{author}{Wu, S.}, \bibinfo{author}{Qian, S.}, \bibinfo{author}{Wu, H.}, \bibinfo{author}{Xu, H.}, \bibinfo{author}{Zhang, X.}, et~al., \bibinfo{year}{2024}.
\newblock \bibinfo{title}{The ustc-nercslip systems for the chime-8 notsofar-1 challenge}, in: \bibinfo{booktitle}{CHiME Workshop}.
%Type = Inproceedings
\bibitem[{Novitasari et~al.(2022)Novitasari, Fukuda and Kurata}]{novitasari2022improving}
\bibinfo{author}{Novitasari, S.}, \bibinfo{author}{Fukuda, T.}, \bibinfo{author}{Kurata, G.}, \bibinfo{year}{2022}.
\newblock \bibinfo{title}{Improving asr robustness in noisy condition through vad integration.}, in: \bibinfo{booktitle}{Interspeech}, pp. \bibinfo{pages}{3784--3788}.
%Type = Article
\bibitem[{Ogawa et~al.(2024)Ogawa, Kamo, Matsuura, Ashihara, Moriya, Kano, Tawara and Delcroix}]{ogawa2024applying}
\bibinfo{author}{Ogawa, A.}, \bibinfo{author}{Kamo, N.}, \bibinfo{author}{Matsuura, K.}, \bibinfo{author}{Ashihara, T.}, \bibinfo{author}{Moriya, T.}, \bibinfo{author}{Kano, T.}, \bibinfo{author}{Tawara, N.}, \bibinfo{author}{Delcroix, M.}, \bibinfo{year}{2024}.
\newblock \bibinfo{title}{Applying {LLMs} for rescoring n-best asr hypotheses of casual conversations: Effects of domain adaptation and context carry-over}.
\newblock \bibinfo{journal}{CHiME Workshop} .
%Type = Misc
\bibitem[{OpenAI(2024)}]{openai-2024-gpt4o}
\bibinfo{author}{OpenAI}, \bibinfo{year}{2024}.
\newblock \bibinfo{title}{{GPT-4o System Card}}.
\newblock \URLprefix \url{https://cdn.openai.com/gpt-4o-system-card.pdf}.
%Type = Inproceedings
\bibitem[{Ozaki et~al.(2020)Ozaki, Tanigaki, Watanabe and Onishi}]{ozaki2020multiobjective}
\bibinfo{author}{Ozaki, Y.}, \bibinfo{author}{Tanigaki, Y.}, \bibinfo{author}{Watanabe, S.}, \bibinfo{author}{Onishi, M.}, \bibinfo{year}{2020}.
\newblock \bibinfo{title}{Multiobjective tree-structured parzen estimator for computationally expensive optimization problems}, in: \bibinfo{booktitle}{Proceedings of the 2020 genetic and evolutionary computation conference}, pp. \bibinfo{pages}{533--541}.
%Type = Inproceedings
\bibitem[{Panayotov et~al.(2015)Panayotov, Chen, Povey and Khudanpur}]{panayotov2015librispeech}
\bibinfo{author}{Panayotov, V.}, \bibinfo{author}{Chen, G.}, \bibinfo{author}{Povey, D.}, \bibinfo{author}{Khudanpur, S.}, \bibinfo{year}{2015}.
\newblock \bibinfo{title}{{LibriSpeech}: an {ASR} corpus based on public domain audio books}, in: \bibinfo{booktitle}{Proc. of ICASSP}.
%Type = Inproceedings
\bibitem[{Park et~al.(2020)Park, Zhang, Chiu, Chen, Li, Chan, Le and Wu}]{park2020specaugment}
\bibinfo{author}{Park, D.S.}, \bibinfo{author}{Zhang, Y.}, \bibinfo{author}{Chiu, C.C.}, \bibinfo{author}{Chen, Y.}, \bibinfo{author}{Li, B.}, \bibinfo{author}{Chan, W.}, \bibinfo{author}{Le, Q.V.}, \bibinfo{author}{Wu, Y.}, \bibinfo{year}{2020}.
\newblock \bibinfo{title}{Specaugment on large scale datasets}, in: \bibinfo{booktitle}{Proc. of ICASSP}, \bibinfo{organization}{IEEE}. pp. \bibinfo{pages}{6879--6883}.
%Type = Article
\bibitem[{Park et~al.(2024)Park, Medennikov, Dhawan, Wang, Huang, Koluguri, Puvvada, Balam and Ginsburg}]{park2024sortformer}
\bibinfo{author}{Park, T.}, \bibinfo{author}{Medennikov, I.}, \bibinfo{author}{Dhawan, K.}, \bibinfo{author}{Wang, W.}, \bibinfo{author}{Huang, H.}, \bibinfo{author}{Koluguri, N.R.}, \bibinfo{author}{Puvvada, K.C.}, \bibinfo{author}{Balam, J.}, \bibinfo{author}{Ginsburg, B.}, \bibinfo{year}{2024}.
\newblock \bibinfo{title}{Sortformer: Seamless integration of speaker diarization and asr by bridging timestamps and tokens}.
\newblock \bibinfo{journal}{Submitted to IEEE/ACM Transaction of Audio, Speech and Language Processing} .
%Type = Article
\bibitem[{Park et~al.(2023a)Park, Dhawan, Koluguri and Balam}]{park2023enhancing}
\bibinfo{author}{Park, T.J.}, \bibinfo{author}{Dhawan, K.}, \bibinfo{author}{Koluguri, N.}, \bibinfo{author}{Balam, J.}, \bibinfo{year}{2023}a.
\newblock \bibinfo{title}{Enhancing speaker diarization with large language models: A contextual beam search approach}.
\newblock \bibinfo{journal}{arXiv preprint arXiv:2309.05248} .
%Type = Article
\bibitem[{Park et~al.(2019)Park, Han, Kumar and Narayanan}]{park2019auto}
\bibinfo{author}{Park, T.J.}, \bibinfo{author}{Han, K.J.}, \bibinfo{author}{Kumar, M.}, \bibinfo{author}{Narayanan, S.}, \bibinfo{year}{2019}.
\newblock \bibinfo{title}{Auto-tuning spectral clustering for speaker diarization using normalized maximum eigengap}.
\newblock \bibinfo{journal}{IEEE Signal Processing Letters} \bibinfo{volume}{27}, \bibinfo{pages}{381--385}.
%Type = Article
\bibitem[{Park et~al.(2023b)Park, Huang, Hooper, Koluguri, Dhawan, Jukic, Balam and Ginsburg}]{park2023property}
\bibinfo{author}{Park, T.J.}, \bibinfo{author}{Huang, H.}, \bibinfo{author}{Hooper, C.}, \bibinfo{author}{Koluguri, N.}, \bibinfo{author}{Dhawan, K.}, \bibinfo{author}{Jukic, A.}, \bibinfo{author}{Balam, J.}, \bibinfo{author}{Ginsburg, B.}, \bibinfo{year}{2023}b.
\newblock \bibinfo{title}{Property-aware multi-speaker data simulation: A probabilistic modelling technique for synthetic data generation}.
\newblock \bibinfo{journal}{Interspeech} .
%Type = Article
\bibitem[{Park et~al.(2023c)Park, Huang, Jukic, Dhawan, Puvvada, Koluguri, Karpov, Laptev, Balam and Ginsburg}]{park2023chime}
\bibinfo{author}{Park, T.J.}, \bibinfo{author}{Huang, H.}, \bibinfo{author}{Jukic, A.}, \bibinfo{author}{Dhawan, K.}, \bibinfo{author}{Puvvada, K.C.}, \bibinfo{author}{Koluguri, N.}, \bibinfo{author}{Karpov, N.}, \bibinfo{author}{Laptev, A.}, \bibinfo{author}{Balam, J.}, \bibinfo{author}{Ginsburg, B.}, \bibinfo{year}{2023}c.
\newblock \bibinfo{title}{The {CHiME-7 Challenge}: System description and performance of nemo team's dasr system}.
\newblock \bibinfo{journal}{CHiME Workshop} .
%Type = Article
\bibitem[{Park et~al.(2022)Park, Koluguri, Balam and Ginsburg}]{park2022multi}
\bibinfo{author}{Park, T.J.}, \bibinfo{author}{Koluguri, N.R.}, \bibinfo{author}{Balam, J.}, \bibinfo{author}{Ginsburg, B.}, \bibinfo{year}{2022}.
\newblock \bibinfo{title}{Multi-scale speaker diarization with dynamic scale weighting}.
\newblock \bibinfo{journal}{Interspeech} .
%Type = Inproceedings
\bibitem[{Paul and Baker(1992)}]{paul1992design}
\bibinfo{author}{Paul, D.B.}, \bibinfo{author}{Baker, J.}, \bibinfo{year}{1992}.
\newblock \bibinfo{title}{The design for the wall street journal-based {CSR} corpus}, in: \bibinfo{booktitle}{Speech and Natural Language Workshop}.
%Type = Article
\bibitem[{Peng et~al.(2024)}]{slm_survey2}
\bibinfo{author}{Peng, J.}, et~al., \bibinfo{year}{2024}.
\newblock \bibinfo{title}{A survey on speech large language models}.
\newblock \bibinfo{journal}{arXiv preprint arXiv:2410.18908} .
%Type = Inproceedings
\bibitem[{Peng et~al.(2022)Peng, Dalmia, Lane and Watanabe}]{peng2022branchformer}
\bibinfo{author}{Peng, Y.}, \bibinfo{author}{Dalmia, S.}, \bibinfo{author}{Lane, I.}, \bibinfo{author}{Watanabe, S.}, \bibinfo{year}{2022}.
\newblock \bibinfo{title}{Branchformer: Parallel mlp-attention architectures to capture local and global context for speech recognition and understanding}, in: \bibinfo{booktitle}{International Conference on Machine Learning}, \bibinfo{organization}{PMLR}. pp. \bibinfo{pages}{17627--17643}.
%Type = Inproceedings
\bibitem[{Peng et~al.(2023)Peng, Tian, Yan, Berrebbi, Chang, Li, Shi, Arora, Chen, Sharma et~al.}]{peng2023reproducing}
\bibinfo{author}{Peng, Y.}, \bibinfo{author}{Tian, J.}, \bibinfo{author}{Yan, B.}, \bibinfo{author}{Berrebbi, D.}, \bibinfo{author}{Chang, X.}, \bibinfo{author}{Li, X.}, \bibinfo{author}{Shi, J.}, \bibinfo{author}{Arora, S.}, \bibinfo{author}{Chen, W.}, \bibinfo{author}{Sharma, R.}, et~al., \bibinfo{year}{2023}.
\newblock \bibinfo{title}{Reproducing {Whisper}-style training using an open-source toolkit and publicly available data}, in: \bibinfo{booktitle}{Proc. of ASRU}, \bibinfo{organization}{IEEE}. pp. \bibinfo{pages}{1--8}.
%Type = Inproceedings
\bibitem[{Polok et~al.(2024a)Polok, Klement, Han, Sedlacek, Yusuf, Maciejewski, Wiesner and Burget}]{maciejewskibut}
\bibinfo{author}{Polok, A.}, \bibinfo{author}{Klement, D.}, \bibinfo{author}{Han, J.}, \bibinfo{author}{Sedlacek, S.}, \bibinfo{author}{Yusuf, B.}, \bibinfo{author}{Maciejewski, M.}, \bibinfo{author}{Wiesner, M.}, \bibinfo{author}{Burget, L.}, \bibinfo{year}{2024}a.
\newblock \bibinfo{title}{But/jhu system description for chime-8 notsofar-1 challenge}, in: \bibinfo{booktitle}{CHiME Workshop}.
%Type = Inproceedings
\bibitem[{Polok et~al.(2024b)Polok, Klement, Wiesner, Khudanpur, {\v{C}}ernock{\`y} and Burget}]{polok2024target}
\bibinfo{author}{Polok, A.}, \bibinfo{author}{Klement, D.}, \bibinfo{author}{Wiesner, M.}, \bibinfo{author}{Khudanpur, S.}, \bibinfo{author}{{\v{C}}ernock{\`y}, J.}, \bibinfo{author}{Burget, L.}, \bibinfo{year}{2024}b.
\newblock \bibinfo{title}{Target speaker asr with whisper}, in: \bibinfo{booktitle}{submitted to ICASSP 2025}.
%Type = Misc
\bibitem[{Povey(2020)}]{poveyk2}
\bibinfo{author}{Povey, D.}, \bibinfo{year}{2020}.
\newblock \bibinfo{title}{k2}.
\newblock \bibinfo{howpublished}{\url{https://github.com/k2-fsa/k2}}.
\newblock \URLprefix \url{https://github.com/k2-fsa/k2}. \bibinfo{note}{[Accessed 12-19-2024]}.
%Type = Inproceedings
\bibitem[{Povey et~al.(2011)Povey, Ghoshal, Boulianne, Burget, Glembek, Goel, Hannemann, Motlicek, Qian, Schwarz et~al.}]{povey2011kaldi}
\bibinfo{author}{Povey, D.}, \bibinfo{author}{Ghoshal, A.}, \bibinfo{author}{Boulianne, G.}, \bibinfo{author}{Burget, L.}, \bibinfo{author}{Glembek, O.}, \bibinfo{author}{Goel, N.}, \bibinfo{author}{Hannemann, M.}, \bibinfo{author}{Motlicek, P.}, \bibinfo{author}{Qian, Y.}, \bibinfo{author}{Schwarz, P.}, et~al., \bibinfo{year}{2011}.
\newblock \bibinfo{title}{The kaldi speech recognition toolkit}, in: \bibinfo{booktitle}{Proc. of ASRU}.
%Type = Article
\bibitem[{Prabhavalkar et~al.(2023)Prabhavalkar, Hori, Sainath, Schl{\"u}ter and Watanabe}]{prabhavalkar2023end}
\bibinfo{author}{Prabhavalkar, R.}, \bibinfo{author}{Hori, T.}, \bibinfo{author}{Sainath, T.N.}, \bibinfo{author}{Schl{\"u}ter, R.}, \bibinfo{author}{Watanabe, S.}, \bibinfo{year}{2023}.
\newblock \bibinfo{title}{End-to-end speech recognition: A survey}.
\newblock \bibinfo{journal}{arXiv preprint arXiv:2303.03329} .
%Type = Inproceedings
\bibitem[{Prisyach et~al.(2023)Prisyach, Khokhlov, Korenevsky, Mitrofanov, Timofeeva, Odegov, Nasretdinov, Lezhenin, Miroshnichenko, Karelin et~al.}]{prisyach2023stcon}
\bibinfo{author}{Prisyach, T.}, \bibinfo{author}{Khokhlov, Y.}, \bibinfo{author}{Korenevsky, M.}, \bibinfo{author}{Mitrofanov, A.}, \bibinfo{author}{Timofeeva, T.}, \bibinfo{author}{Odegov, I.}, \bibinfo{author}{Nasretdinov, R.}, \bibinfo{author}{Lezhenin, I.}, \bibinfo{author}{Miroshnichenko, D.}, \bibinfo{author}{Karelin, A.}, et~al., \bibinfo{year}{2023}.
\newblock \bibinfo{title}{Stcon system for the chime-7 challenge}, in: \bibinfo{booktitle}{CHiME Workshop}.
%Type = Article
\bibitem[{Rabiner(1989)}]{rabiner1989tutorial}
\bibinfo{author}{Rabiner, L.R.}, \bibinfo{year}{1989}.
\newblock \bibinfo{title}{A tutorial on hidden markov models and selected applications in speech recognition}.
\newblock \bibinfo{journal}{Proceedings of the IEEE} \bibinfo{volume}{77}, \bibinfo{pages}{257--286}.
%Type = Article
\bibitem[{Radford et~al.(2022)Radford, Kim, Xu, Brockman, McLeavey and Sutskever}]{radford2022robust}
\bibinfo{author}{Radford, A.}, \bibinfo{author}{Kim, J.W.}, \bibinfo{author}{Xu, T.}, \bibinfo{author}{Brockman, G.}, \bibinfo{author}{McLeavey, C.}, \bibinfo{author}{Sutskever, I.}, \bibinfo{year}{2022}.
\newblock \bibinfo{title}{Robust speech recognition via large-scale weak supervision}.
\newblock \bibinfo{journal}{ArXiv} .
%Type = Article
\bibitem[{Raffel et~al.(2020)Raffel, Shazeer, Roberts, Lee, Narang, Matena, Zhou, Li and Liu}]{raffel2020exploring}
\bibinfo{author}{Raffel, C.}, \bibinfo{author}{Shazeer, N.}, \bibinfo{author}{Roberts, A.}, \bibinfo{author}{Lee, K.}, \bibinfo{author}{Narang, S.}, \bibinfo{author}{Matena, M.}, \bibinfo{author}{Zhou, Y.}, \bibinfo{author}{Li, W.}, \bibinfo{author}{Liu, P.J.}, \bibinfo{year}{2020}.
\newblock \bibinfo{title}{Exploring the limits of transfer learning with a unified text-to-text transformer}.
\newblock \bibinfo{journal}{Journal of machine learning research} \bibinfo{volume}{21}, \bibinfo{pages}{1--67}.
%Type = Inproceedings
\bibitem[{Raj et~al.(2021a)Raj, Denisov, Chen, Erdogan, Huang, He, Watanabe, Du, Yoshioka, Luo et~al.}]{raj2021integration}
\bibinfo{author}{Raj, D.}, \bibinfo{author}{Denisov, P.}, \bibinfo{author}{Chen, Z.}, \bibinfo{author}{Erdogan, H.}, \bibinfo{author}{Huang, Z.}, \bibinfo{author}{He, M.}, \bibinfo{author}{Watanabe, S.}, \bibinfo{author}{Du, J.}, \bibinfo{author}{Yoshioka, T.}, \bibinfo{author}{Luo, Y.}, et~al., \bibinfo{year}{2021}a.
\newblock \bibinfo{title}{Integration of speech separation, diarization, and recognition for multi-speaker meetings: System description, comparison, and analysis}, in: \bibinfo{booktitle}{IEEE SLT}.
%Type = Inproceedings
\bibitem[{Raj et~al.(2021b)Raj, Garcia-Perera, Huang, Watanabe, Povey, Stolcke and Khudanpur}]{raj2021dover}
\bibinfo{author}{Raj, D.}, \bibinfo{author}{Garcia-Perera, L.P.}, \bibinfo{author}{Huang, Z.}, \bibinfo{author}{Watanabe, S.}, \bibinfo{author}{Povey, D.}, \bibinfo{author}{Stolcke, A.}, \bibinfo{author}{Khudanpur, S.}, \bibinfo{year}{2021}b.
\newblock \bibinfo{title}{{DOVER-Lap}: A method for combining overlap-aware diarization outputs}, in: \bibinfo{booktitle}{Proc. of SLT}, \bibinfo{organization}{IEEE}. pp. \bibinfo{pages}{881--888}.
%Type = Inproceedings
\bibitem[{Raj et~al.(2023)Raj, Povey and Khudanpur}]{Raj2022GPUacceleratedGS}
\bibinfo{author}{Raj, D.}, \bibinfo{author}{Povey, D.}, \bibinfo{author}{Khudanpur, S.}, \bibinfo{year}{2023}.
\newblock \bibinfo{title}{{GPU}-accelerated guided source separation for meeting transcription}, in: \bibinfo{booktitle}{Interspeech}.
%Type = Article
\bibitem[{Reynolds et~al.(2000)Reynolds, Quatieri and Dunn}]{reynolds2000speaker}
\bibinfo{author}{Reynolds, D.A.}, \bibinfo{author}{Quatieri, T.F.}, \bibinfo{author}{Dunn, R.B.}, \bibinfo{year}{2000}.
\newblock \bibinfo{title}{Speaker verification using adapted gaussian mixture models}.
\newblock \bibinfo{journal}{Digital Signal Processing} \bibinfo{volume}{10}, \bibinfo{pages}{19--41}.
%Type = Inproceedings
\bibitem[{Richey et~al.(2018)Richey, Barrios, Armstrong, Bartels, Franco, Graciarena, Lawson, Nandwana, Stauffer, van Hout et~al.}]{richey2018voices}
\bibinfo{author}{Richey, C.}, \bibinfo{author}{Barrios, M.A.}, \bibinfo{author}{Armstrong, Z.}, \bibinfo{author}{Bartels, C.}, \bibinfo{author}{Franco, H.}, \bibinfo{author}{Graciarena, M.}, \bibinfo{author}{Lawson, A.}, \bibinfo{author}{Nandwana, M.K.}, \bibinfo{author}{Stauffer, A.}, \bibinfo{author}{van Hout, J.}, et~al., \bibinfo{year}{2018}.
\newblock \bibinfo{title}{Voices obscured in complex environmental settings ({VOiCES}) corpus}, in: \bibinfo{booktitle}{Interspeech}.
%Type = Inproceedings
\bibitem[{Ryant et~al.(2019)Ryant, Church, Cieri, Cristia, Du and Ganapathy}]{ryant2019second}
\bibinfo{author}{Ryant, N.}, \bibinfo{author}{Church, K.}, \bibinfo{author}{Cieri, C.}, \bibinfo{author}{Cristia, A.}, \bibinfo{author}{Du, J.}, \bibinfo{author}{Ganapathy, M.L.}, \bibinfo{year}{2019}.
\newblock \bibinfo{title}{The second dihard diarization challenge: Dataset, task, and baselines}, in: \bibinfo{booktitle}{Interspeech}.
%Type = Article
\bibitem[{Snyder et~al.(2015)Snyder, Chen and Povey}]{snyder2015musan}
\bibinfo{author}{Snyder, D.}, \bibinfo{author}{Chen, G.}, \bibinfo{author}{Povey, D.}, \bibinfo{year}{2015}.
\newblock \bibinfo{title}{{MUSAN}: A music, speech, and noise corpus}.
\newblock \bibinfo{journal}{ArXiv} .
%Type = Article
\bibitem[{Stupakov et~al.(2012)Stupakov, Hanusa, Vijaywargi, Fox and Bilmes}]{stupakov2012design}
\bibinfo{author}{Stupakov, A.}, \bibinfo{author}{Hanusa, E.}, \bibinfo{author}{Vijaywargi, D.}, \bibinfo{author}{Fox, D.}, \bibinfo{author}{Bilmes, J.}, \bibinfo{year}{2012}.
\newblock \bibinfo{title}{The design and collection of cosine, a multi-microphone in situ speech corpus recorded in noisy environments}.
\newblock \bibinfo{journal}{Computer Speech \& Language} \bibinfo{volume}{26}, \bibinfo{pages}{52--66}.
%Type = Inproceedings
\bibitem[{Szyma{\'n}ski et~al.(2020)Szyma{\'n}ski, {\.Z}elasko, Morzy, Szymczak, {\.Z}y{\l}a-Hoppe, Banaszczak, Augustyniak, Mizgajski and Carmiel}]{szymanski2020we}
\bibinfo{author}{Szyma{\'n}ski, P.}, \bibinfo{author}{{\.Z}elasko, P.}, \bibinfo{author}{Morzy, M.}, \bibinfo{author}{Szymczak, A.}, \bibinfo{author}{{\.Z}y{\l}a-Hoppe, M.}, \bibinfo{author}{Banaszczak, J.}, \bibinfo{author}{Augustyniak, L.}, \bibinfo{author}{Mizgajski, J.}, \bibinfo{author}{Carmiel, Y.}, \bibinfo{year}{2020}.
\newblock \bibinfo{title}{{WER} we are and {WER} we think we are}, in: \bibinfo{booktitle}{Findings of EMNLP}.
%Type = Article
\bibitem[{Touvron et~al.(2023)Touvron, Martin, Stone, Albert, Almahairi, Babaei, Bashlykov, Batra, Bhargava, Bhosale et~al.}]{touvron2023llama}
\bibinfo{author}{Touvron, H.}, \bibinfo{author}{Martin, L.}, \bibinfo{author}{Stone, K.}, \bibinfo{author}{Albert, P.}, \bibinfo{author}{Almahairi, A.}, \bibinfo{author}{Babaei, Y.}, \bibinfo{author}{Bashlykov, N.}, \bibinfo{author}{Batra, S.}, \bibinfo{author}{Bhargava, P.}, \bibinfo{author}{Bhosale, S.}, et~al., \bibinfo{year}{2023}.
\newblock \bibinfo{title}{{Llama 2}: Open foundation and fine-tuned chat models}.
\newblock \bibinfo{journal}{arXiv preprint arXiv:2307.09288} .
%Type = Article
\bibitem[{Van~Segbroeck et~al.(2019)Van~Segbroeck, Zaid, Kutsenko, Huerta, Nguyen, Luo, Hoffmeister, Trmal, Omologo and Maas}]{van2019dipco}
\bibinfo{author}{Van~Segbroeck, M.}, \bibinfo{author}{Zaid, A.}, \bibinfo{author}{Kutsenko, K.}, \bibinfo{author}{Huerta, C.}, \bibinfo{author}{Nguyen, T.}, \bibinfo{author}{Luo, X.}, \bibinfo{author}{Hoffmeister, B.}, \bibinfo{author}{Trmal, J.}, \bibinfo{author}{Omologo, M.}, \bibinfo{author}{Maas, R.}, \bibinfo{year}{2019}.
\newblock \bibinfo{title}{{DiPCo} -- dinner party corpus}.
\newblock \bibinfo{journal}{ArXiv} .
%Type = Inproceedings
\bibitem[{Vaswani et~al.(2017)Vaswani, Shazeer, Parmar, Uszkoreit, Jones, Gomez, Kaiser and Polosukhin}]{vaswani2017attention}
\bibinfo{author}{Vaswani, A.}, \bibinfo{author}{Shazeer, N.}, \bibinfo{author}{Parmar, N.}, \bibinfo{author}{Uszkoreit, J.}, \bibinfo{author}{Jones, L.}, \bibinfo{author}{Gomez, A.N.}, \bibinfo{author}{Kaiser, {\L}.}, \bibinfo{author}{Polosukhin, I.}, \bibinfo{year}{2017}.
\newblock \bibinfo{title}{Attention is all you need}, in: \bibinfo{booktitle}{NIPS}.
%Type = Inproceedings
\bibitem[{Vincent et~al.(2013)Vincent, Barker, Watanabe, Le~Roux, Nesta and Matassoni}]{vincent2013second}
\bibinfo{author}{Vincent, E.}, \bibinfo{author}{Barker, J.}, \bibinfo{author}{Watanabe, S.}, \bibinfo{author}{Le~Roux, J.}, \bibinfo{author}{Nesta, F.}, \bibinfo{author}{Matassoni, M.}, \bibinfo{year}{2013}.
\newblock \bibinfo{title}{The second {CHiME} speech separation and recognition challenge: An overview of challenge systems and outcomes}, in: \bibinfo{booktitle}{IEEE ASRU}.
%Type = Article
\bibitem[{Vincent et~al.(2006)Vincent, Gribonval and F{\'e}votte}]{vincent2006performance}
\bibinfo{author}{Vincent, E.}, \bibinfo{author}{Gribonval, R.}, \bibinfo{author}{F{\'e}votte, C.}, \bibinfo{year}{2006}.
\newblock \bibinfo{title}{Performance measurement in blind audio source separation}.
\newblock \bibinfo{journal}{IEEE Trans. Audio, Speech, Lang. Process.} \bibinfo{volume}{14}, \bibinfo{pages}{1462--1469}.
%Type = Article
\bibitem[{Vincent et~al.(2016)Vincent, Watanabe, Barker and Marxer}]{vincent20164th}
\bibinfo{author}{Vincent, E.}, \bibinfo{author}{Watanabe, S.}, \bibinfo{author}{Barker, J.}, \bibinfo{author}{Marxer, R.}, \bibinfo{year}{2016}.
\newblock \bibinfo{title}{The 4th chime speech separation and recognition challenge}.
\newblock \bibinfo{journal}{CHiME Workshop} .
%Type = Article
\bibitem[{Vinnikov et~al.(2024)Vinnikov, Ivry, Hurvitz, Abramovski, Koubi, Gurvich, Peer, Xiao, Elizalde, Kanda et~al.}]{vinnikov2024notsofar}
\bibinfo{author}{Vinnikov, A.}, \bibinfo{author}{Ivry, A.}, \bibinfo{author}{Hurvitz, A.}, \bibinfo{author}{Abramovski, I.}, \bibinfo{author}{Koubi, S.}, \bibinfo{author}{Gurvich, I.}, \bibinfo{author}{Peer, S.}, \bibinfo{author}{Xiao, X.}, \bibinfo{author}{Elizalde, B.M.}, \bibinfo{author}{Kanda, N.}, et~al., \bibinfo{year}{2024}.
\newblock \bibinfo{title}{{NOTSOFAR-1} challenge: New datasets, baseline, and tasks for distant meeting transcription}.
\newblock \bibinfo{journal}{Interspeech} .
%Type = Inproceedings
\bibitem[{Wang et~al.(2021)Wang, Riviere, Lee, Wu, Talnikar, Haziza, Williamson, Pino and Dupoux}]{wang2021voxpopuli}
\bibinfo{author}{Wang, C.}, \bibinfo{author}{Riviere, M.}, \bibinfo{author}{Lee, A.}, \bibinfo{author}{Wu, A.}, \bibinfo{author}{Talnikar, C.}, \bibinfo{author}{Haziza, D.}, \bibinfo{author}{Williamson, M.}, \bibinfo{author}{Pino, J.}, \bibinfo{author}{Dupoux, E.}, \bibinfo{year}{2021}.
\newblock \bibinfo{title}{Voxpopuli: A large-scale multilingual speech corpus for representation learning, semi-supervised learning and interpretation}, in: \bibinfo{booktitle}{Proceedings of the 59th Annual Meeting of the Association for Computational Linguistics and the 11th International Joint Conference on Natural Language Processing (Volume 1: Long Papers)}, pp. \bibinfo{pages}{993--1003}.
%Type = Inproceedings
\bibitem[{Wang et~al.(2024a)Wang, Xiao, Kanda, Yousefi, Yoshioka and Wu}]{wang2024profile}
\bibinfo{author}{Wang, D.}, \bibinfo{author}{Xiao, X.}, \bibinfo{author}{Kanda, N.}, \bibinfo{author}{Yousefi, M.}, \bibinfo{author}{Yoshioka, T.}, \bibinfo{author}{Wu, J.}, \bibinfo{year}{2024}a.
\newblock \bibinfo{title}{Profile-error-tolerant target-speaker voice activity detection}, in: \bibinfo{booktitle}{Proc. of ICASSP}, \bibinfo{organization}{IEEE}. pp. \bibinfo{pages}{11906--11910}.
%Type = Article
\bibitem[{Wang et~al.(2024b)Wang, Huang, Zhao, Clark, Xia and Liao}]{wang2024diarizationlm}
\bibinfo{author}{Wang, Q.}, \bibinfo{author}{Huang, Y.}, \bibinfo{author}{Zhao, G.}, \bibinfo{author}{Clark, E.}, \bibinfo{author}{Xia, W.}, \bibinfo{author}{Liao, H.}, \bibinfo{year}{2024}b.
\newblock \bibinfo{title}{{DiarizationLM}: Speaker diarization post-processing with large language models}.
\newblock \bibinfo{journal}{Interspeech} .
%Type = Article
\bibitem[{Wang et~al.(2023a)Wang, He, Du, Zhou, Niu, Chen, Yue, Yang, Wu, Sun et~al.}]{wang2023ustc}
\bibinfo{author}{Wang, R.}, \bibinfo{author}{He, M.}, \bibinfo{author}{Du, J.}, \bibinfo{author}{Zhou, H.}, \bibinfo{author}{Niu, S.}, \bibinfo{author}{Chen, H.}, \bibinfo{author}{Yue, Y.}, \bibinfo{author}{Yang, G.}, \bibinfo{author}{Wu, S.}, \bibinfo{author}{Sun, L.}, et~al., \bibinfo{year}{2023}a.
\newblock \bibinfo{title}{The {USTC-NERCSLIP} systems for the {CHiME-7 DASR} challenge}.
\newblock \bibinfo{journal}{CHiME Workshop} .
%Type = Inproceedings
\bibitem[{Wang et~al.(2023b)Wang, Wu, Chen, He, Du, Lee, Chen, Watanabe, Siniscalchi, Scharenborg, Liu, Yin, Pan, Gao and Liu}]{Wang2023misp}
\bibinfo{author}{Wang, Z.}, \bibinfo{author}{Wu, S.}, \bibinfo{author}{Chen, H.}, \bibinfo{author}{He, M.K.}, \bibinfo{author}{Du, J.}, \bibinfo{author}{Lee, C.H.}, \bibinfo{author}{Chen, J.}, \bibinfo{author}{Watanabe, S.}, \bibinfo{author}{Siniscalchi, S.}, \bibinfo{author}{Scharenborg, O.}, \bibinfo{author}{Liu, D.}, \bibinfo{author}{Yin, B.}, \bibinfo{author}{Pan, J.}, \bibinfo{author}{Gao, J.}, \bibinfo{author}{Liu, C.}, \bibinfo{year}{2023}b.
\newblock \bibinfo{title}{The multimodal information based speech processing (misp) 2022 challenge: {Audio}-visual diarization and recognition}, in: \bibinfo{booktitle}{Proc. of ICASSP}, pp. \bibinfo{pages}{1--5}.
%Type = Article
\bibitem[{Warsitz and Haeb-Umbach(2007)}]{warsitz2007blind}
\bibinfo{author}{Warsitz, E.}, \bibinfo{author}{Haeb-Umbach, R.}, \bibinfo{year}{2007}.
\newblock \bibinfo{title}{Blind acoustic beamforming based on generalized eigenvalue decomposition}.
\newblock \bibinfo{journal}{IEEE Transactions on audio, speech, and language processing} \bibinfo{volume}{15}, \bibinfo{pages}{1529--1539}.
%Type = Inproceedings
\bibitem[{Watanabe et~al.(2018)Watanabe, Hori, Karita, Hayashi, Nishitoba, Unno, {Enrique Yalta Soplin}, Heymann, Wiesner, Chen, Renduchintala and Ochiai}]{watanabe2018espnet}
\bibinfo{author}{Watanabe, S.}, \bibinfo{author}{Hori, T.}, \bibinfo{author}{Karita, S.}, \bibinfo{author}{Hayashi, T.}, \bibinfo{author}{Nishitoba, J.}, \bibinfo{author}{Unno, Y.}, \bibinfo{author}{{Enrique Yalta Soplin}, N.}, \bibinfo{author}{Heymann, J.}, \bibinfo{author}{Wiesner, M.}, \bibinfo{author}{Chen, N.}, \bibinfo{author}{Renduchintala, A.}, \bibinfo{author}{Ochiai, T.}, \bibinfo{year}{2018}.
\newblock \bibinfo{title}{{ESPnet}: End-to-end speech processing toolkit}, in: \bibinfo{booktitle}{Interspeech}.
%Type = Article
\bibitem[{Watanabe et~al.(2017)Watanabe, Hori, Kim, Hershey and Hayashi}]{Watanabe2017}
\bibinfo{author}{Watanabe, S.}, \bibinfo{author}{Hori, T.}, \bibinfo{author}{Kim, S.}, \bibinfo{author}{Hershey, J.R.}, \bibinfo{author}{Hayashi, T.}, \bibinfo{year}{2017}.
\newblock \bibinfo{title}{Hybrid {C}{T}{C}/attention architecture for end-to-end speech recognition}.
\newblock \bibinfo{journal}{IEEE J. Sel. Top. Signal Process.} \bibinfo{volume}{11}, \bibinfo{pages}{1240--1253}.
%Type = Article
\bibitem[{Watanabe et~al.(2020)Watanabe, Mandel, Barker, Vincent, Arora, Chang, Khudanpur, Manohar, Povey, Raj et~al.}]{watanabe2020chime}
\bibinfo{author}{Watanabe, S.}, \bibinfo{author}{Mandel, M.}, \bibinfo{author}{Barker, J.}, \bibinfo{author}{Vincent, E.}, \bibinfo{author}{Arora, A.}, \bibinfo{author}{Chang, X.}, \bibinfo{author}{Khudanpur, S.}, \bibinfo{author}{Manohar, V.}, \bibinfo{author}{Povey, D.}, \bibinfo{author}{Raj, D.}, et~al., \bibinfo{year}{2020}.
\newblock \bibinfo{title}{{CHiME-6} challenge: Tackling multispeaker speech recognition for unsegmented recordings}.
\newblock \bibinfo{journal}{CHiME Workshop} .
%Type = Article
\bibitem[{Wisdom et~al.(2020)Wisdom, Tzinis, Erdogan, Weiss, Wilson and Hershey}]{wisdom2020unsupervised}
\bibinfo{author}{Wisdom, S.}, \bibinfo{author}{Tzinis, E.}, \bibinfo{author}{Erdogan, H.}, \bibinfo{author}{Weiss, R.J.}, \bibinfo{author}{Wilson, K.}, \bibinfo{author}{Hershey, J.R.}, \bibinfo{year}{2020}.
\newblock \bibinfo{title}{Unsupervised sound separation using mixture invariant training}.
\newblock \bibinfo{journal}{arXiv preprint arXiv:2006.12701} .
%Type = Article
\bibitem[{Wolf and Nadeu(2014)}]{WOLF_EV_2014}
\bibinfo{author}{Wolf, M.}, \bibinfo{author}{Nadeu, C.}, \bibinfo{year}{2014}.
\newblock \bibinfo{title}{Channel selection measures for multi-microphone speech recognition}.
\newblock \bibinfo{journal}{Speech Communication} \bibinfo{volume}{57}.
%Type = Inproceedings
\bibitem[{wen Yang et~al.(2021)wen Yang, Chi, Chuang, Lai, Lakhotia, Lin, Liu, Shi, Chang, Lin, Huang, Tseng, tik Lee, Liu, Huang, Dong, Li, Watanabe, Mohamed and yi~Lee}]{yang21c_interspeech}
\bibinfo{author}{wen Yang, S.}, \bibinfo{author}{Chi, P.H.}, \bibinfo{author}{Chuang, Y.S.}, \bibinfo{author}{Lai, C.I.J.}, \bibinfo{author}{Lakhotia, K.}, \bibinfo{author}{Lin, Y.Y.}, \bibinfo{author}{Liu, A.T.}, \bibinfo{author}{Shi, J.}, \bibinfo{author}{Chang, X.}, \bibinfo{author}{Lin, G.T.}, \bibinfo{author}{Huang, T.H.}, \bibinfo{author}{Tseng, W.C.}, \bibinfo{author}{tik Lee, K.}, \bibinfo{author}{Liu, D.R.}, \bibinfo{author}{Huang, Z.}, \bibinfo{author}{Dong, S.}, \bibinfo{author}{Li, S.W.}, \bibinfo{author}{Watanabe, S.}, \bibinfo{author}{Mohamed, A.}, \bibinfo{author}{yi~Lee, H.}, \bibinfo{year}{2021}.
\newblock \bibinfo{title}{{SUPERB: Speech Processing Universal PERformance Benchmark}}, in: \bibinfo{booktitle}{Interspeech 2021}, pp. \bibinfo{pages}{1194--1198}.
\newblock \DOIprefix\doi{10.21437/Interspeech.2021-1775}.
%Type = Article
\bibitem[{Yang et~al.(2024)Yang, Chang, Huang, Liu, Lai, Wu, Shi, Chang, Tsai, Huang et~al.}]{yang2024large}
\bibinfo{author}{Yang, S.w.}, \bibinfo{author}{Chang, H.J.}, \bibinfo{author}{Huang, Z.}, \bibinfo{author}{Liu, A.T.}, \bibinfo{author}{Lai, C.I.}, \bibinfo{author}{Wu, H.}, \bibinfo{author}{Shi, J.}, \bibinfo{author}{Chang, X.}, \bibinfo{author}{Tsai, H.S.}, \bibinfo{author}{Huang, W.C.}, et~al., \bibinfo{year}{2024}.
\newblock \bibinfo{title}{A large-scale evaluation of speech foundation models}.
\newblock \bibinfo{journal}{IEEE/ACM Transactions on Audio, Speech, and Language Processing} .
%Type = Inproceedings
\bibitem[{Yang et~al.(2021)Yang, Chi, Chuang, Lai, Lakhotia, Lin, Liu, Shi, Chang, Lin et~al.}]{yang2021superb}
\bibinfo{author}{Yang, S.w.}, \bibinfo{author}{Chi, P.H.}, \bibinfo{author}{Chuang, Y.S.}, \bibinfo{author}{Lai, C.I.J.}, \bibinfo{author}{Lakhotia, K.}, \bibinfo{author}{Lin, Y.Y.}, \bibinfo{author}{Liu, A.T.}, \bibinfo{author}{Shi, J.}, \bibinfo{author}{Chang, X.}, \bibinfo{author}{Lin, G.T.}, et~al., \bibinfo{year}{2021}.
\newblock \bibinfo{title}{{SUPERB}: Speech processing universal performance benchmark}, in: \bibinfo{booktitle}{Interspeech}.
%Type = Inproceedings
\bibitem[{Yao et~al.(2023)Yao, Guo, Yang, Kang, Kuang, Yang, Jin, Lin and Povey}]{yao2023zipformer}
\bibinfo{author}{Yao, Z.}, \bibinfo{author}{Guo, L.}, \bibinfo{author}{Yang, X.}, \bibinfo{author}{Kang, W.}, \bibinfo{author}{Kuang, F.}, \bibinfo{author}{Yang, Y.}, \bibinfo{author}{Jin, Z.}, \bibinfo{author}{Lin, L.}, \bibinfo{author}{Povey, D.}, \bibinfo{year}{2023}.
\newblock \bibinfo{title}{Zipformer: A faster and better encoder for automatic speech recognition}, in: \bibinfo{booktitle}{International Conference on Learning Representations}.
%Type = Inproceedings
\bibitem[{Ye et~al.(2023)Ye, Lu, Cheng, Chen, Shang and Li}]{ye2023iacas}
\bibinfo{author}{Ye, L.}, \bibinfo{author}{Lu, H.}, \bibinfo{author}{Cheng, G.}, \bibinfo{author}{Chen, Y.}, \bibinfo{author}{Shang, Z.}, \bibinfo{author}{Li, X.}, \bibinfo{year}{2023}.
\newblock \bibinfo{title}{The iacas-thinkit system for chime-7 challenge}, in: \bibinfo{booktitle}{CHiME Workshop}.
%Type = Inproceedings
\bibitem[{Yoshioka et~al.(2019)Yoshioka, Abramovski, Aksoylar, Chen, David, Dimitriadis, Gong, Gurvich, Huang, Huang et~al.}]{yoshioka2019advances}
\bibinfo{author}{Yoshioka, T.}, \bibinfo{author}{Abramovski, I.}, \bibinfo{author}{Aksoylar, C.}, \bibinfo{author}{Chen, Z.}, \bibinfo{author}{David, M.}, \bibinfo{author}{Dimitriadis, D.}, \bibinfo{author}{Gong, Y.}, \bibinfo{author}{Gurvich, I.}, \bibinfo{author}{Huang, X.}, \bibinfo{author}{Huang, Y.}, et~al., \bibinfo{year}{2019}.
\newblock \bibinfo{title}{Advances in online audio-visual meeting transcription}, in: \bibinfo{booktitle}{Proc. of ASRU}, \bibinfo{organization}{IEEE}. pp. \bibinfo{pages}{276--283}.
%Type = Article
\bibitem[{Yoshioka and Nakatani(2012)}]{yoshioka2012generalization}
\bibinfo{author}{Yoshioka, T.}, \bibinfo{author}{Nakatani, T.}, \bibinfo{year}{2012}.
\newblock \bibinfo{title}{Generalization of multi-channel linear prediction methods for blind mimo impulse response shortening}.
\newblock \bibinfo{journal}{IEEE Transactions on Audio, Speech, and Language Processing} \bibinfo{volume}{20}, \bibinfo{pages}{2707--2720}.
%Type = Inproceedings
\bibitem[{You et~al.(2021)You, Feng, Su and Yu}]{you2021speechmoe}
\bibinfo{author}{You, Z.}, \bibinfo{author}{Feng, S.}, \bibinfo{author}{Su, D.}, \bibinfo{author}{Yu, D.}, \bibinfo{year}{2021}.
\newblock \bibinfo{title}{Speechmoe: Scaling to large acoustic models with dynamic routing mixture of experts}, in: \bibinfo{booktitle}{Interspeech}.
%Type = Inproceedings
\bibitem[{Yu et~al.(2022)Yu, Zhang, Fu, Xie, Zheng, Du, Huang, Guo, Yan, Ma et~al.}]{yu2022m2met}
\bibinfo{author}{Yu, F.}, \bibinfo{author}{Zhang, S.}, \bibinfo{author}{Fu, Y.}, \bibinfo{author}{Xie, L.}, \bibinfo{author}{Zheng, S.}, \bibinfo{author}{Du, Z.}, \bibinfo{author}{Huang, W.}, \bibinfo{author}{Guo, P.}, \bibinfo{author}{Yan, Z.}, \bibinfo{author}{Ma, B.}, et~al., \bibinfo{year}{2022}.
\newblock \bibinfo{title}{{M2MeT}: The {ICASSP} 2022 multi-channel multi-party meeting transcription challenge}, in: \bibinfo{booktitle}{Proc. of ICASSP}.
%Type = Article
\bibitem[{{\.Z}elasko et~al.(2021){\.Z}elasko, Povey, Trmal, Khudanpur et~al.}]{zelasko2021lhotse}
\bibinfo{author}{{\.Z}elasko, P.}, \bibinfo{author}{Povey, D.}, \bibinfo{author}{Trmal, J.}, \bibinfo{author}{Khudanpur, S.}, et~al., \bibinfo{year}{2021}.
\newblock \bibinfo{title}{Lhotse: a speech data representation library for the modern deep learning ecosystem}.
\newblock \bibinfo{journal}{arXiv preprint arXiv:2110.12561} .
%Type = Inproceedings
\bibitem[{Zhang et~al.(2021)Zhang, Negrinho, Ghosh, Jagannathan, Hassanzadeh, Schaaf and Gormley}]{zhang2021leveraging}
\bibinfo{author}{Zhang, L.}, \bibinfo{author}{Negrinho, R.}, \bibinfo{author}{Ghosh, A.}, \bibinfo{author}{Jagannathan, V.}, \bibinfo{author}{Hassanzadeh, H.R.}, \bibinfo{author}{Schaaf, T.}, \bibinfo{author}{Gormley, M.R.}, \bibinfo{year}{2021}.
\newblock \bibinfo{title}{Leveraging pretrained models for automatic summarization of doctor-patient conversations}, in: \bibinfo{booktitle}{Findings of the Association for Computational Linguistics: EMNLP 2021}, pp. \bibinfo{pages}{3693--3712}.
%Type = Inproceedings
\bibitem[{Zhang et~al.(2023)Zhang, Chen, Bukharin, He, Cheng, Chen and Zhao}]{zhangadaptive}
\bibinfo{author}{Zhang, Q.}, \bibinfo{author}{Chen, M.}, \bibinfo{author}{Bukharin, A.}, \bibinfo{author}{He, P.}, \bibinfo{author}{Cheng, Y.}, \bibinfo{author}{Chen, W.}, \bibinfo{author}{Zhao, T.}, \bibinfo{year}{2023}.
\newblock \bibinfo{title}{Adaptive budget allocation for parameter-efficient fine-tuning}, in: \bibinfo{booktitle}{International Conference on Learning Representations}.
%Type = Inproceedings
\bibitem[{Zhong et~al.(2022)Zhong, Liu, Yin, Mao, Jiao, Liu, Zhu, Ji and Han}]{zhong2022towards}
\bibinfo{author}{Zhong, M.}, \bibinfo{author}{Liu, Y.}, \bibinfo{author}{Yin, D.}, \bibinfo{author}{Mao, Y.}, \bibinfo{author}{Jiao, Y.}, \bibinfo{author}{Liu, P.}, \bibinfo{author}{Zhu, C.}, \bibinfo{author}{Ji, H.}, \bibinfo{author}{Han, J.}, \bibinfo{year}{2022}.
\newblock \bibinfo{title}{Towards a unified multi-dimensional evaluator for text generation}, in: \bibinfo{booktitle}{Proceedings of the 2022 Conference on Empirical Methods in Natural Language Processing}, pp. \bibinfo{pages}{2023--2038}.
%Type = Inproceedings
\bibitem[{Zhong et~al.(2021)Zhong, Yin, Yu, Zaidi, Mutuma, Jha, Hassan, Celikyilmaz, Liu, Qiu et~al.}]{zhong2021qmsum}
\bibinfo{author}{Zhong, M.}, \bibinfo{author}{Yin, D.}, \bibinfo{author}{Yu, T.}, \bibinfo{author}{Zaidi, A.}, \bibinfo{author}{Mutuma, M.}, \bibinfo{author}{Jha, R.}, \bibinfo{author}{Hassan, A.}, \bibinfo{author}{Celikyilmaz, A.}, \bibinfo{author}{Liu, Y.}, \bibinfo{author}{Qiu, X.}, et~al., \bibinfo{year}{2021}.
\newblock \bibinfo{title}{Qmsum: A new benchmark for query-based multi-domain meeting summarization}, in: \bibinfo{booktitle}{Proceedings of the 2021 Conference of the North American Chapter of the Association for Computational Linguistics: Human Language Technologies}, pp. \bibinfo{pages}{5905--5921}.
%Type = Article
\bibitem[{{\v{Z}}mol{\'\i}kov{\'a} et~al.(2019){\v{Z}}mol{\'\i}kov{\'a}, Delcroix, Kinoshita, Ochiai, Nakatani, Burget and {\v{C}}ernock{\`y}}]{vzmolikova2019speakerbeam}
\bibinfo{author}{{\v{Z}}mol{\'\i}kov{\'a}, K.}, \bibinfo{author}{Delcroix, M.}, \bibinfo{author}{Kinoshita, K.}, \bibinfo{author}{Ochiai, T.}, \bibinfo{author}{Nakatani, T.}, \bibinfo{author}{Burget, L.}, \bibinfo{author}{{\v{C}}ernock{\`y}, J.}, \bibinfo{year}{2019}.
\newblock \bibinfo{title}{Speakerbeam: Speaker aware neural network for target speaker extraction in speech mixtures}.
\newblock \bibinfo{journal}{IEEE Journal of Selected Topics in Signal Processing} \bibinfo{volume}{13}, \bibinfo{pages}{800--814}.
%Type = Inproceedings
\bibitem[{Zmolikova et~al.(2024)Zmolikova, Merello, Kalgaonkar, Lin, Moritz, Ma, Sun, Chen, Saliou, Petridis et~al.}]{zmolikova2024chime}
\bibinfo{author}{Zmolikova, K.}, \bibinfo{author}{Merello, S.}, \bibinfo{author}{Kalgaonkar, K.}, \bibinfo{author}{Lin, J.}, \bibinfo{author}{Moritz, N.}, \bibinfo{author}{Ma, P.}, \bibinfo{author}{Sun, M.}, \bibinfo{author}{Chen, H.}, \bibinfo{author}{Saliou, A.}, \bibinfo{author}{Petridis, S.}, et~al., \bibinfo{year}{2024}.
\newblock \bibinfo{title}{The {CHiME-8 MMCSG Challenge}: Multi-modal conversations in smart glasses}, in: \bibinfo{booktitle}{CHiME Workshop}.

\end{thebibliography}





% \bibitem[Lamport(1994)]{lamport94}
%   \textit{\LaTeX: a document preparation system},
%   Addison Wesley, Massachusetts,

% \end{thebibliography}
\end{document}